%% file: main.tex
\documentclass[VANCOUVER,Times1COL]{WileyNJDv5} %

\articletype{Research Article}%

\journal{}
\volume{}
\copyyear{}
\startpage{1}

\usepackage{cleveref}
\usepackage{subcaption}
\usepackage{hyphenat}
\usepackage{upgreek}
\usepackage{siunitx}
\usepackage{circuitikz}

\usepackage{color}
\definecolor{colUniBwOr}{rgb}{0.929,0.431,0.0} %
\definecolor{colUniBwGr}{RGB}{113,112,114} %
\definecolor{matlabBlue}{rgb}{0 0.4470 0.7410} %
\definecolor{matlabOrange}{rgb}{0.8500 0.3250 0.0980} %
\definecolor{matlabYellow}{rgb}{0.9290 0.6940 0.1250} %
\definecolor{matlabGreen}{rgb}{0.4660 0.6740 0.1880} %

\usepackage{tikz}
\usepackage{tikz-3dplot}
\usepackage{pgfplots}
\usepackage{pgfplotstable}
\usetikzlibrary{shapes,arrows,calc,fit,backgrounds,shapes.multipart,decorations.pathreplacing}
\usepgfplotslibrary{groupplots}
\pgfplotsset{compat = newest}
\usetikzlibrary{intersections}
\usepgfplotslibrary{fillbetween}

\algdef{SE}[VARIABLES]{Variables}{EndVariables}
   {\algorithmicvariables}
   {\algorithmicend\ \algorithmicvariables}
\algnewcommand{\algorithmicvariables}{\textbf{variables}}

\graphicspath{{figures/}}

\begin{document}

\include{defines.tex}

\title{Smoothed aggregation algebraic multigrid for problems with heterogeneous and anisotropic materials}

\author[1]{Max Firmbach}

\author[2]{Malachi Phillips}

\author[2]{Christian Glusa}

\author[1]{Alexander Popp}

\author[2]{Christopher M. Siefert}

\author[1,3]{Matthias Mayr}

\authormark{Firmbach \textsc{et al.}}
\titlemark{Smoothed aggregation for heterogeneous and anisotropic materials}

\address[1]{\orgdiv{Institute for Mathematics and Computer-Based Simulation}, \orgname{Universit\"{a}t der Bundeswehr M\"{u}nchen}, \orgaddress{\state{Bavaria}, \country{Germany}}}

\address[2]{\orgdiv{Computer Science Research Institute}, \orgname{Sandia National Laboratories}, \orgaddress{\state{New Mexico}, \country{United States}}}

\address[3]{\orgdiv{Data Science \& Computing Lab}, \orgname{Universit\"{a}t der Bundeswehr M\"{u}nchen}, \orgaddress{\state{Bavaria}, \country{Germany}}}

\corres{Corresponding author: Max Firmbach \email{max.firmbach@unibw.de}}

\presentaddress{Institute for Mathematics and Computer-Based Simulation (IMCS), \orgname{Universit\"{a}t der Bundeswehr M\"{u}nchen}, Werner-Heisenberg-Weg 39, 85577 Neubiberg, \orgaddress{\state{Bavaria}, \country{Germany}}}

\abstract[Abstract]{
This paper introduces a material-aware {\soc} measure for smoothed aggregation algebraic multigrid
methods, aimed at improving robustness for scalar partial differential equations with heterogeneous
and anisotropic material properties. Classical {\soc} measures typically rely only on matrix entries
or geometric distances, which often fail to capture weak couplings across material interfaces or align
with anisotropy directions, ultimately leading to poor convergence. The proposed approach directly incorporates
material tensor information into the coarsening process, enabling a reliable detection of weak connections
and ensuring that coarse levels preserve the true structure of the underlying problem.
As a result, smooth error components are represented properly and sharp coefficient jumps or directional
anisotropies are handled consistently. A wide range of academic tests and real-world applications,
including thermally activated batteries and solar cells, demonstrate that the proposed method maintains
robustness across material contrasts, anisotropies, and mesh variations. Scalability and parallel performance
of the algebraic multigrid method highlight the suitability for large-scale, high-performance computing
environments.
}

\keywords{algebraic multigrid, smoothed aggregation, heterogeneous materials, anisotropic materials, {\soc}}

\jnlcitation{\cname{%
\author{Firmbach M},
\author{Phillips M},
\author{Glusa C},
\author{Popp A},
\author{Siefert C}, and
\author{Mayr M}}.
\ctitle{Smoothed aggregation for heterogeneous and anisotropic materials.}
\cjournal{}
\cvol{}.
}

\maketitle

\footnotetext{\textbf{Abbreviations:} AMG, algebraic multigrid; SA, smoothed aggregation; CG, conjugate gradient.}

\section{Introduction}
\label{sec:introduction}

Multimaterial elliptic boundary value problems arise in a variety of different contexts, such as
computational electromagnetics \cite{Bochev2003, Robinson, Robinson2024}, steady-state thermal diffusion \cite{Cole2010, Sheikh2003},
pressure-projection schemes for fluid flow \cite{Chorin1968a,Temam1969a,Patankar1972a},
or thermally activated batteries \cite{Voskuilen2021a}.
In particular, the simulation of thermally activated batteries requires the solution of complex
multiphysics systems. A key subproblem is the computation of the voltage field, governed by Ohm's law,
which presents a challenging scalar multimaterial elliptic boundary value problem due to the highly heterogeneous
material properties involved. For example, the difference in electrical conductivity spans as many as ten
orders of magnitude between the cathode, separator, and anode layers. These sharp material interfaces are
then repeated many times throughout the battery stack, requiring a fine resolution and thus many elements
in the stackwise direction. This often leads to meshes that exhibit a high degree of anisotropy.
Therefore, solution methods that are robust with respect to mesh anisotropy and heterogeneous material
interfaces are required to solve this type of problem efficiently.

In this paper, we specifically consider the material-weighted Laplacian problem on a domain $\dom \subset \REalSp{\ndim}$,
\begin{align}
\begin{split}
-\divergence{(\material(\coord)\grad\unknown)} &= \anyRhs \quad \quad \text{in} \; \dom, \\
u &= g \quad \quad \text{on} \; \indexedDirichlet{\boundary}, \\
-(\material(\coord)\grad\unknown)\cdot n &= h \quad \quad \text{on} \; \indexedNeumann{\boundary},
\end{split}
\label{eq:model_problem}
\end{align}
where $\Gamma_D$ and $\Gamma_N$ denote the Dirichlet and Neumann parts of the boundary \(\partial\Omega=\indexedDirichlet{\boundary}\cup\indexedNeumann{\boundary}\), respectively.
The material tensor $\material(\coord)\in \REalSp{\ndim\times\ndim}$, which can vary
substantially over space and might feature anisotropies, is assumed to be symmetric and positive definite.
In preparation for a finite element discretization, we transform \eqref{eq:model_problem} into the associated
variational formulation: Find
$\unknown \in H^1_D(\dom):=\{v\in H^{1}(\dom) \mid \operatorname{trace}_{\indexedDirichlet{\boundary}}v=g\}$
such that
\begin{equation}
a(\unknown,\test) = b(\test) \quad \quad \forall \test \in H^1_0(\dom):=\{v\in H^{1}(\dom) \mid \operatorname{trace}_{\indexedDirichlet{\boundary}}v=0\},
\end{equation}
with the bilinear form $a(\unknown,\test)$ and the linear functional $b(\test)$ reading
\begin{equation}
\quad a(\unknown,\test) = \int_{\dom} (\material(\coord)\grad\unknown) \cdot \grad\test \; \mathrm{d}\coord \quad \text{and} \quad b(\test) = \int_{\dom} \anyRhs\test \; \mathrm{d}\coord.
\end{equation}
Discretizing the problem with finite elements yields the linear system of equations to be solved
\begin{equation}
A\unknown=b,
\label{eq:linear_system}
\end{equation}
with the stiffness matrix $A$, the right-hand side load vector $b$ and the discrete solution vector $\unknown$
in the finite element space $V_h(\dom)$. For simplicity of notation, we omit specific highlighting of discrete
quantities. While the presented methods are applicable to arbitrary meshes, we solely focus on discretizations
based on quadrilateral meshes in 2D and hexahedral meshes in 3D.

Algebraic multigrid (AMG) methods \cite{BrMcRu84,RuSt85} have been successfully applied to our model problem, but can
struggle in the case of a highly heterogeneous or highly anisotropic material tensor $\sigma(\coord)$.  In these
cases, the AMG method tends to generate poor coarse grids by not resolving the material properties of the underlying
problem correctly, which then leads to a poorly converging method. While solution approaches other than AMG for highly
heterogeneous materials exist, based on special coarse spaces used for domain decomposition methods \cite{Heinlein, Kim2017}
or block preconditioning \cite{Fang2019}, changing the coarse grids generated by AMG to better accommodate the materials is
arguably a more common and user-friendly approach in practical applications. These methods often use
a so-called {\soc} measure to identify and drop ``unimportant'' entries of matrix $A$ as a way of improving coarse grid quality.
While being a fundamental part of the coarsening process of AMG, there is no ideal {\soc} measure that is suitable in all situations.
The ``classical'' {\soc} approaches are based directly on the matrix stencil itself \cite{BrMcRu84,RuSt85} to determine a
strong coupling of neighboring nodes. For smoothed-aggregation (SA) AMG the strongly coupled neighborhood is defined in a
similar manner \cite{Vanek1996a}, also using matrix entries to
determine weak connections. As these measures are generally designed for
$M$-matrices with a dominant diagonal, the convergence of AMG for our
model problem usually degrades quickly, as increasing
material contrast pushes $A$ away from being an $M$-matrix.
To overcome these issues, energy-based approaches have been introduced \cite{Brannick2006}, which try to measure
the ``sizes'' of entries in the inverse of $A$. An extension to those methods is given by taking into account the evolution of
a $\delta$-function to form the {\soc} measure \cite{Olson2010a}. These metrics capture anisotropy better, yet add computational
complexity to the overall multigrid algorithm. In the case of stretched meshes, methods based on geometric distances have been
proposed to mimic the coarsening of geometric multigrid methods \cite{ddproc06,ML}, while other approaches depend on an
algebraic distance calculation to handle anisotropy properly \cite{Brandt2015a}.

This paper presents a material-based
coarsening approach as an alternative to the existing {\soc} measures. Rather than restricting the {\soc} calculation solely to the matrix
entries, this approach draws inspiration from methods such as the {\DistanceLaplacian} {\soc} measure \cite{ddproc06,ML},
which uses auxiliary information (nodal coordinates in the case of the {\DistanceLaplacian}) to improve coarse grid quality.
By using the material tensor of the underlying problem as distance metric for the {\DistanceLaplacian}, we are able to construct
a {\soc} measure that is ``material-aware''. The proposed method is able to detect material jumps and anisotropy directions,
while being computationally lightweight.

This paper is structured as follows. A short introduction to AMG and a brief review of the state of the art and frequently used {\soc}
measures will be given in \secref{sec:sa_multigrid}. In addition, we discuss the interplay with smoothed aggregation considering
prolongator smoothing and matrix filtering. \secref{sec:new_dropping_scheme} introduces a material-aware {\soc} measure build on a
{\DistanceLaplacian} with a material tensor based distance metric. Computational experiments on a variety of academic test problems as
well as real-world applications representing thermally activated batteries and solar cells can be found in \secref{sec:numerical_results}.
We compare the proposed method to classical SA-AMG, investigate parameter robustness and show weak and strong scalability of the
multigrid method. \secref{sec:conclusion} will summarize our findings and hint at future research directions.

\section{Smoothed aggregation algebraic multigrid methods}
\label{sec:sa_multigrid}

A multigrid V-cycle, as given in Algorithm~\ref{alg:Vcycle},
is used to solve a linear system of equations as stated in \eqref{eq:linear_system}.
This requires the construction of a hierarchy of matrices
$\indexedLevel{A}{\level}\in\REalSp{\indexedLevel{n}{\level}\times\indexedLevel{n}{\level}}$
where $\level$ designates the level of the V-cycle and
$\indexedLevel{n}{\level}$ denotes the number of unknowns on level~$\level$.
We define a relaxation-based smoothing method $\indexedLevel{\smoother}{\level}$ with
pre- and post-iterations on levels $\level = 1, ..., \noLevels$, with
the coarsest level, $\level=L$,  being solved
directly. In addition, the rectangular transfer operators of size
$\indexedLevel{n}{\level}\times\indexedLevel{n}{\level+1}$ are specified as
$\indexedLevel{P}{\level}:\REalSp{\indexedLevel{n}{\level+1}}\rightarrow\REalSp{\indexedLevel{n}{\level}}$
on $\level = 1, ..., \noLevels-1$, which interpolate the solution updates from level $\level+1$
to level $\level$, while $\trans{\indexedLevel{P}{\level}}:\REalSp{\indexedLevel{n}{\level}}\rightarrow\REalSp{\indexedLevel{n}{\level+1}}$
restrict residuals from level $\level$ to level $\level+1$. For
AMG, which we consider herein,
the coarse level operators
$\indexedLevel{A}{\level+1}$ are constructed through the Galerkin product
\begin{equation}
\indexedLevel{A}{\level+1} = \trans{\indexedLevel{P}{\level}}\indexedLevel{A}{\level}\indexedLevel{P}{\level} \quad \text{for} \quad 1 \leq \level < L,
\end{equation}
with \(\indexedLevel{A}{1}:=A\) denoting the matrix of the original linear
system \eqref{eq:linear_system}.

\begin{algorithm}
\caption{Multigrid V-cycle with $\noLevels$ levels to solve $\indexedLevel{A}{1}u=b$.}
\label{alg:Vcycle}
\begin{algorithmic}
\Procedure{Vcycle}{$\indexedLevel{A}{\level}$,$\indexedLevel{u}{\level}$,$\indexedLevel{b}{\level}$,$\level$}
\If{$\level \neq \noLevels$}
\State $\indexedLevel{u}{\level}=\indexedLevel{\smoother_{\text{pre}}}{\level}(\indexedLevel{A}{\level},\indexedLevel{u}{\level},\indexedLevel{b}{\level})$
\Comment pre-smoothing
\State $\text{Vcycle}(\indexedLevel{A}{\level+1}, \trans{\indexedLevel{P}{\level}}(\indexedLevel{b}{\level}-\indexedLevel{A}{\level}\indexedLevel{u}{\level}), 0, \level+1)$
\State $\indexedLevel{u}{\level}=\indexedLevel{u}{\level}+\indexedLevel{P}{\level}\indexedLevel{u}{\level+1}$
\Comment prolongation
\State $\indexedLevel{u}{\level}=\indexedLevel{\smoother_{\text{post}}}{\level}(\indexedLevel{A}{\level},\indexedLevel{u}{\level},\indexedLevel{b}{\level})$
\Comment post-smoothing
\Else
\State $\indexedLevel{u}{\level}=\inv{\indexedLevel{A}{\level}}\indexedLevel{b}{\level}$
\Comment coarse solve
\EndIf
\EndProcedure
\end{algorithmic}
\end{algorithm}
Constructing the prolongators \(\indexedLevel{P}{\level}\) from
\(\indexedLevel{A}{\level}\) and providing auxiliary quantities is the
core task of AMG. For smoothed aggregation AMG, we
begin by defining a set of aggregates,
$\indexedLevel{\aggregate}{\level}$, which form a disjoint covering of the index set of all nodes on
the current level~$\level$ following~\cite{Vanek1996a,Vanek2001a}
\begin{equation}
\bigcup_{i=1}^{\indexedLevel{n}{\level+1}} \indexedLevel{\aggregate}{\level}_{i} = \{1, ..., \indexedLevel{n}{\level}\}.
\end{equation}

The aggregates are chosen by first creating a {\soc} matrix $\strength(\indexedLevel{A}{\level})$,
derived from the matrix $\indexedLevel{A}{\level}$ at that level~$\level$. A dropping
criterion $\dropCriterion$ is then applied to $\strength(\indexedLevel{A}{\level})$ to create a filtered graph,
$\graph_C$, at that level.  These two components are often
conflated in the literature (and collectively referred to as {\soc} approach),
but for the purpose of this work, we separate them. Both the {\soc} measure
and the dropping criterion will be discussed in \secref{sec:soc}.
Once we have the filtered graph $\graph_C$, we apply an
aggregation algorithm to it.
The simplest aggregation algorithm can be stated as: ``Find a node that does not
neighbor any node in an existing aggregate and call that and all of
its neighbors an aggregate.''  There are subtleties here (e.g. what to do
with leftover nodes) which will not be discussed in this paper, but
are discussed in detail elsewhere \cite{Vanek1996a}.

Once the aggregates are chosen, a tentative prolongator,
$\indexedLevel{\approxMat{P}}{\level}$, is constructed
such that  it accurately interpolates the near null space~$\indexedLevel{B}{\level}\in\REalSp{\indexedLevel{n}{\level}\times 1}$
of operator~$\indexedLevel{A}{\level}$.  Thus,
$\indexedLevel{A}{\level}\indexedLevel{B}{\level}\approx0$ holds on each level. The entries of the tentative prolongator
$\indexedLevel{\approxMat{P}}{\level} = \indexedLevel{\approxMat{P}}{\level}_{ij}$
are determined by partitioning the near null space over the aggregates.
For the considered scalar elliptic model problem, the near null space is a vector containing only ones, resulting in tentative
prolongator entries of either one or zero and thus a piecewise constant interpolation,
\begin{equation}
\indexedLevel{\approxMat{P}}{\level}_{ij} =
\begin{cases}
1 \quad \text{if} \; i \in \indexedLevel{\aggregate}{\level}_{j}, \\
0 \quad \text{otherwise}.
\end{cases}
\end{equation}
Finally, the tentative prolongator undergoes smoothing (hence the name
``smoothed aggregation''), which is described in more detail in \secref{sec:smoothing_filtering},

\subsection{Strength-of-connection measures and dropping criteria}
\label{sec:soc}
For isotropic problems, it is common to apply the aggregation algorithm directly to the
graph~$\graphOf{\indexedLevel{A}{\level}}$ of the matrix at any given level~$\level$,
and ignore both {\soc} measure and dropping criterion. However, this generally does not work well for anisotropic or
heterogeneous problems. In an anisotropic setting, the geometrically smooth error components
follow the strong direction, whereas the error in the weak direction shows oscillations. Applying a smoother
results in poor convergence, as only the smooth error components are reduced efficiently. A more detailed
discussion will be given in \secref{sec:new_dropping_scheme}. This behavior can be improved by instead using
a modified graph with ``unimportant'' entries removed, $\graph_C(\indexedLevel{A}{\level}) := \graphOf{\dropCriterion(\strength(\indexedLevel{A}{\level}))}$,
where $\strength(\cdot)$ is the {\soc} function and $\dropCriterion(\cdot)$ is the dropping criterion function.
In the anisotropic case, connections that are not aligned with the anisotropy direction are considered weak and the
respective entries are deemed to be unimportant.
The resulting operator, $\dropCriterion(\strength(A))$, is a matrix of
ones and zeroes which matches the sparsity pattern of the
corresponding operator~$\indexedLevel{A}{\level}$ at that level,
but with the respective entries removed.

The most common dropping criterion is what is sometimes referred to as
pointwise dropping.
For a given {\soc} matrix $\strength$, and a user-chosen drop
tolerance \(0\leq\dropTol\leq1\), pointwise dropping is defined as
\begin{equation}
\dropCriterion^{\text{pw}}(\strength)_{ij} =
\begin{cases}
1 \quad \text{if} \; |\strength_{ij}| \geq \dropTol, \\
0 \quad \text{otherwise}.
\end{cases}
\end{equation}
This gives a criterion where each entry is considered independently.

An alternative dropping criterion is the cut-drop criterion \cite{cut-drop}
which works row-by-row rather than
entry-by-entry.

Let \(j(i,\cdot)\) be a permutation that orders the
entries of the \(i\)-th row of \(\strength\) in descending
order, i.e. \(|\strength_{i,j(i,1)}|\geq |\strength_{i,j(i,2)}| \geq \dots \geq |\strength_{i,j(i,n)}|\), and let
\(k(i)\in \{1,\dots,n-1\}\) be the smallest index such that \(\dropTol |\strength_{i,j(i,k(i))}| \geq |\strength_{i,j(i,k(i)+1)}|\)
and set \(\hat{j}(i)=j(i,k(i))\). The cut-drop criterion is defined as
\begin{equation}
\dropCriterion^{\text{cut-drop}}(\strength)_{ij} =
\begin{cases}
1 \quad \text{if} \; |\strength_{ij}| \geq |\strength_{i,\hat{j}(i)}|, \\
0 \quad \text{otherwise}.
\end{cases}
\end{equation}
In other words, the cut-drop criterion orders all entries $\strength_{ij}$ of row $i$ in descending order,
identifies the first pair of adjacent ordered entries with a gap larger than $\dropTol$, and then drops all
values after this gap.

Traditional smoothed aggregation\cite{Vanek1996a} uses
\(\strength^{sa}(A) =D^{-1/2}AD^{-1/2}\) with matrix diagonal
\(D:=\diag{A}\) as {\soc} measure, coupled with the pointwise dropping criterion.
If we expand these expressions for \(\dropCriterion^{\text{pw}}(\strength^{sa})\), we recover the more familiar form
\begin{equation}
  \frac{|A_{ij}|}{\sqrt{|A_{ii}| |A_{jj}|}} \geq \dropTol.
\end{equation}
While this criterion is robust in many situations, its application to highly heterogeneous and anisotropic material coefficients remains a challenge~\cite{Olson2010a}.

Alternatively, the {\DistanceLaplacian} \cite{ddproc06,ML}, can be used as a {\soc} measure.
It uses a distance metric~\(\distance\) associated with a sparse matrix~\(A\) and
coordinates~\(\{\coord_{i} \in\mathbb{R}^{\ndim}\}_{j}\), reading
\begin{equation}
L_{\distance}(A, \coord)_{ij} :=
\begin{cases}
-\frac{1}{\distance(\coord_{i},\coord_{j})^{2}} & \text{if } i\neq j \text{ and } A_{ij}\neq 0,\\
0 & \text{if } i\neq j \text{ and } A_{ij}= 0,\\
-\sum_{j\neq i}L_{\distance}(A, \coord)_{ij} & \text{if } i=j.
\end{cases}
\end{equation}
Thus, \(L_{\distance}(A, \coord)\) matches the pattern of \(A\),
while its values are given in terms of the distance between the
associated coordinates.
If~\(A\) has a symmetric sparsity pattern, then \(L_{\distance}(A, x)\) is symmetric due to the fact
that \(\distance(\coord_{i},\coord_{j})=\distance(\coord_{j},\coord_{i})\).
The {\DistanceLaplacian} {\soc} function is then defined as \(\strength^{dlap}=D^{-1/2}LD^{-1/2}\)
where \(L=L_{\normTwo{\cdot}}(A, x)\), \(D=\diag{L}\) and \(x\) are the coordinates of the
degrees of freedom of the finite element discretization \cite{Gee2009a}.
When combined with pointwise dropping, the {\DistanceLaplacian} \(\dropCriterion^{\text{pw}}(\strength^{dlap})\) has already shown to work
well for problems related to stretched meshes and coefficient jumps~\cite{Hu2022a}, yet lacks material information due to the fact that only
the sparsity pattern of \(A\) is used, but not its values.

\subsection{Prolongator smoothing and matrix filtering}
\label{sec:smoothing_filtering}

Traditional smoothed aggregation introduces a smoothing procedure for the tentative prolongator \(\indexedLevel{\approxMat{P}}{\level}\)
to reduce the energy of the coarse level basis functions~\cite{Vanek1996a,Vanek2001a}
with the goal of improving the convergence
properties of the algebraic multigrid method. To this end, one step of a weighted Jacobi iteration is applied to the tentative
prolongator, which yields the final, smoothed transfer operator
\begin{equation}
\indexedLevel{P}{\level} = (I-\omega\inv{D}\indexedLevel{A}{\level})\indexedLevel{\approxMat{P}}{\level} \label{eq:operator_smoothing}
\end{equation}
with $D=\diag{A}$ and a damping factor $\omega \geq 0$. As our model problem is of symmetric nature, the
damping factor is usually chosen to be $\omega = \frac{\omega_{sym}}{\tilde{\lambda}}$ with $\tilde{\lambda} \geq \rho(\inv{D}A)$
as approximation of the spectral radius, determined by a few
iterations of the power method and $\omega_{sym}=\frac{4}{3}$ \cite{Vanek2001a}.

The smoothing step given in \eqref{eq:operator_smoothing} increases the density of \(\indexedLevel{P}{\level}\) compared
to \(\indexedLevel{\approxMat{P}}{\level}\). This density growth is propagated to coarse level operators via the Galerkin product.
The so-called operator complexity
\begin{equation}
\opComplexity = \frac{\sum_{\level=1}^{\noLevels}\text{nnz}(\indexedLevel{A}{\level})}{\text{nnz}(\indexedLevel{A}{1})}
\end{equation}
measures the number of nonzeros (nnz) across all levels of the hierarchy relative to the number of nonzeros of the fine-level
system matrix. Naturally, \(\opComplexity\geq1\).
A smaller operator complexity is desirable as it corresponds to the
computational cost of  operators within the AMG preconditioner.
In cases with many weak connections, as they appear in anisotropic and heterogeneous problems, and thus a great number of entries removed
from $\graphOf{\indexedLevel{A}{\level}}$, the prolongator smoothing introduces undesired overlap of basis
functions between non-neighboring aggregates. This results in an prohibitively high fill-in and therefore a unnecessary high
operator complexity \cite{Gee2009a, Hu2022a}, which can make the multigrid method infeasible to apply.
In order to circumvent this problem and reduce the operator complexity, one can replace the matrix \(\indexedLevel{A}{\level}\) used in
the smoothing step \eqref{eq:operator_smoothing}
with a filtered matrix $\indexedLevel{A}{\level}_{F}:=\mathcal{F}_{\dropCriterion}(\indexedLevel{A}{\level})$.
This filtered matrix is constructed from the dropping criterion $\dropCriterion$:
\begin{equation}
  \mathcal{F}_{\dropCriterion}(\indexedLevel{A}{\level})_{ij} =
  \begin{cases}
    \indexedLevel{A}{\level}_{ij} & \text{if } i\neq j \text{ and } \dropCriterion_{ij} = 1, \\
    0 & \text{if } i\neq j \text{ and } \dropCriterion_{ij} = 0, \\
    \indexedLevel{A}{\level}_{ii} + \sum_{k \neq i} (\indexedLevel{A}{\level}_{ik}-\mathcal{F}_{\dropCriterion}(\indexedLevel{A}{\level})_{ik}) & \text{if } i=j.
  \end{cases}
  \label{eq:filtered_matrix}
\end{equation}
The diagonal entries of $\indexedLevel{A}{\level}_{F}$ are modified by lumping off-diagonal entries such that the sum of entries
in a row of the filtered matrix is zero whenever the sum of the entries in a row of $\indexedLevel{A}{\level}$ is zero. This is
necessary to maintain the correct interpolation of the near null space after prolongator smoothing. The final prolongator using
the filtered matrix for smoothing is given by
\begin{equation}
\indexedLevel{P}{\level} = (I-\omega\inv{D_{F}}\indexedLevel{A}{\level}_{F})\indexedLevel{\approxMat{P}}{\level}
\end{equation}
with $D_{F}:=\diag{A_{F}}$.
However, in the case of excessive dropping, the lumping procedure
\eqref{eq:filtered_matrix} can yield near-zero or even
negative diagonal entries in $A_{F}$.  This can render the prolongator smoothing procedure ineffective or
even counterproductive. One way to circumvent this problem is the introduction of a diagonal matrix $\approxMat{D}_{F} = (\approxMat{D}_F)_{ii}$
based on the absolute row sum of $A_{F}$ defined as
\begin{equation}
(\approxMat{D}_F)_{ii} = \sum_j |(A_F)_{ij}|.
\label{eq:1_norm_diagonal}
\end{equation}
The expression given in \eqref{eq:1_norm_diagonal} can be interpreted as a 1-norm approximation of the diagonal, which better captures
the scaling of an entire row and thus will be less sensitive to the diagonal dominance properties of the filtered matrix~\cite{Hu2022a}.
While distributed lumping techniques preserve row sums by keeping nonzero entries in proportion to their respective magnitudes, we will
not consider such approaches in the scope of this publication.

\section{A material based Strength-of-connection measure}
\label{sec:new_dropping_scheme}

We now turn to the construction of a novel strength-of-connection measure designed to overcome the limitations of the approaches
discussed in \secref{sec:soc}, particularly with respect to material heterogeneity and anisotropy.

Consider our model problem \eqref{eq:model_problem} on a unit square $\dom = \{(x,y)\in\REalSp{2}:-1<x<1,-1<y<1\}$
with the following material tensor\cite{Falgout2006a}

\begin{align}
\material(\coord) &=
\begin{cases}
  I &  \text{for} \; \xaxis<0,
\\[10pt]
\left(
  \begin{array}{cc}
    \kappa&\\
    &1
  \end{array}
  \right)
&  \text{for} \; \xaxis \geq 0.
\end{cases}
\label{eq:academic_test_problem}
\end{align}

\begin{figure}
\centering
\begin{subfigure}[b]{0.49\textwidth}
\centering
\includegraphics[width=\textwidth]{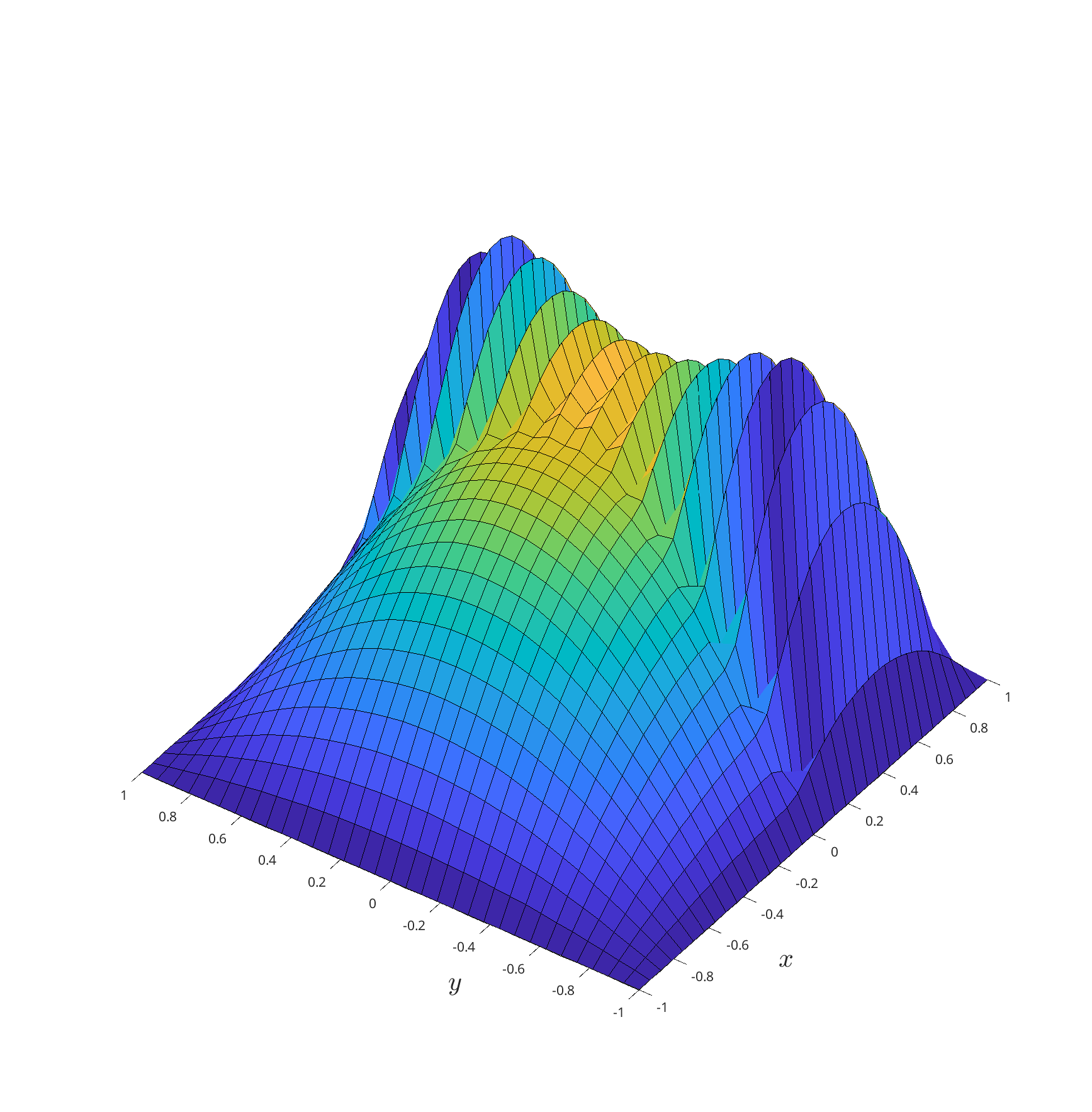}
\caption{\footnotesize Error $e_k$ of the solution on $\Omega$.}
\label{fig:domain_error}
\end{subfigure}
\hfill
\begin{subfigure}[b]{0.49\textwidth}
	\begin{subfigure}[b]{0.49\textwidth}
	\centering
    \includegraphics[width=\textwidth]{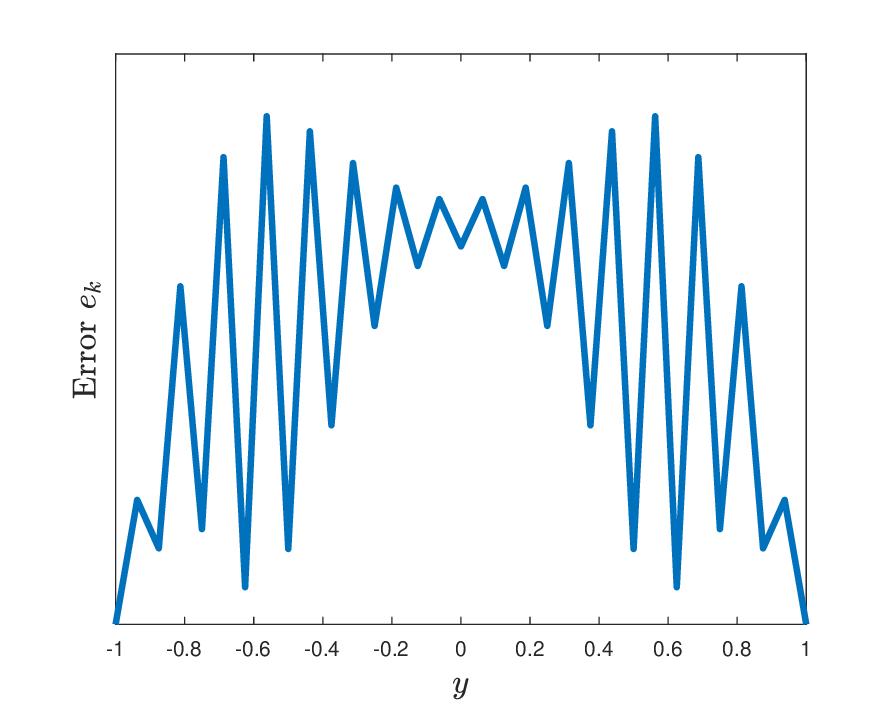}
    \caption{\footnotesize Error $e_k$ at $x=0.5$.}
    \label{fig:y_error_at_x_0_5}
	\end{subfigure}
	\begin{subfigure}[b]{0.49\textwidth}
    \centering
    \includegraphics[width=\textwidth]{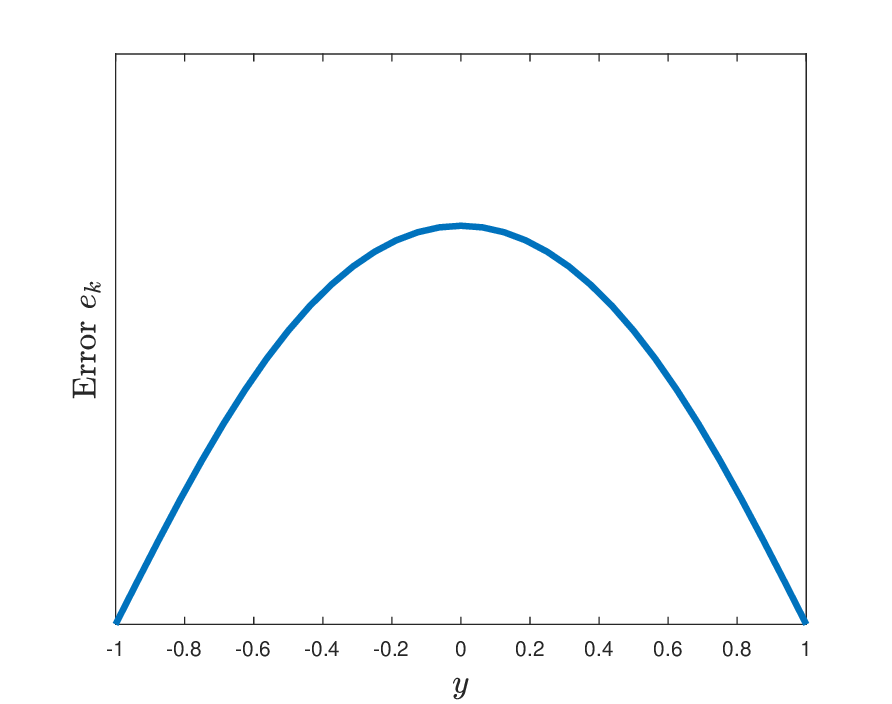}
    \caption{\footnotesize Error $e_k$ at $x=-0.5$.}
    \label{fig:y_error_at_x_-0_5}
	\end{subfigure}
	\\
	\begin{subfigure}[b]{0.49\textwidth}
	\centering
    \includegraphics[width=\textwidth]{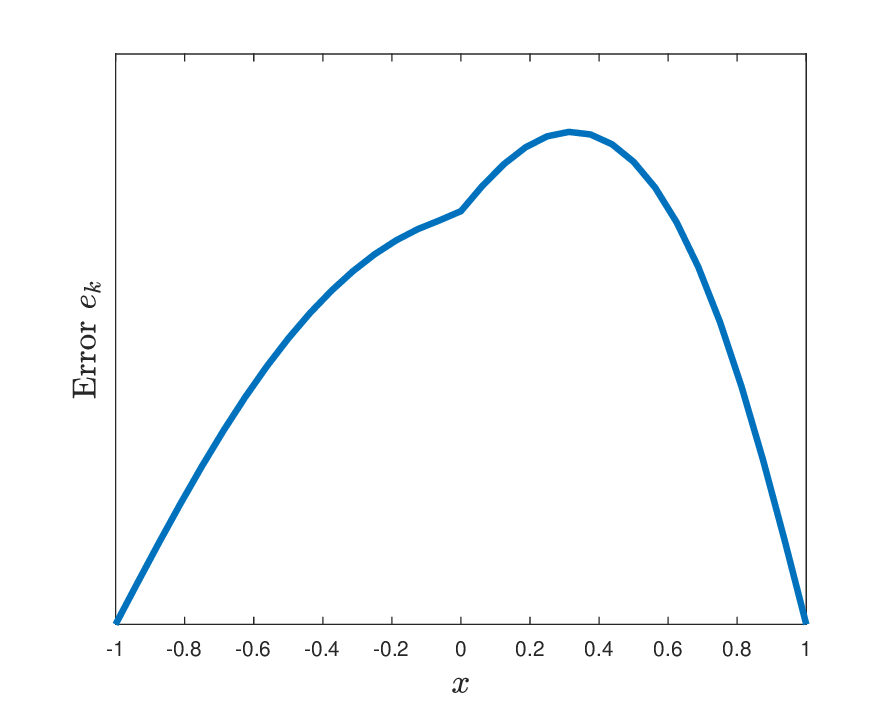}
    \caption{\footnotesize Error $e_k$ at $y=0.0$.}
    \label{fig:x_error_at_y_0_0}
	\end{subfigure}
	\begin{subfigure}[b]{0.49\textwidth}
	\centering
    \includegraphics[width=\textwidth]{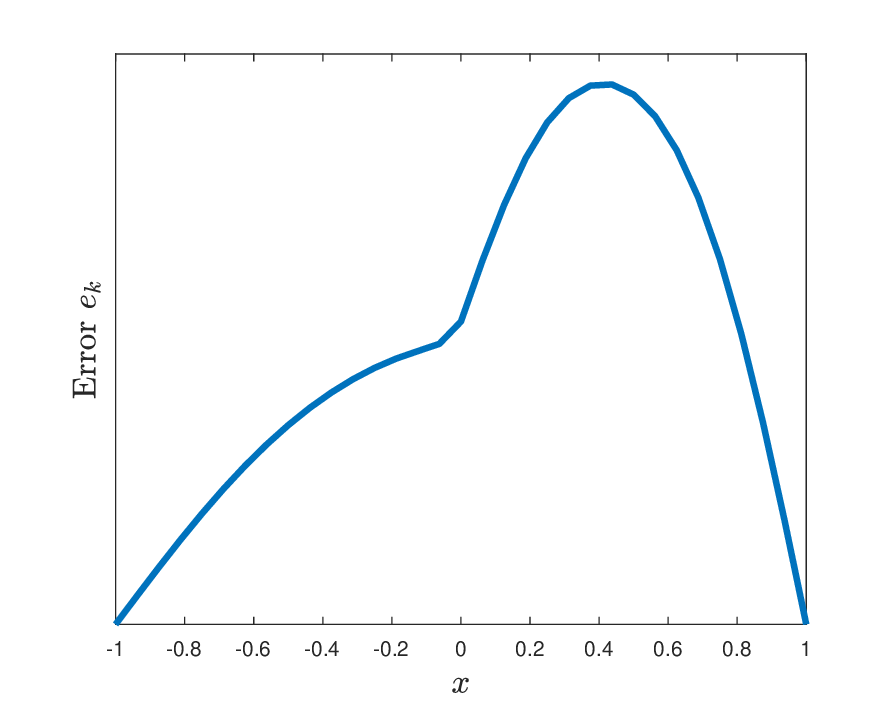}
    \caption{\footnotesize Error $e_k$ at $y=0.5$.}
    \label{fig:x_error_at_y_0_5}
	\end{subfigure}
\end{subfigure}
\caption{Visual representation of the error $e_k$ over the domain $\Omega$ with directional components at different
$x$- and $y$-coordinates after a few steps $k$ of a Jacobi iteration.}
\label{fig:error}
\end{figure}

In the case of a large material contrast $\kappa$ the $\xaxis$-direction is weakly coupled to the $\yaxis$-direction for $\xaxis \geq 0$,
and the subdomains on either side of $x=0$ are weakly connected to
each other.
This causes the smooth error $e_k$ to align with the anisotropy direction while becoming non-smooth at $x=0$. Pointwise
smoothers like a Jacobi or Gauss-Seidel iteration are therefore inefficient in reducing the oscillatory error component in the weakly
coupled directions. \figref{fig:error} illustrates this behavior showing a smooth error for $\xaxis<0$, a discontinuity in
the derivative of the error at $x=0$ and an oscillatory error in $\yaxis$-direction for $\xaxis \geq 0$. An effective {\soc} measure therefore
needs to detect the weak connections related to the material properties and must drop these unimportant entries from
$\graphOf{\indexedLevel{A}{\level}}$, resulting in a coarsening in direction of smooth error components. While this behavior
might hold true in some cases for the {\soc} measures introduced in \secref{sec:soc}, it is not guaranteed, as these measures
solely rely on matrix entries or coordinate information, but do not make explicit use of the material property of the
underlying problem.

To the authors' knowledge, material heterogeneity as well as anisotropy information have not been included in the calculation of
{\soc} measures and thus in the construction of the multigrid hierarchy so far. The novelty of this work consists in taking into
account the material tensor into the {\soc} calculation to remove weak connections due to material property changes. For a constant
tensor coefficient \(\material\), the Green's function \(G(x,y)\) associated with the scalar elliptic problem
\eqref{eq:model_problem} can be stated as
\begin{align}
  G(x,y)
  &=
    \begin{cases}
      \frac{1}{2\pi \sqrt{\operatorname{det}(\material)}}\log \frac{1}{\|x-y\|_{\material}} & \text{if } \ndim=2, \\
      \frac{1}{4\pi \sqrt{\operatorname{det}(\material)}} \frac{1}{\|x-y\|_{\material}}& \text{if } \ndim\geq2,
    \end{cases}
\label{eq:greens_function}
\end{align}
with $\|x-y\|_{\material} = \sqrt{(x-y)^{T}\material^{-1}(x-y)}$.
This motivates the definition of a material-weighted distance for spatially variant \(\material\)
\begin{align*}
  \distance_{\material}(x,y) &:= \max\{\tilde{\distance}_{\material}(x,y),\tilde{\distance}_{\material}(x,y)\}
                               &\text{with}&&\tilde{\distance}_{\material}(x,y) := \sqrt{(x-y)^{T}\material(x)^{-1}(x-y)}.
\end{align*}
We will assume for the moment that \(\sigma\) allows for pointwise evaluation.
The distance \(\distance_{\material}\) can then be used in the weighted {\DistanceLaplacian} $L_{\material}:=L_{\distance_{\material}}(A,\coord)$, where again \(\coord\) are the coordinates
of the degrees of freedom of the discretization.
Similar to the classical {\DistanceLaplacian} dropping criterion, we use {\soc} \(\strength_{\material}^{dlap}:=D_{\material}^{-1/2}L_{\material}D_{\material}^{-1/2}\) with diagonal \(D_{\material}=\diag{L_{\material}}\) and \(\dropCriterion^{\text{pw}}(\strength_{\material}^{dlap})\) or \(\dropCriterion^{\text{cut-drop}}(\strength_{\material}^{dlap})\).

We note that there are different ways of symmetrizing the function \(\tilde{\distance}_{\material}(x,y)\) to obtain a distance.
Consider the following simple scenario to illustrate why using the maximum is preferable.
Assume a regular mesh and a scalar material coefficient \(\sigma\) and that \(\coord_{i_{\text{low}}}\) is in a region of low coefficient
\(\material_{\text{low}}\) and \(\coord_{i_{\text{high}}}\) is in a region of high coefficient
\(\material_{\text{high}}\) and that degrees of freedom \(i_{\text{low}}\) and \(i_{\text{high}}\) are connected by an edge,
as illustrated in \figref{fig:model_problem_discretization}, taking into account node 12 and 13.

\input{figures/test_problem}

Then
\begin{align*}
  \tilde{\distance}_{\material}(x_{i_{\text{low}}},\cdot) &\sim \material_{\text{low}}^{-1/2},
  &  \tilde{\distance}_{\material}(x_{i_{\text{high}}},\cdot) &\sim \material_{\text{high}}^{-1/2},
  & \distance_{\material}(x_{i_{\text{low}}},x_{i_{\text{high}}}) &\sim \max\{\material_{\text{low}}^{-1/2},\material_{\text{high}}^{-1/2}\}\sim \material_{\text{low}}^{-1/2},\\
  |(L_{\material})_{i_{\text{low}}i_{\text{low}}}| &\sim \material_{\text{low}},
  & |(L_{\material})_{i_{\text{high}}i_{\text{high}}}| &\sim \material_{\text{low}}+\material_{\text{high}} ,
  & |(L_{\material})_{i_{\text{low}}i_{\text{high}}}| &\sim \material_{\text{low}} ,
\end{align*}
and hence
\begin{align}
(\strength_{\material}^{dlap})_{i_{\text{low}}i_{\text{high}}}
=\frac{|(L_{\material})_{i_{\text{low}}i_{\text{high}}}|}{\sqrt{|(L_{\material})_{i_{\text{low}}i_{\text{low}}} |(L_{\material})_{i_{\text{high}}i_{\text{high}}}|}}
\sim \frac{\material_{\text{low}}}{\sqrt{\material_{\text{low}}} \sqrt{\material_{\text{low}} + \material_{\text{high}}} }
\sim \kappa^{-1/2}
&&\text{with}&&\kappa:=\frac{\material_{\text{high}}}{\material_{\text{low}}}.
\label{eq:low_high_soc_parameter_relation}
\end{align}
By choosing a symmetrization of \(\tilde{\distance}_{\material}\) that gives \((L_{\material})_{i_{\text{low}}i_{\text{high}}}\sim \material_{\text{low}}\)
we encourage dropping of edges between low and high coefficient regions, provided the value of the drop tolerance \(\dropTol\) is small enough compared to the coefficient contrast \(\kappa\).
Subsequent aggregation algorithms will therefore not create aggregates that straddle the interface between low and high coefficient regions.

\begin{figure}
\centering
\begin{subfigure}[t]{0.32\textwidth}
\centering
\includegraphics[scale=0.19]{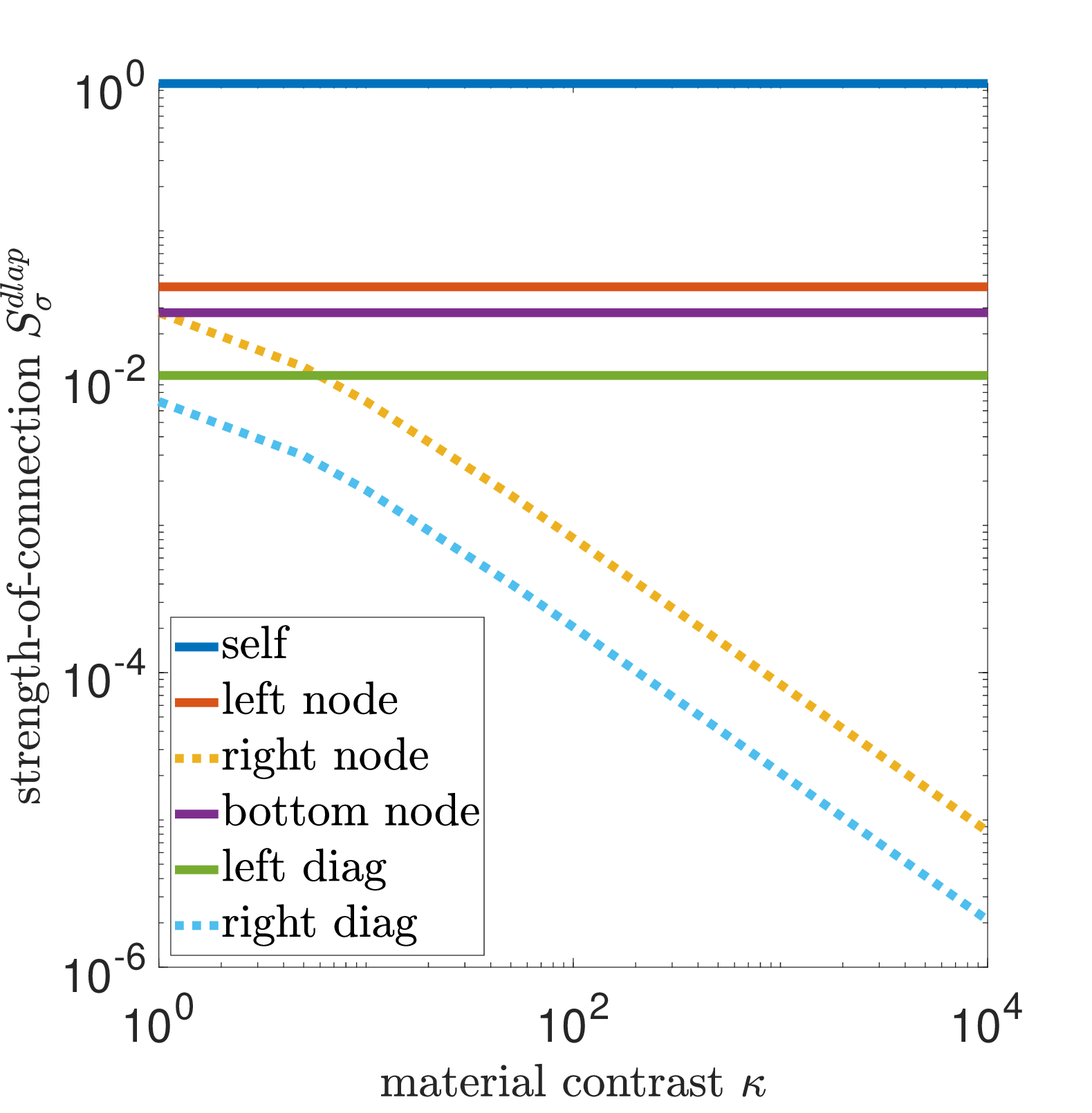}
\caption{\footnotesize $\strength^{dlap}_{\material}$ of node 12 in the low material parameter region $\material_{\text{low}}$.}
\label{fig:problem_1_soc_node_12}
\end{subfigure}
\begin{subfigure}[t]{0.32\textwidth}
\centering
\includegraphics[scale=0.19]{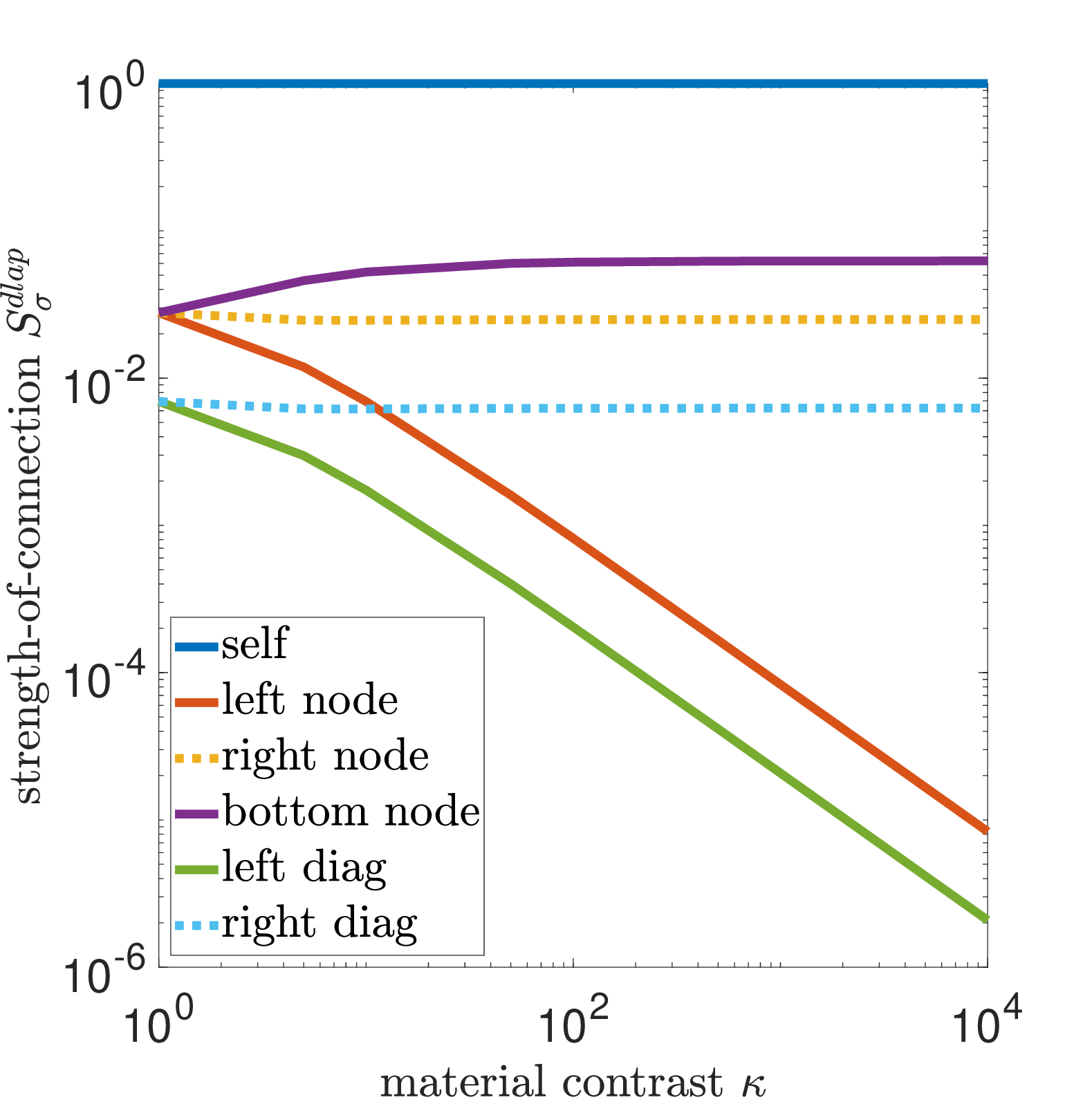}
\caption{\footnotesize $\strength^{dlap}_{\material}$ of node 13 at the material interface boundary.}
\label{fig:problem_1_soc_node_13}
\end{subfigure}
\begin{subfigure}[t]{0.32\textwidth}
\centering
\includegraphics[scale=0.19]{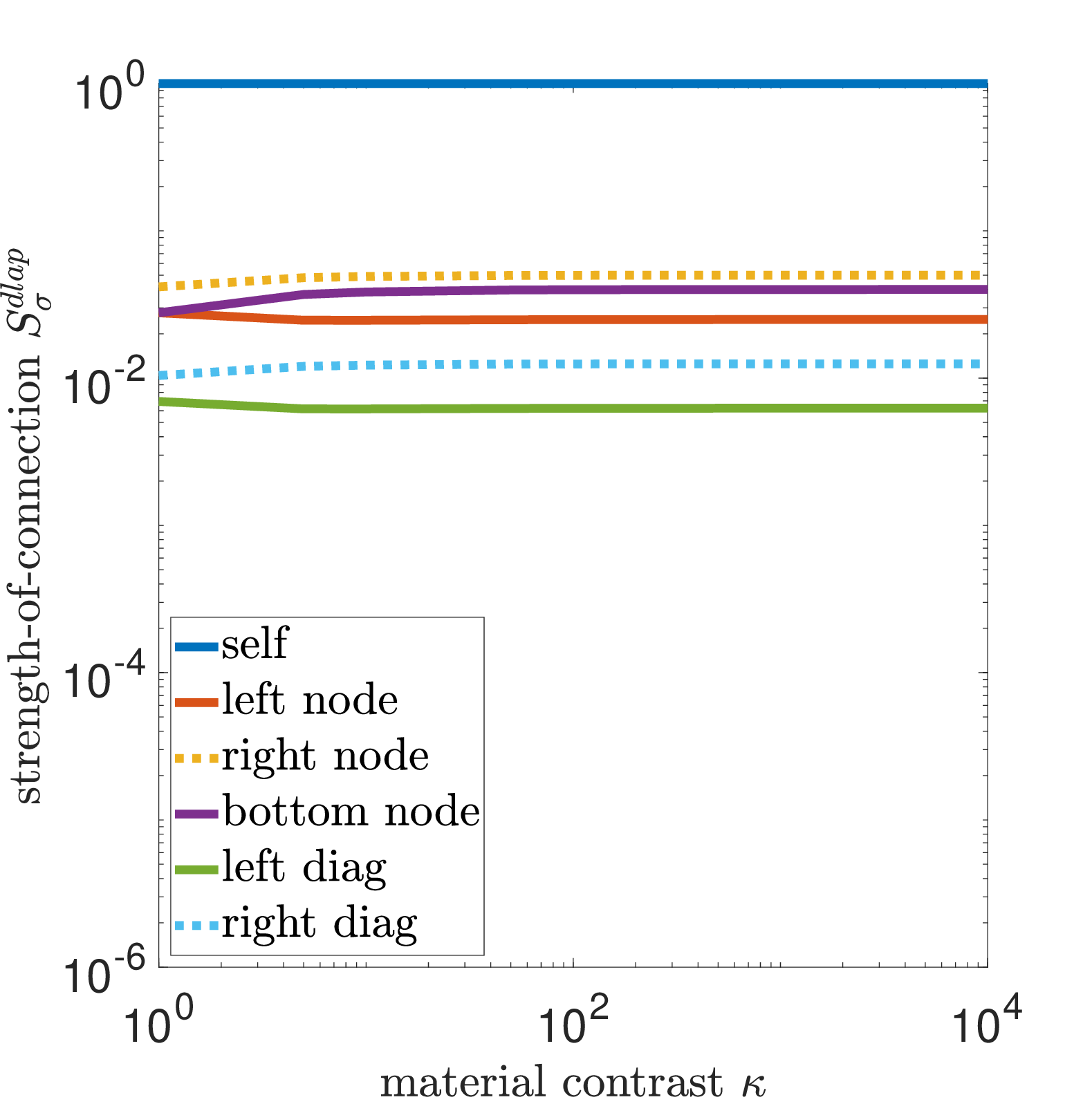}
\caption{\footnotesize $\strength^{dlap}_{\material}$ of node 14 in the high material parameter region $\material_{\text{high}}$.}
\label{fig:problem_1_soc_node_14}
\end{subfigure}
\caption{Visualization of the {\soc} $\strength$ of a node to its direct neighbors based on the material-based {\DistanceLaplacian}
measure for different material contrast ratios $\kappa$.}
\label{fig:problem_1_soc}
\end{figure}

This behavior is illustrated in \figref{fig:problem_1_soc}, showing the {\soc} values over different material
contrasts $\kappa$ from the direct neighborhood for node 12, 13 and 14 from \figref{fig:model_problem_discretization}
respectively. The direct neighborhood of a node is defined as given in \figref{fig:neighbor_node_naming}, taking into
account the direct left and right neighbors as well as the bottom and related diagonal nodes. The upper part of the
neighborhood is omitted, due to the symmetry of the problem and the {\soc} measure.
Considering node 12 first, as shown in \figref{fig:problem_1_soc_node_12}, the left and bottom neighbors as
well as node 12 itself are related to $\material_{\text{low}}$, whereas the remaining nodes in the neighborhood
are located in a region of high coefficient $\material_{\text{high}}$. All edges to nodes with $\coord_{i_{\text{low}}}$
and thus $(\strength_{\material}^{dlap})_{i_{\text{low}}i_{\text{low}}}$ show the expected constant {\soc} with increasing $\kappa$.
Connections to neighbors related to $\material_{\text{high}}$, therefore featuring $(\strength_{\material}^{dlap})_{i_{\text{low}}i_{\text{high}}}$,
exhibit a decreasing {\soc} as stated in \eqref{eq:low_high_soc_parameter_relation}
wanting to separate from  $\material_{\text{low}}$. In contrary, this behavior is flipped for node 13, illustrated
in \figref{fig:problem_1_soc_node_13}. This is to be expected, as the left neighborhood is related to
$\material_{\text{low}}$ and thus $(\strength_{\material}^{dlap})_{i_{\text{low}}i_{\text{high}}}$, while all other nodes are
located in the region with $\material_{\text{high}}$. The {\soc} value of edges to neighbors of node 14 is rather trivial as
all nodes are located in a region of high coefficient with $\coord_{i_{\text{high}}}$ and thus showing constant values for
$\strength_{\material}^{dlap}$ with increasing $\kappa$.

\begin{figure}
\centering
\begin{subfigure}[b]{0.45\textwidth}
\centering
\includegraphics[trim={1.7cm 4.7cm 0.5cm 3.5cm}, clip, width=\textwidth]{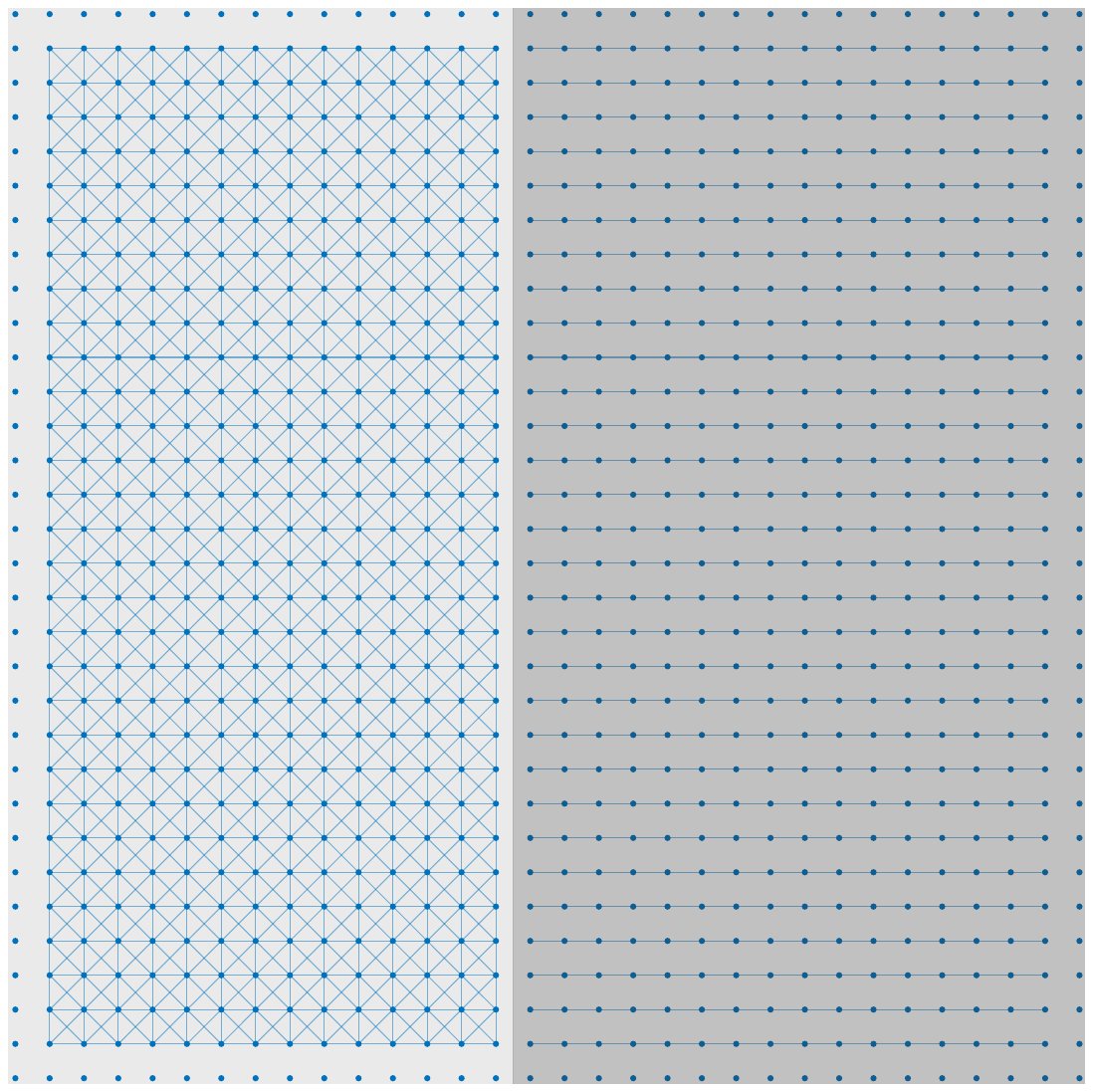}
\caption{\footnotesize Matrix graph $\graphOf{\dropCriterion^{pw}(\strength_{\material}^{dlap})}$.}
\label{fig:academic_test_problem_graph}
\end{subfigure}
\hfill
\begin{subfigure}[b]{0.45\textwidth}
\centering
\includegraphics[trim={3.5cm 3.0cm 3.0cm 2.5cm}, clip, width=\textwidth]{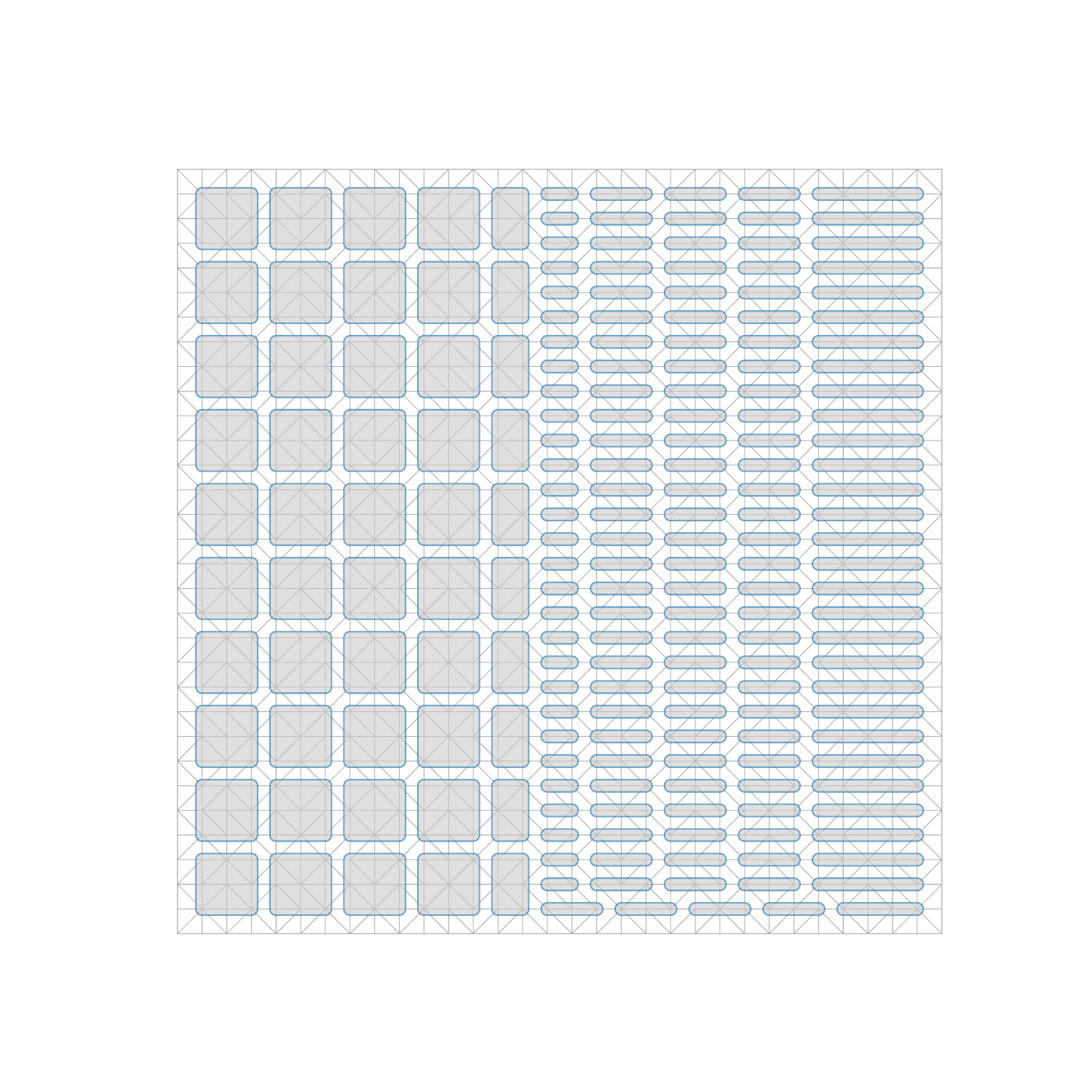}
\caption{\footnotesize Aggregates $\aggregate$.}
\label{fig:academic_test_problem_aggregates}
\end{subfigure}
\caption{
Graphical illustration of the modified matrix graph $\graphOf{\dropCriterion^{pw}(\strength_{\material}^{dlap})}$ with
``unimportant'' entries removed and the respective aggregates $\aggregate$ build on it. Nodes associated with the Dirichlet
boundary have been isolated in $\graphOf{\dropCriterion^{pw}(\strength_{\material}^{dlap})}$ as well and will not be propagated
to coarser levels. It can be clearly seen that the material information enters the aggregation process as intended.
}
\label{fig:academic_test_problem_coarsening}
\end{figure}

We confirm the correct dropping of edges in the matrix graph and thus a proper construction of aggregates by performing coarsening using $\dropCriterion^{pw}(\strength_{\material}^{dlap})$ and an appropriate value for the drop tolerance $\dropTol$ on the test problem given by \eqref{eq:academic_test_problem}.
The corresponding matrix graph, mapped to the node coordinates of the underlying
mesh, and respective aggregates are shown in Figure \ref{fig:academic_test_problem_coarsening}. The matrix graph
of the isotropic part of the domain is still
well connected, forming the expected regularly shaped aggregates. All graph edges connecting regions with different
material properties have been removed. In addition, due to the anisotropic material property featured in the
part of the domain with a high value of $\kappa$, only graph edges in the strong material direction are kept, resulting in
a semi-coarsening in the $\xaxis$-axis direction following the geometrically smooth error component. This is in
agreement with the error plot shown in \figref{fig:domain_error}.

Equation~\eqref{eq:model_problem} is well-defined even when \(\material\) has low regularity.
Point evaluation at the locations \(\{\coord_{i}\}\) of the degrees of freedom might not make sense in this case.
For example, a common use case is a elementwise constant tensor coefficient.
In this case, we can replace \(\material(\coord_{i})\) with the Clément interpolant~\cite{Clement1975a}
\((\mathcal{I}^{C}_{h}\material)(\coord_{i})\) that is given in terms of integrals over the patches associated with degrees of freedom,
resulting in an averaging of the material property around a node.
The inputs to the {\soc} are therefore \(\{\coord_{i}\}_{i}\) with \(\coord_{i}\in\mathbb{R}^{\ndim}\) and
\(\{\material_{i}\}_{i}\), \(\material_{i}\in\mathbb{R}^{\ndim\times\ndim}\).

Additionally, for the recursive construction of multigrid levels we need to transfer the material data
vector given at level \(\level\) to the next coarser level \(\level+1\).
We construct coarse coordinates and coefficient tensors by averaging over aggregates.
We assume that the coordinates and coefficients on level \(\level\) are given by
\(\indexedLevel{\coord}{\level}_{i}\) and \(\indexedLevel{\material}{\level}_{i}\) respectively.
For the coarse unknown \(j\), associated with an aggregate \(\indexedLevel{\aggregate}{\level}_{j}\), we set
\begin{align}
  \indexedLevel{\coord}{\level+1}_{j} = \frac{1}{|\indexedLevel{\aggregate}{\level}_{j}|} \sum_{i\in\indexedLevel{\aggregate}{\level}_{j}} \indexedLevel{\coord}{\level}_{i}\, \quad \quad \text{and} \quad \quad
  \indexedLevel{\material}{\level+1}_{j} = \frac{1}{|\indexedLevel{\aggregate}{\level}_{j}|} \sum_{i\in\indexedLevel{\aggregate}{\level}_{j}} \indexedLevel{\material}{\level}_{i}.
\label{eq:auxiliary_transfer}
\end{align}
We remark that there exist other normalization approaches to preserve certain quantities through the multigrid hierarchy
than averaging over aggregates, yet we will consider \eqref{eq:auxiliary_transfer} for the remainder of this manuscript.
Since our algorithm prefers to drop edges between low and high coefficient regions, we preserve the sharp interface between
regions on coarse levels as illustrated in \figref{fig:academic_test_problem_coarsening}.

\section{Numerical results}
\label{sec:numerical_results}

We conduct several experiments in 2D and 3D to examine the behavior of the proposed coarsening scheme
used in the context of smoothed aggregation AMG preconditioning using the
{\trilinos}\cite{Heroux2005a}\cite{Trilinos}\cite{Mayr2025a} package {\muelu}\cite{BergerVergiat2023a}.
The first test scenarios are academic test problems, whereas the latter ones are actual
application cases. The first problem consists of a two dimensional domain featuring a material jump
and anisotropy, identical to the one introduced in \secref{sec:new_dropping_scheme}. Afterwards
we investigate a thermal diffusion problem on an annulus geometry \cite{Green2024a}, which highlights
a strong anisotropic material behavior. Problems with rotated anisotropies are commonly used
in literature as academic test cases \cite{Brezina2006a} \cite{Brandt2015a}. The first application
example is based on a thermal battery \cite{Voskuilen2021a}, which features a stretched mesh and a
highly heterogeneous, scalar material distribution. The second application case focuses on a
unit volume of a solar cell with a very localized, highly anisotropic material part.

If not specified differently, a conjugate gradient method (CG) from {\belos}\cite{Bavier2012a} is used to
solve the arising linear systems, with the introduced algebraic AMG method from {\muelu} as preconditioner.
The iterative solver is assumed to be converged, when the residual is reduced by a factor of $10^{8}$.
We use a 2nd-order {\Chebychev} polynomial for smoothing, which corresponds to two sweeps as pre- and post
smoother on each level, whereas the coarsest level is solved with a direct method. The filtered matrix~$A_{F}$
from \eqref{eq:filtered_matrix} is used during smoothing of the tentative prolongator to control the
operator complexity, as well as the 1-norm diagonal approximation \eqref{eq:1_norm_diagonal} to counteract
possible small values on the matrix diagonal. Coarse levels of the multigrid hierarchy are repartitioned by
a coordinate based approach\cite{Deveci2015} implemented in {\zoltan2}.

\subsection{Two domain problem}
\label{subsec:Example_1}

So far, the behavior of the material-weighted {\DistanceLaplacian} {\soc} $\strength_{\material}^{dlap}$
has been investigated in a purely isolated setting, omitting its interplay with other components of the
multigrid hierarchy. We again consider our two domain test problem of \secref{sec:new_dropping_scheme} based on a
quadrilateral mesh with first-order (bilinear) {\Lagrangian} finite element shape functions on a unit-square
$\dom = \{(x,y)\in\REalSp{2}:-1<x<1,-1<y<1\}$  with a material tensor following

\begin{align*}
\material(\coord) &=
\begin{cases}
  I &  \text{for} \; \xaxis<0,
\\[10pt]
\left(
  \begin{array}{cc}
    \kappa&\\
    &1
  \end{array}
  \right)
&  \text{for} \; \xaxis \geq 0.
\end{cases}
 & \kappa \in \{1,100,\num{10000}\}
\end{align*}
The domain is assumed to be fully enclosed by a homogeneous Dirichlet boundary.
The base mesh consists of $\num{1024}$ elements, with $\num{32}$ elements along each spatial direction.

To confirm proper working of the coarsening scheme in a full multigrid cycle and to get a first
estimate of robust {\soc} measure and dropping criterion combinations, we study the
behavior of the method based on the iteration count and the cost of preconditioner application,
which is defined as the product of operator complexity times number of linear iterations taken by the iterative
solver. We use different mesh sizes $h \in \{1/32, 1/128, 1/512\}$ and material
contrasts $\kappa$ based on the given material tensor $\material(\coord)$. In addition, we consider
a variable drop tolerance of $\dropTol \in \{0.0, 0.0025, 0.005, 0.01, 0.02, 0.04, 0.08, 0.16, 0.32, 0.64\}$
for the coarsening process for different combinations of dropping criteria $\dropCriterion^{\text{pw}}$,
$\dropCriterion^{\text{cut-drop}}$ and {\soc} measures $\strength^{sa}$, $\strength^{dlap}$,
$\strength_{\material}^{dlap}$.

\begin{figure}
\centering
\includegraphics[width=0.9\linewidth]{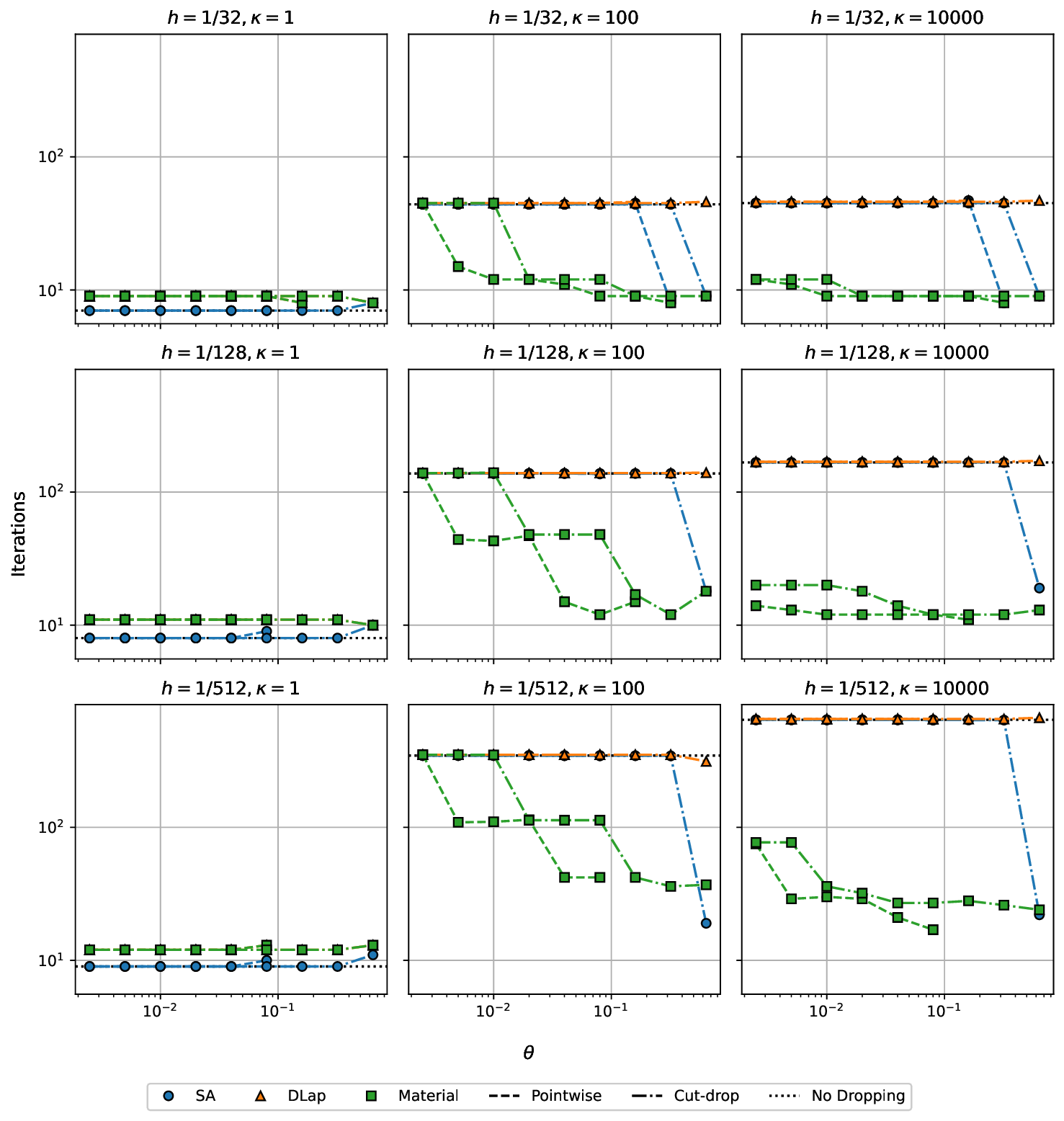}
\caption{Number of iterations shown over different drop tolerances $\dropTol$ for combinations of
dropping criteria $\dropCriterion^{\text{pw}}$, $\dropCriterion^{\text{cut-drop}}$ and {\soc} measures
$\strength^{sa}$ , $\strength^{dlap}$ and $\strength_{\sigma}^{dlap}$ shown for different material
contrasts $\kappa$ and mesh refinements $h$. Configurations that fail to reach the designated coarse
grid size or fail to converge are not shown, as for example, if the algebraic multigrid coarsening stagnates.
The material based dropping shows robustness and performance across a wide range of dropping tolerances $\dropTol$.}
\label{fig:problem_1_iteration_results}
\end{figure}

The iteration count of the linear solver for increasing values of $\dropTol$ and different
combinations of dropping criteria and {\soc} measures are shown in \figref{fig:problem_1_iteration_results},
with each diagram being related to a unique combination of mesh size $h$ and material contrast
$\kappa$, with configurations that fail to coarsen or converge omitted.
For the purely homogeneous material case $\kappa=1$, a {\soc} measure based on $\strength^{sa}$ delivers
the best results with $\strength^{dlap}$ and $\strength_{\material}^{dlap}$ performing slightly worse, but
still being competitive. In this case, the dropping criterion does not have a major influence on the overall
results as there is no material jump or anisotropy to be resolved on coarse levels. This behavior is expected
and holds true for all homogeneous material cases with the iteration count staying nearly constant over all mesh
refinement levels. The iteration count of these cases can be seen as baseline or reference to compare against.
With increasing $\kappa$, {\soc} measures based on $\strength^{sa}$ and $\strength^{dlap}$ start
to perform poorly, not being able to resolve the material jump and/or material anisotropy properly on coarse
levels and thus show a highly increased iteration count compared to the homogeneous case. In contrast,
a coarsening based on $\strength_{\material}^{dlap}$  maintains good convergence properties resulting in a
low iteration count, assuming a proper value of $\dropTol$ is chosen. Overall, an effective dropping with the
material-weighted {\DistanceLaplacian} {\soc} is achieved roughly at a drop tolerance of $\dropTol > 0.01$,
which can be seen in a rapid drop of linear iterations taken to find the respective solution of the linear system.
The dropping criterion $\dropCriterion^{\text{cut-drop}}$ shows to be more robust for high drop tolerance
values $\dropTol>0.08$, reaching similar iteration counts as the
pointwise dropping criterion for $\theta$ where the pointwise
criterion does not fail.
It isn't feasible to use
$\strength^{sa}$ and $\strength^{dlap}$ on meshes more refined than
those shown as those methods are
neither robust in regards to the material contrast $\kappa$ nor the
mesh size $h$. Only at the highest $\theta$  can the traditional
smoothed aggregation measure effects of the underlying material
property, recovering the proper convergence behavior at the cost of
very aggressive dropping.
The newly
introduced material-weighted {\DistanceLaplacian} {\soc} measure shows to be robust for both the material contrast
and the mesh size, reaching iteration counts similar to the homogeneous reference case and thus outperforming
the state of the art {\soc} measures. Especially for aggressive dropping and thus over-sparsification of the matrix graph,
$\dropCriterion^{\text{cut-drop}}$ shows a robust behavior. Overall,
$\dropCriterion^{\text{cut-drop}}(\strength_{\material}^{dlap})$ is able to perform well for most combinations of
$\kappa$ and $h$ for a wide range of $\dropTol$. If performance matters most,
$\dropCriterion^{\text{pw}}(\strength_{\material}^{dlap})$ is able to achieve the lowest iteration count, but is
quite sensitive with respect to the drop tolerance.

\begin{figure}
\centering
\includegraphics[width=0.9\linewidth]{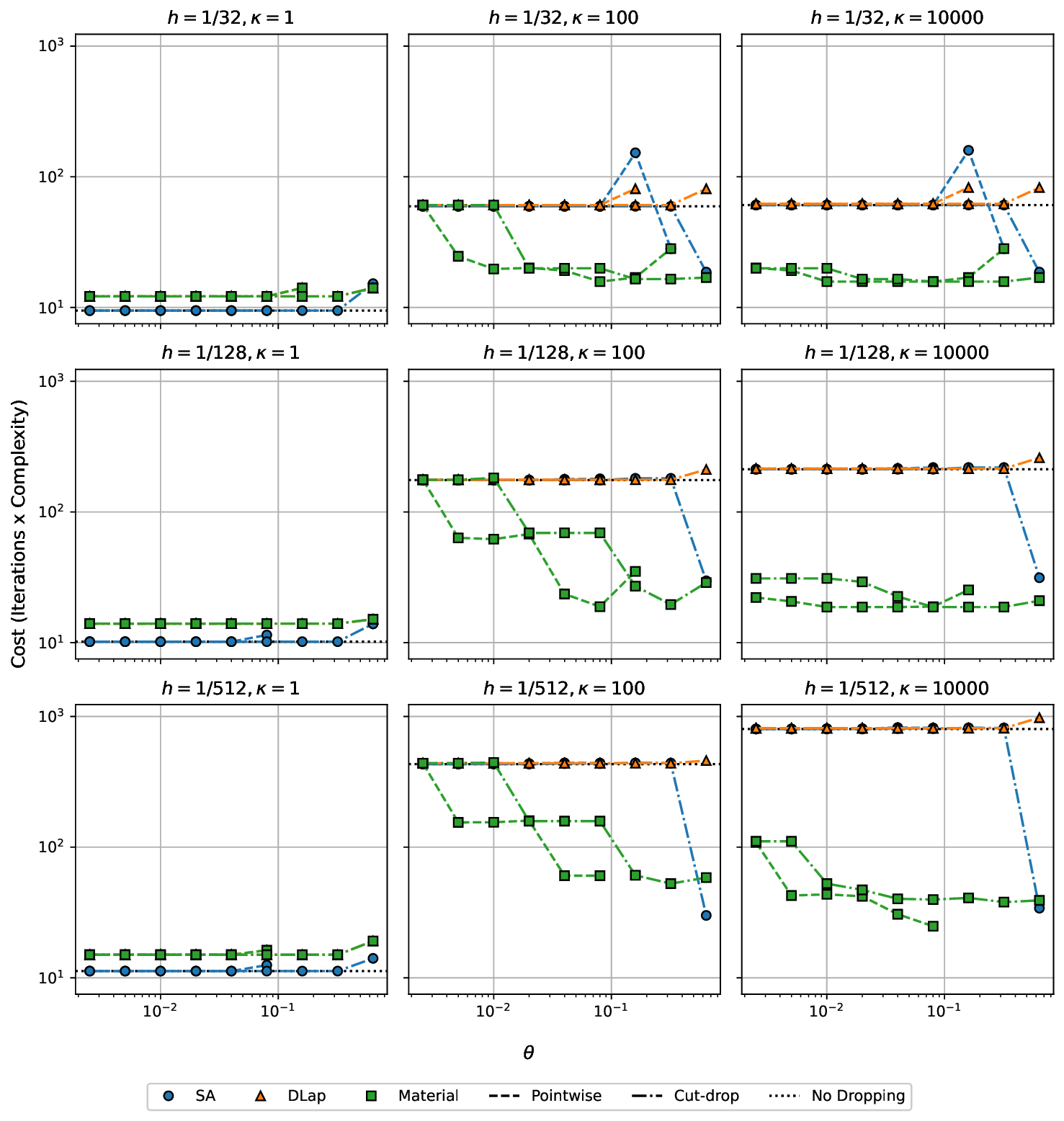}
\caption{Cost of preconditioner application (iteration count times operator complexity \(\opComplexity\)) shown over different drop tolerances $\dropTol$ for combinations of dropping
criteria $\dropCriterion^{\text{pw}}$, $\dropCriterion^{\text{cut-drop}}$ and {\soc} measures $\strength^{sa}$, $\strength^{dlap}$
and $\strength_{\sigma}^{dlap}$ shown for different material contrasts $\kappa$ and mesh refinements $h$. Cost points lying too
far away from other points are not shown. The material based dropping shows robustness and performance across a wide range of
dropping tolerances $\dropTol$.}
\label{fig:problem_1_cost_results}
\end{figure}

Next, we discuss the computational cost of application of the multigrid
preconditioner, as shown in
\figref{fig:problem_1_cost_results}. Again, each diagram is related to a fixed combination of $h$ and $\kappa$, while the cost
is plotted over a varying value of $\dropTol$ with different combinations of drop criterion and {\soc} measures.
As the application cost related to the iteration count, the results for the homogeneous material
cases are similar, showing a slight advantage for the smoothed aggregation {\soc}.
As this problem is homogeneous and has very little dropping and
thus cut-drop and pointwise dropping are comparable to each other.
Increasing the material
contrast $\kappa$ shows the benefit of using
$\strength_{\material}^{dlap}$, with an overall lower application cost
compared to the other methods.
Similar to before,
the dropping criterion $\dropCriterion^{\text{cut-drop}}$ is more robust for a wider range of $\dropTol$ than
$\dropCriterion^{\text{sa}}$. While the results related to the iteration count indicate that a higher drop
tolerance results in a better method, it does not necessarily directly relate to a lower application cost and thus a faster
scheme. This can especially be seen for an increasing material contrast.
For pointwise dropping, the  cost decreases with increasing values of $\dropTol$, till the operator complexity dominates and the
behavior reverses by increasing the overall cost for high values of $\dropTol$, with the lowest application cost lying
around $\dropTol = 0.08$. While the iteration count might decrease with more dropped edges from the matrix graph during
coarsening, the operator complexity increases equivalently due to denser coarse level representations of the matrix as
discussed in \secref{sec:smoothing_filtering} and therefore increasing the overall cost. This also highlights that
caution should be taken when applying aggressive dropping.

In conclusion, robustness of the multigrid preconditioner with respect to the material contrast $\kappa$ and the mesh size $h$
can be achieved by properly selecting the drop tolerance, dropping criterion and {\soc} measure for an accurate coarsening
of the matrix graph to properly represent the material property on coarse levels. The newly introduced {\soc}
measure $\strength_{\material}^{dlap}$ is able to capture these features properly and in combination with the cut-drop criterion
$\dropCriterion^{\text{cut-drop}}$ manages to provide a robust method in terms of linear iterations
over the range of parameters investigated. The best overall performance can be achieved by carefully adjusting
$\dropCriterion^{\text{pw}}$. Additionally, a drop tolerance of $\dropTol = 0.08$ proved to be a good choice and
resulted in the lowest application cost. Thus $\dropCriterion^{\text{cut-drop}}(\strength_{\material}^{dlap})$
with $\dropTol=0.08$ is recommended by the authors as first guess when tackling these type of problems, as it gives
a good tradeoff between performance and robustness.

\subsection{Anisotropic thermal diffusion}
\label{subsec:Example_2}

As second numerical example, we investigate the robustness and especially scalability of the new
dropping scheme with respect to an anisotropic material behavior in a thermal diffusion setting.
The problem is solved on a three dimensional annulus domain, which is defined by
an inner radius of $r_i=0.5$, an outer radius of $r_o=1.0$ and a thickness of $t=0.1$.
In this scenario, the material tensor features a strong anisotropy due to a high
conductivity in circumferential direction, which is defined as follows,
\begin{equation*}
\material(x) \leftarrow Q^T\hat{\material}(x)Q \quad \text{with} \quad
Q =
\begin{pmatrix}
\frac{y}{r} & \frac{x}{r} & 0\\
-\frac{x}{r} & \frac{y}{r} & 0 \\
0 & 0 & 1
\end{pmatrix}
\quad \text{and} \quad
\hat{\material}(x) =
\begin{pmatrix}
\kappa & ~ & ~ \\
~ & 1 & ~ \\
~ & ~ & 1 \\
\end{pmatrix}.
\end{equation*}

We employ $\num{3000}$ hexahedral finite elements with trilinear {\Lagrangean} shape functions
for the discretization of the base problem using $n_\radial=20$ elements in radial,
$n_\tangential=150$ elements in circumferential and $n_\zaxis=1$ element in thickness
direction. All other meshes are generated by uniform refinement of the base problem.
The inner and outer boundary of the annulus are subject to inhomogeneous Dirichlet boundary
conditions, while the front and bottom face are constrained by natural boundary conditions.
The dropping of weak connections during the multigrid hierarchy construction
is performed with the presented material-weighted {\DistanceLaplacian} approach.
The $\mathcal{QR}$ orthogonalization during the creation of the tentative prolongator
is omitted as the kernel is described by a constant. The system matrix is
coarsened until $\num{5000}$ or fewer unknowns remain. All simulations
are run using the multi-physics code 4C \cite{4C}.
on a cluster provided by the Data Science \& Computing Lab at the Institute for Mathematics and
Computer-Based Simulation of the University of the Bundeswehr Munich. One CPU node features 2x Intel Cascadelake CPUs with 26 cores each.

\begin{figure}
\centering
\begin{subfigure}[b]{0.3\textwidth}
\includegraphics[trim={2.5cm 4cm 1.5cm 2cm}, clip, width=\textwidth]{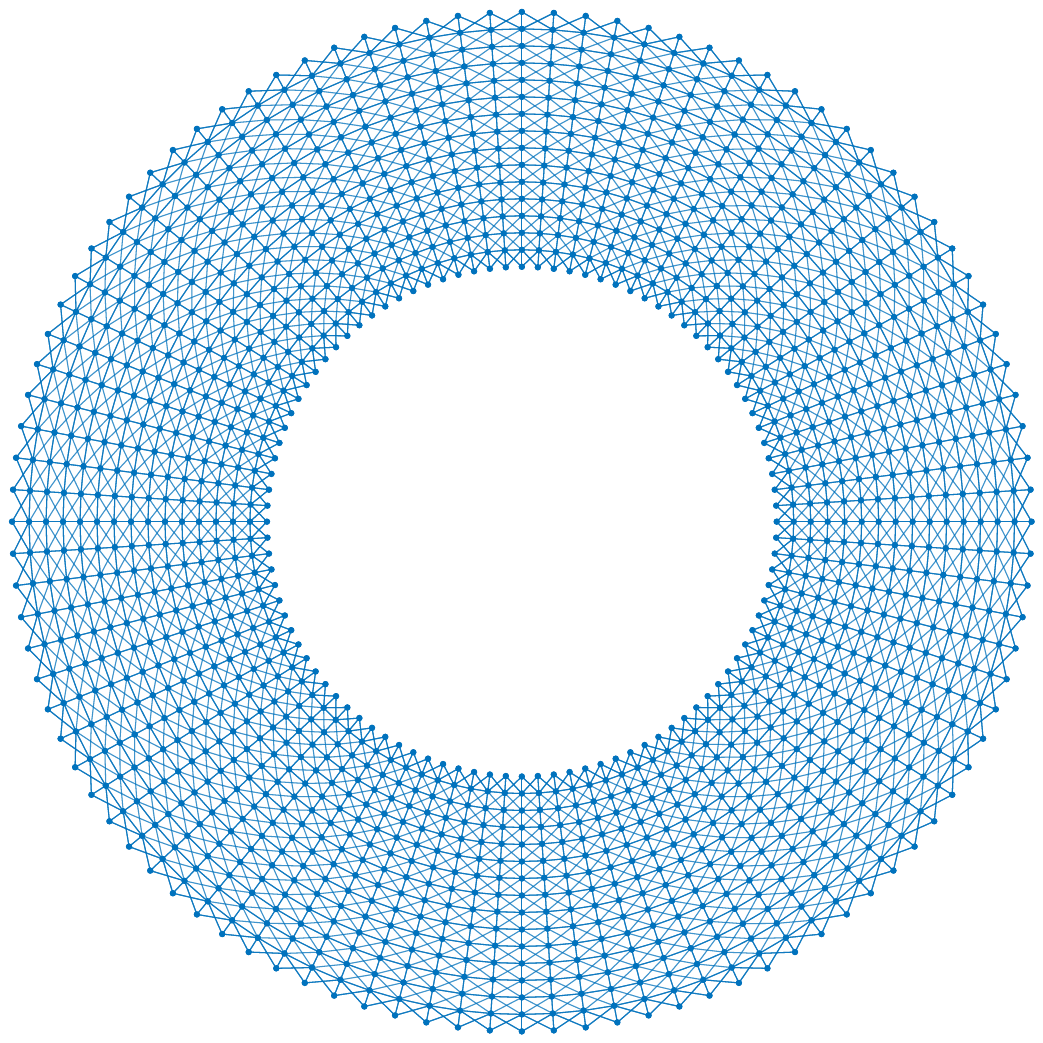}
\caption{\footnotesize Fine level matrix graph $\graphOf{\indexedLevel{A}{1}}$ mapped onto its spatial coordinates $x$.}
\end{subfigure}
\hfill
\begin{subfigure}[b]{0.3\textwidth}
\includegraphics[trim={2.5cm 4cm 1.5cm 2cm}, clip, width=\textwidth]{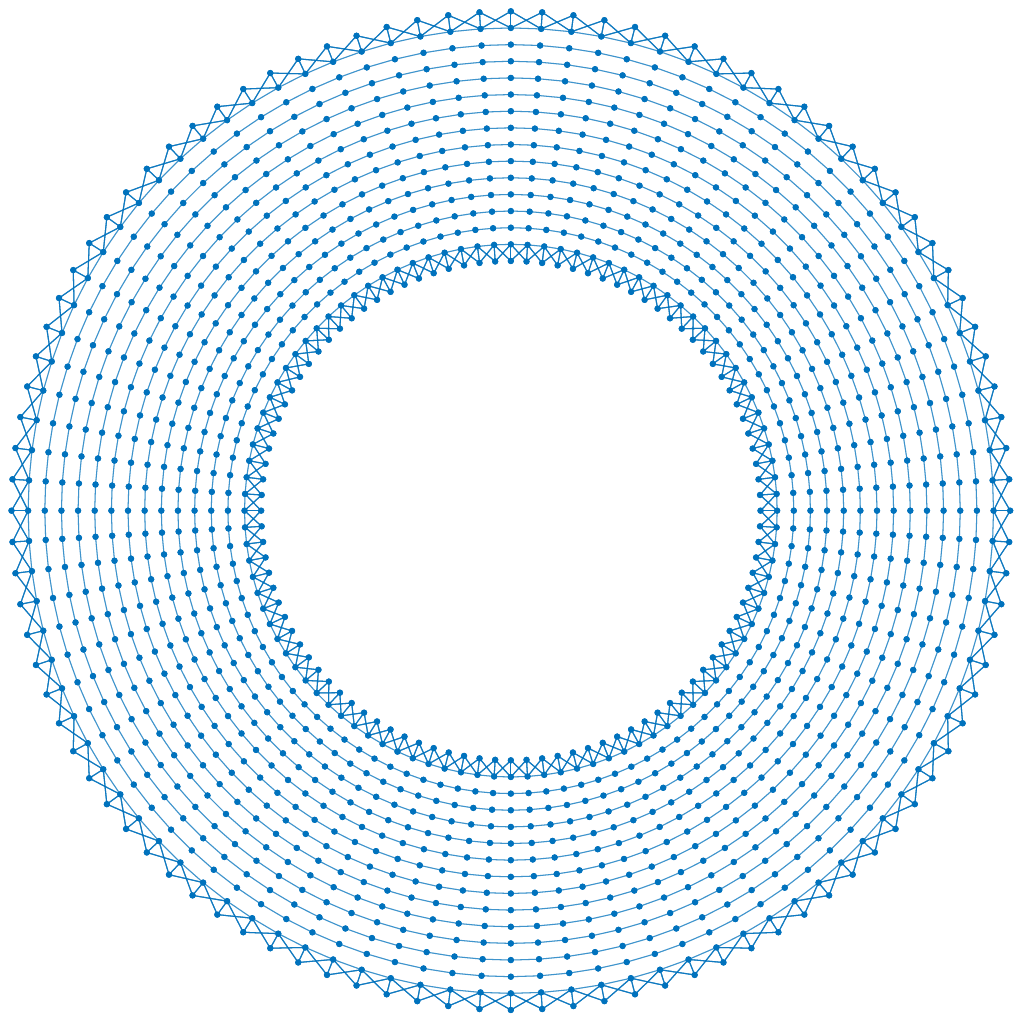}
\caption{\footnotesize Filtered graph $\graphOf{C^{pw}(S^{dlap}_{\material})}$ for aggregation on level $\level=1$.}
\end{subfigure}
\hfill
\begin{subfigure}[b]{0.3\textwidth}
\includegraphics[trim={2.5cm 4cm 1.5cm 2cm}, clip, width=\textwidth]{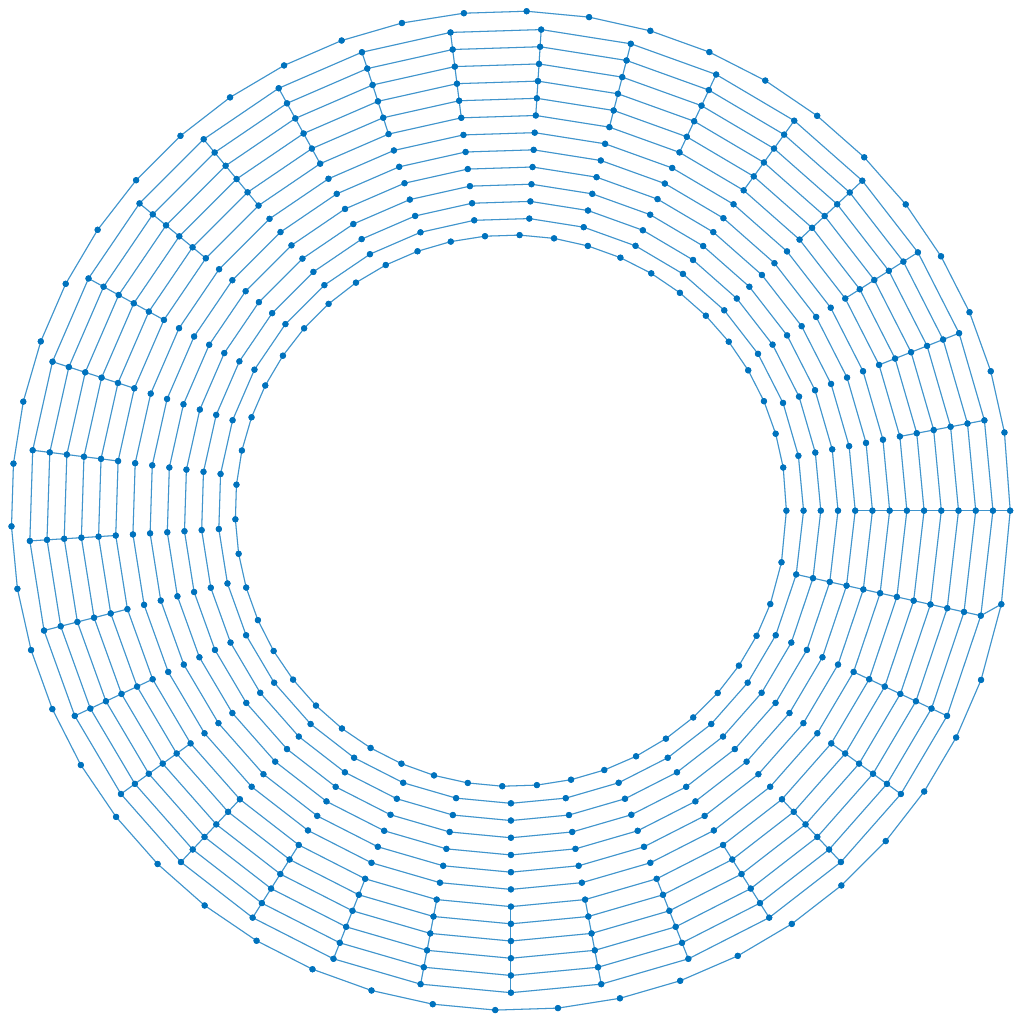}
\caption{\footnotesize Filtered graph $\graphOf{C^{pw}(S^{dlap}_{\material})}$ for aggregation on level $\level=2$.}
\end{subfigure}
\caption{Visualization of the matrix graph $\graphOf{\indexedLevel{A}{1}}$ of the fine level operator as well as on the first coarse level during the multigrid coarsening process.}
\label{fig:Graphs}
\end{figure}

\begin{figure}
\centering
\begin{subfigure}[b]{0.33\textwidth}
\centering
\includegraphics[trim={3cm 3cm 3cm 3cm}, clip, width=\textwidth]{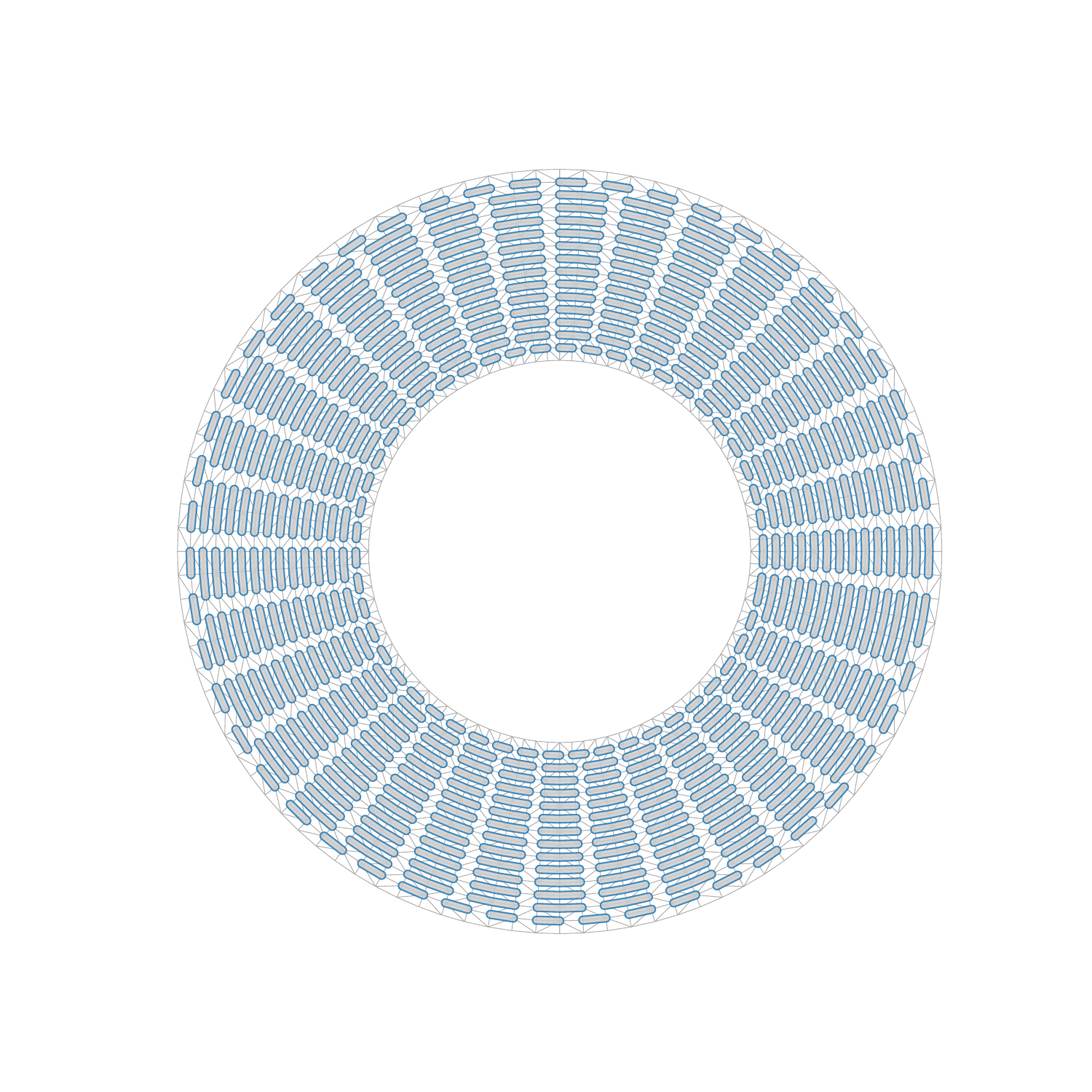}
\caption{\footnotesize Aggregates $\indexedLevel{\aggregate}{\level}$ on level $\level=1$.}
\end{subfigure}
\hfill
\begin{subfigure}[b]{0.33\textwidth}
\centering
\includegraphics[trim={3cm 3cm 3cm 3cm}, clip, width=\textwidth]{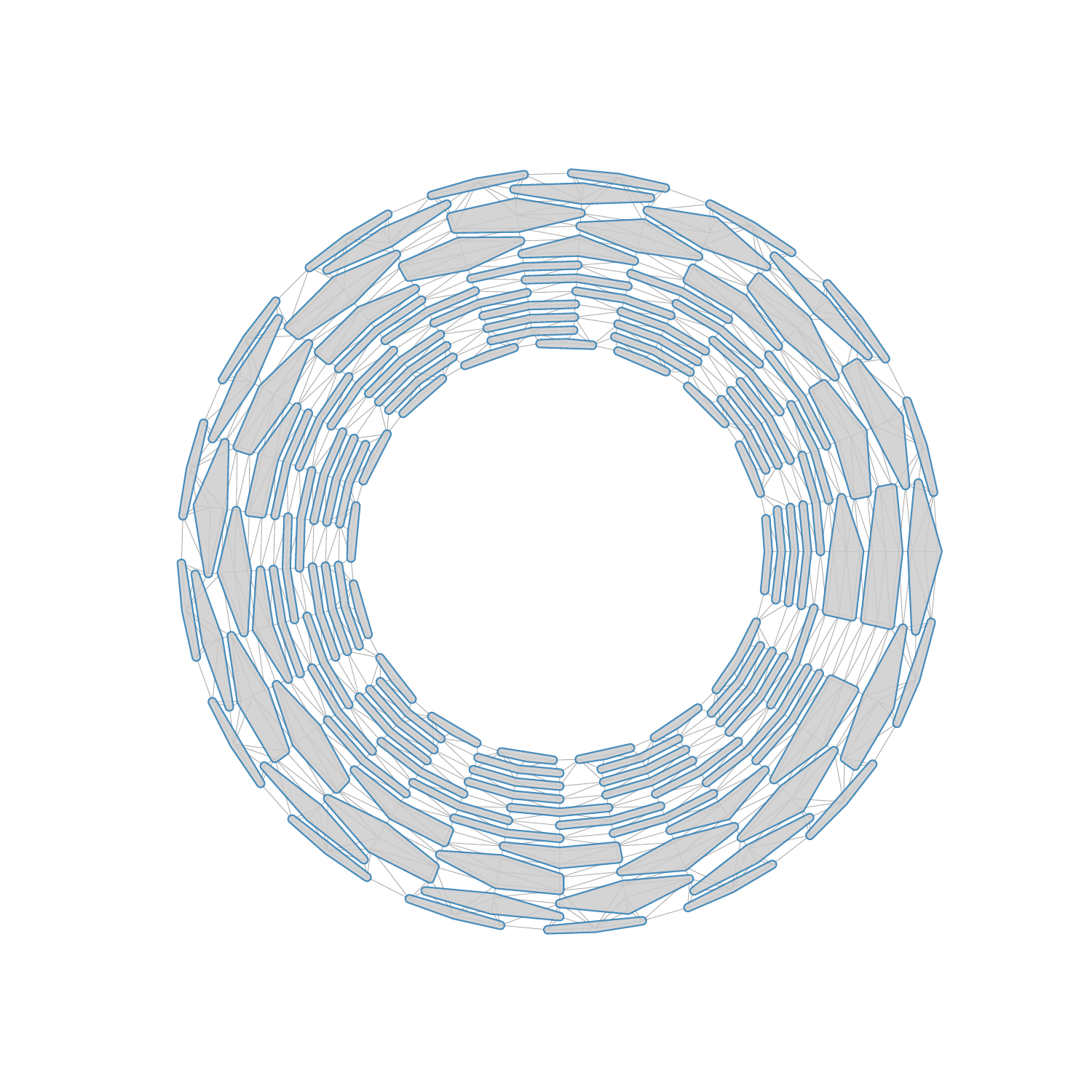}
\caption{\footnotesize Aggregates $\indexedLevel{\aggregate}{\level}$ on level $\level=2$ with $\dropTol=0.05$.}
\end{subfigure}
\begin{subfigure}[b]{0.33\textwidth}
\centering
\includegraphics[trim={3cm 2.5cm 3cm 3cm}, clip, width=\textwidth]{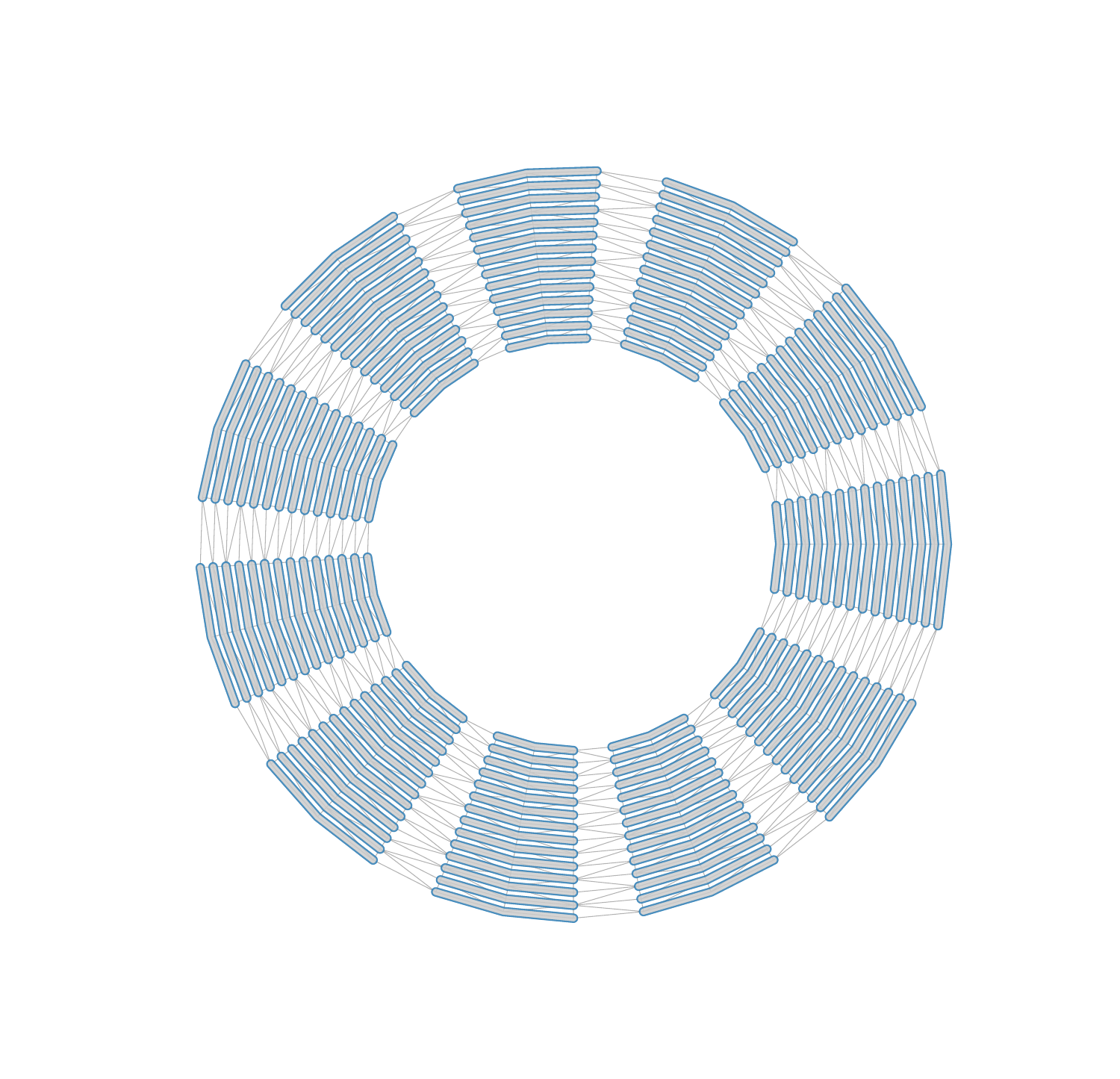}
\caption{\footnotesize Aggregates $\indexedLevel{\aggregate}{\level}$ on level $\level=2$ with $\dropTol=0.1$.}
\end{subfigure}
\caption{Visualization of different aggregates $\indexedLevel{\aggregate}{\level}$ during the material-weighted coarsening process,
showing how the interplay of material property and spatial distance influences the aggregate creation for different drop tolerances.}
\label{fig:Aggregates}
\end{figure}

First, we example how the drop tolerance $\dropTol$ removes weak
connections of the matrix graph $\graphOf{\indexedLevel{A}{\level}}$ of matrix $A$ on a multigrid
level $\level$ for a simplified two dimensional annulus problem, omitting the thickness direction.
Figure \ref{fig:Graphs} shows the matrix graph of the fine level operator, mapped to the node
coordinates of the underlying mesh, as well as the results after dropping on $\level=1$ and $\level=2$. While
the strong connections of the matrix graph in circumferential direction, due to the high material contrast,
are kept on $\level=1$, the weak connections in radial direction are dropped resulting in the expected
semi-coarsening in the strong anisotropy direction. On the first level however, the chosen drop tolerance is
not sufficient to remove all weak connections leading to an inadequate representation
of the material behavior close to the outer boundary of the annulus.
The result of the aggregation process is given in Figure \ref{fig:Aggregates} featuring the
one-dimensional shaped aggregates due to the semi-coarsening on $\level=1$. On $\level=2$
however, two-dimensional aggregates are constructed based on the graph connectivity, thus
a smearing of the material property is taking place. There is an interesting interplay happening
between the geometric distance and the material-based distance metric. As coordinates are averaged
over the respective aggregates, as stated in \eqref{eq:auxiliary_transfer}, coarse levels might
generate a stretched version of the original problem, slightly shifting coordinates.
While we do not construct explicit coarse grids in AMG methods, an anisotropy related to the aggregate
coordinates can still occur, which can be observed comparing the matrix graph of level
$\level=2$  and $\level=1$. As we semi-coarsen in one direction, the distance between aggregates
is also getting bigger in that direction, resulting in spatially closer radial positions. As our
material-based coarsening scheme is also taking into account the spatial distance in combination with
the gradual increase of this effect towards the outer boundary, the influence of material property
and spatial distance cancel each other out, thus forming two-dimensional aggregates. This
example highlights the importance of the dropping tolerance $\dropTol$
during the coarsening process. The drop tolerance needs to be chosen sufficiently high to be able to
remove all weak connections and preserve the convergence properties of the linear solver.

Next, we study the robustness of the
preconditioner with respect to the element size $h$ and material contrast $\kappa$.
We apply a uniform mesh refinement of the three-dimensional base mesh consisting of
$h/2, h/4, h/8, h/16$ and vary the material contrast with $\kappa\in\{1, 10, 100, \num{1000}, \num{10000}\}$.
As neither {\soc} measures based on traditional smoothed aggregation and {\DistanceLaplacian} lead
to a converging method, we solely focus on the newly introduced material-weighted {\DistanceLaplacian}
measure in combination
with a pointwise dropping criterion. In addition, we consider two values for the
drop tolerance given as $\dropTol\in\{0.05, 0.1\}$ inspired by the results of
\secref{subsec:Example_1}.The performance of the linear solver is
reported in Table~\ref{tab:annulus_parameter_study}
showing the iteration count for a linear solve, the
number of multigrid levels as well as the operator complexity $\opComplexity$ for
combinations of $h$, $\kappa$ and $\dropTol$. The cases with
$\kappa=1$ are of homogeneous nature and can be seen as baseline. With
increasing values of
$\kappa$, the iterative solver also shows a models increase in the iteration count,
except the case with mesh size $h$ and $\dropTol=0.05$, where the
increase is substantial at high $\kappa$.
This is most
likely due to the interplay of material contrast and spatial distance as described earlier.
An increase to $\dropTol=0.1$ solves this issue, which is not as
severe at higher levels of mesh refinement. At $\dropTol=0.1$, the iteration count varies from $~11$ to $~24$
while increasing the material contrast by a factor of $\num{10000}$ and reducing the mesh size
by a factor of $16$.
Increasing the drop tolerance stabilizes the number of iterations, but
substantially increases the operator complexity at small values of
$\kappa$.  However, for the challenging cases with high $\kappa$ and
small $h$, the proposed method does quite well.
This again underlines the robustness of the method and
stays in agreement with the observations made so far, highlighting, that the use of
purely matrix and distance based {\soc} measures does not result in a converging method
for this example, while the proposed material-weighted coarsening reaches convergence.

\begin{table}
\centering
\caption{Averaged number of linear iterations for different drop tolerances $\dropTol$,
mesh refinements and contrast ratios $\log(\kappa)$ with the number of levels $\noLevels$ and
the operator complexity $\opComplexity$ in parentheses.}
\label{tab:annulus_parameter_study}
\begin{tabular}{c c c c c c}
\hline
$\log (\kappa)$ & $h$ & $h/2$ & $h/4$ & $h/8$ & $h/16$ \\
\hline
\multicolumn{6}{c}{$\dropTol = 0.05$} \\
\hline
0 & 11  (2, 1.11) & 11 (2, 1.13) & 13 (3, 1.16) & 14 (3, 1.17) & 14 (3, 1.23) \\
1 & 15  (3, 1.82) & 16 (3, 1.82) & 17 (3, 1.83) & 19 (2, 1.76) & 18 (3, 2.13) \\
2 & 22  (3, 2.36) & 21 (3, 2.49) & 21 (3, 2.76) & 21 (3, 2.82) & 19 (3, 3.57) \\
3 & 41  (3, 2.73) & 29 (3, 2.86) & 20 (3, 3.09) & 21 (3, 3.12) & 18 (3, 3.79) \\
4 & 137 (3, 2.72) & 59 (3, 2.88) & 23 (3, 3.13) & 23 (3, 3.15) & 20 (3, 3.83) \\
\hline
\multicolumn{6}{c}{$\dropTol = 0.1$} \\
\hline
0 & 10 (3, 2.83) & 11 (3, 2.96) & 11 (3, 3.18) & 12 (3, 3.41) & 12 (3, 4.28) \\
1 & 11 (3, 4.17) & 11 (3, 3.72) & 11 (3, 3.68) & 12 (3, 3.06) & 12 (3, 5.58) \\
2 & 14 (3, 2.75) & 15 (3, 2.89) & 14 (3, 3.13) & 16 (3, 3.17) & 15 (3, 3.83) \\
3 & 16 (3, 2.75) & 21 (3, 2.89) & 18 (3, 3.13) & 21 (3, 3.17) & 17 (3, 3.83) \\
4 & 24 (3, 2.75) & 24 (3, 2.89) & 21 (3, 3.13) & 23 (3, 3.17) & 19 (3, 3.83) \\
\hline
\end{tabular}
\end{table}

Finally, we analyze the scalability of the algebraic multigrid method by performing a strong
and weak scaling study up to $P=512$ processors. The material contrast is set to a fixed value
of $\kappa=100$. Coarse levels of the multigrid hierarchy are
rebalanced, such that each active processor
owns at minimum $\num{5000}$ degrees of freedom. Based on the findings in \secref{subsec:Example_1}
and \secref{subsec:Example_2} a pointwise dropping criterion is chosen with a drop tolerance of $\theta=0.1$.

\begin{figure}
\centering
\includegraphics[width=\textwidth]{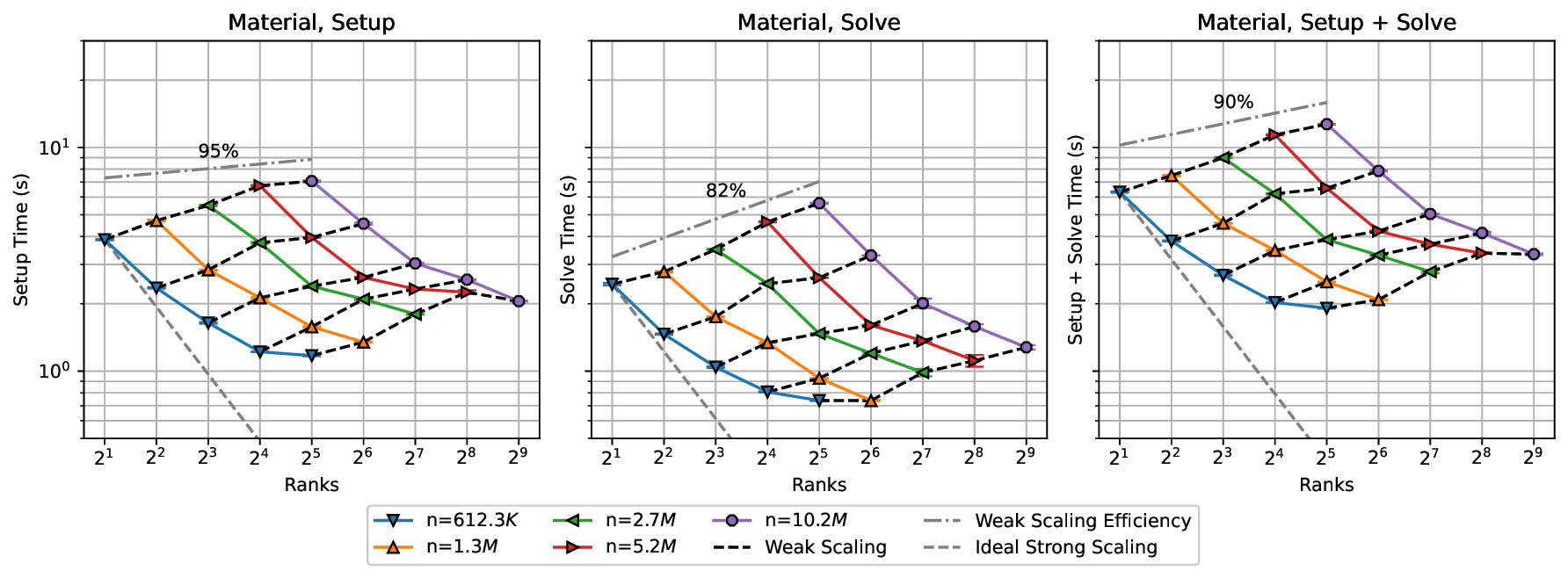}
\caption{Scaling results for the anisotropic diffusion problem for different levels of uniform mesh refinement.
We use a material-based pointwise dropping with $\dropTol=0.08$. Multigrid setup, solve, and combined times are
reported. Ideal strong scaling is shown for reference, along with a line corresponding to the estimated weak scaling
efficiency over the problem scales. Ideal weak scaling is a horizontal line.}
\label{fig:thermal-diffusion-scaling}
\end{figure}

For our scaling study, we consider five different problem sizes,
uniformly refined off of a base mesh, ranging
from $n = 6.12 \times 10^5$ up to
$n = 1.02 \times 10^7$.
Weak scaling (constant number of degrees of freedom per processor) are
shown using dashed lines, while strong scaling (constant problem size)
are shown using colored lines.
The setup and solve phases of the AMG solver are
repeated $\num{20}$ times for each configuration. The mean setup, solve, and combined setup plus
solve times are reported in \figref{fig:thermal-diffusion-scaling}, with the error bars representing
one standard deviation from the mean.  The error bars are fairly tight
and only easily visible on a few data points.
Strong scaling efficiencies are overall above 50\% for cases equal
or larger than $n/P \sim \num{70000}$.  Lower $n/P$ yields decreased scaling
efficiency, which is expected as the relative communication overhead
increases.
The smallest problem size given
by $n = 6.12 \times 10^5$ almost reaches the strong scaling limit at $n/P \sim \num{40000}$ with setup
and solve times only changing slightly for consecutive
configurations. Also, the strong scaling behavior for the setup and solve
are qualitatively similar.
The influence of on-node computations for smaller
configurations compared to distributed ones does not seem to effect the overall results. For
weak scaling efficiency of the solve phase, we reach values above 80\%, while the combined
setup and solve phase surpasses that with efficiencies of 90\% for large problem scales across different
values for $n/P$. The weak scalability of the setup phase achieves an efficiency of up to 95\%.

We conclude the discussion of the anisotropic thermal diffusion problem by stating that the material-based coarsening
results in a robust method for a wide range of mesh refinements and material contrasts. The findings related to the
newly introduced {\soc} measure from \secref{subsec:Example_1} proved to also be valid for three-dimensional problems.
In addition, we showed weak and strong scalability of the multigrid algorithm, emphasizing the applicability for
real-world applications outside of a purely academic setting. While lacking the direct scalability comparison to existing
{\soc} measures, we consider the method to be competitive in those regards, which will be shown with the following
numerical application cases.

\subsection{Thermally activated batteries}\label{sec:battery}

Thermally activated batteries \cite{Crompton1982, Guidotti2006}, also known as thermal or molten-salt batteries, are single-use primary reserve power sources with a molten salt electrolyte that is solid at room temperature.
During the activation process, pyrotechnic pellets are ignited to melt the electrolyte, allowing the battery to be operational until the electrolyte solidifies or the reactants are depleted.
Due to the single-use nature of the batteries, multi-physics modeling is especially important for the design and validation of batteries.

The simulation of thermally activated batteries requires the solution of monolithically coupled linear systems across many physics contributions, including Stefan-Maxwell diffusion, Darcy's law, and Butler-Volmer electrochemical potential~\cite{Voskuilen2021a}.
Due to the complex multi-physics couplings in the linear systems for the thermal battery problems, we employ a block-based Gauss-Seidel preconditioning strategy using {\teko}~\cite{Phillips2026a,Cyr2016a}.
An important sub-step in the preconditioner application requires solving for the solid phase voltage, $\Phi_s$, which is governed by Ohm's law
\begin{equation}\label{eq:solid-phase-voltage}
\nabla \cdot \left(-\material(\coord)\nabla\Phi_s\right) = S_e,
\end{equation}
where $\material$ is the solid phase electrical conductivity and $S_e$ is a source term.
The material tensor $\material$ varies by as much as ten orders of magnitude throughout the domain shown in \Cref{fig:thermal_battery_stack}.
For example, $\material \approx 10^{-4}$ \unit{\siemens\per\meter} in the separator, while the conductivity is as large as $\material \approx 10^{6}$ \unit{\siemens\per\meter} in the anode.
The cathode, separator, and anode layers as depicted in \Cref{fig:thermal_battery_stack} are repeated $N=20$ times throughout the domain.

\input{figures/battery}

We consider the solution to the voltage equation \eqref{eq:solid-phase-voltage} for a fixed snapshot taken during the electrochemical activation for the 2D axisymmetric simulation of thermal batteries for three different mesh resolutions.
The first mesh represents a coarse resolution featuring multiple elements along the axial dimension to resolve the multiple interfaces between the cathode, separator, and anode layers within the battery.
In the radial dimension, however, very few elements are used.
As a consequence, the element aspect ratios are severely stretched, with the aspect ratio varying from 11.28 to 37.03.
The second mesh resolution features a 16-fold refinement in the number of elements in the radial dimension while using the same number of elements in the axial dimension,
In this case, the element aspect ratios are improved, varying from 1.17 to 2.78.
Finally, the third mesh represents a single level of uniform mesh refinement based on the previous mesh.
The drop tolerance is varied from $\dropTol=0$ (no dropping) to $\dropTol=0.64$.

\begin{figure}
\centering
\includegraphics[width=\textwidth]{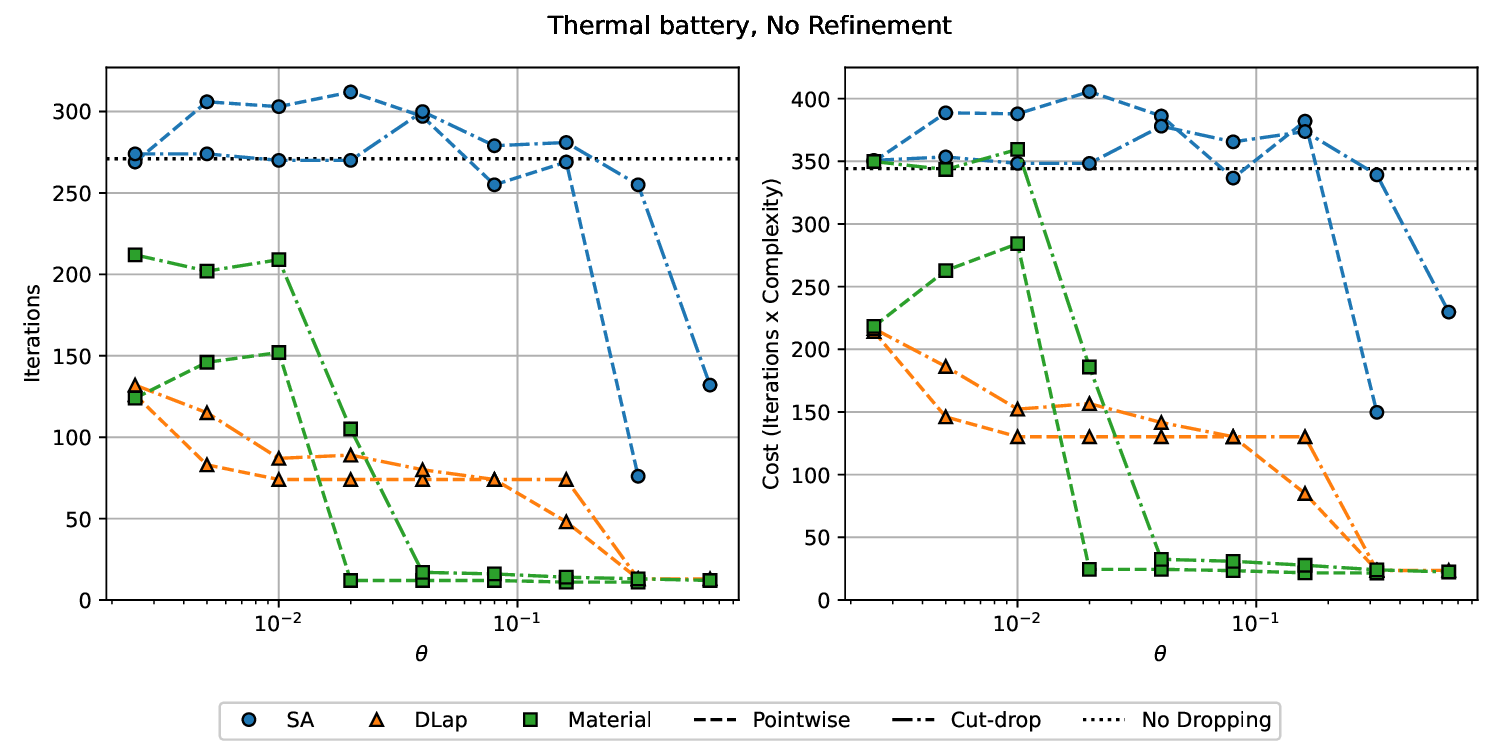}
\caption{Number of iterations and cost of application shown over different drop tolerances $\dropTol$ for
combinations of dropping criteria $\dropCriterion^{\text{pw}}$, $\dropCriterion^{\text{cut-drop}}$ and {\soc}
measures $\strength^{sa}$ , $\strength^{dlap}$ and $\strength_{\sigma}^{dlap}$ for the initial (coarse) mesh
resolution. Configurations that fail to reach the designated coarse grid size or fail to converge are not shown. Distance Laplacian and in particular material based
dropping shows robustness and performance across a wide range of dropping tolerances $\dropTol$.}

\label{fig:thermal-battery-1x}
\end{figure}

\begin{figure}
\centering
\includegraphics[width=\textwidth]{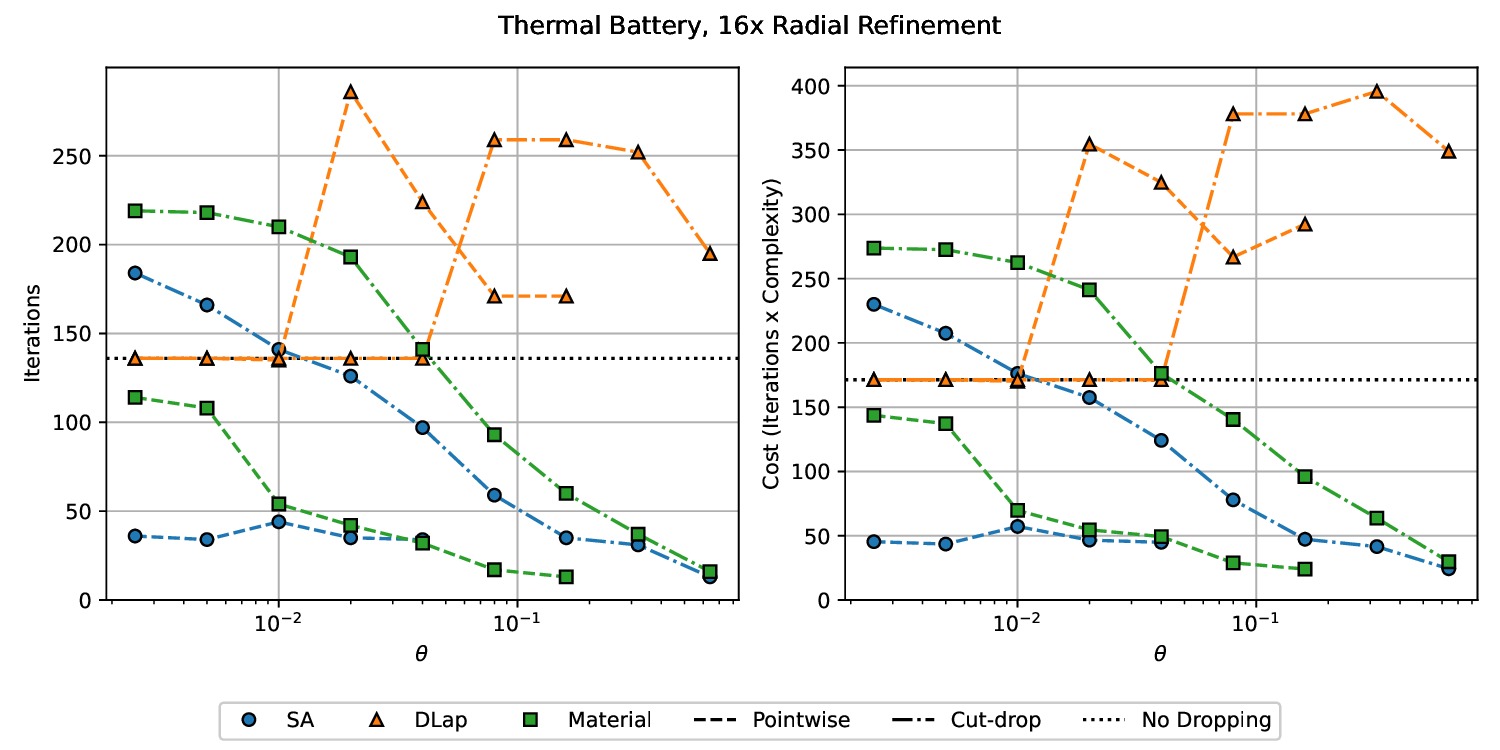}
\caption{Number of iterations and cost of application shown over different drop tolerances $\dropTol$ for
combinations of dropping criteria $\dropCriterion^{\text{pw}}$, $\dropCriterion^{\text{cut-drop}}$ and {\soc}
measures $\strength^{sa}$ , $\strength^{dlap}$ and $\strength_{\sigma}^{dlap}$ for the 16x radial mesh refinement
case. Configurations that fail to reach the designated coarse grid size or fail to converge are not shown. The reductions in the mesh aspect ratio results in a less
pronounced effect of {\DistanceLaplacian} {\soc} and improves the use of smoothed aggregation and material based
schemes.}
\label{fig:thermal-battery-16x}
\end{figure}

\begin{figure}
\centering
\includegraphics[width=\textwidth]{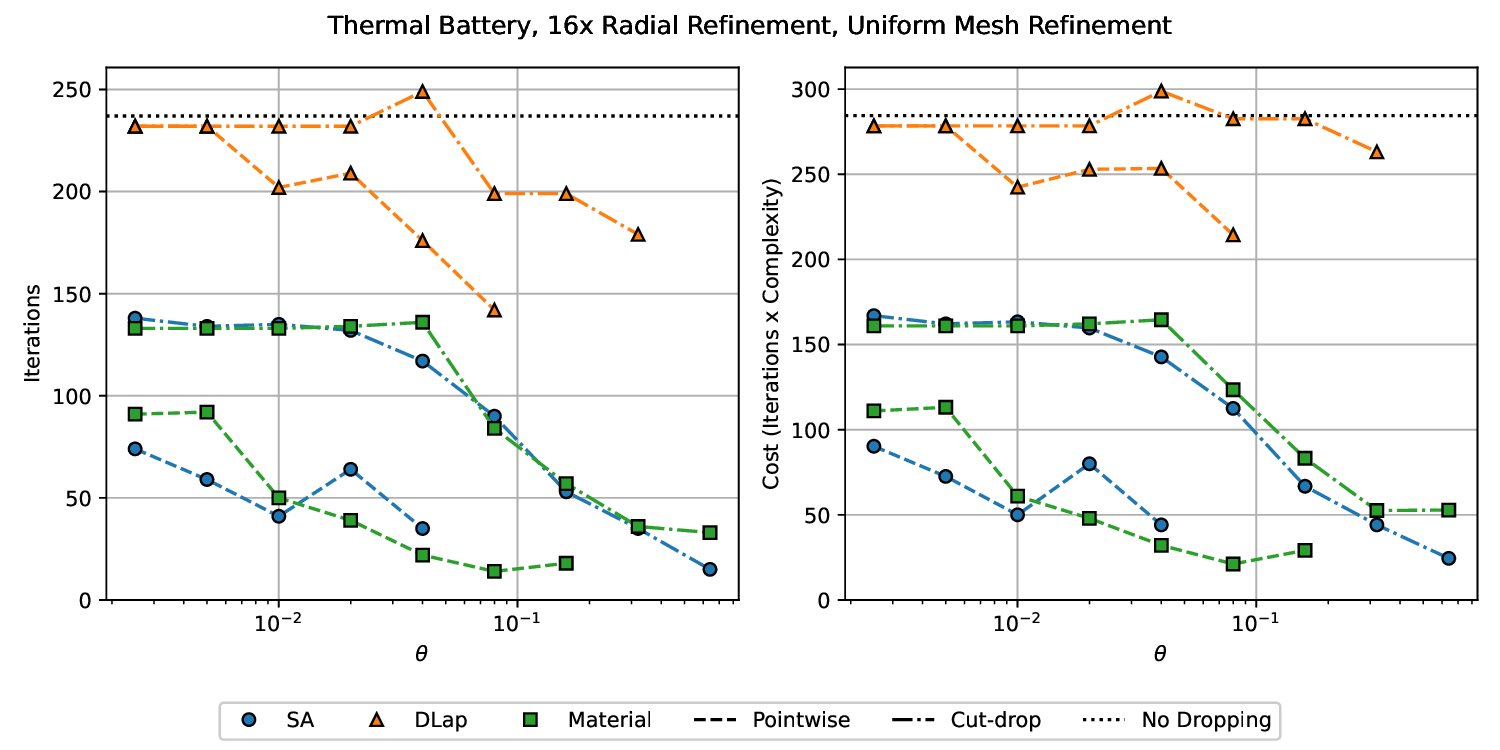}
\caption{Number of iterations and cost of application shown over different drop tolerances $\dropTol$ for
combinations of dropping criteria $\dropCriterion^{\text{pw}}$, $\dropCriterion^{\text{cut-drop}}$ and {\soc}
measures $\strength^{sa}$ , $\strength^{dlap}$ and $\strength_{\sigma}^{dlap}$ for the 16x radial mesh refinement,
plus a single uniform mesh refinement case. Configurations that fail to reach the designated coarse grid size or fail to converge are not shown.
The relative performance of the smoothed aggregation and material based schemes over the {\DistanceLaplacian} {\soc} remains the same as in the 16x radially refined case,
and good choices of the drop tolerance $\dropTol$ remain the same.}
\label{fig:thermal-battery-16x-umr-1}
\end{figure}

Results for the solid-phase voltage solve for the three mesh resolutions are shown in \Cref{fig:thermal-battery-1x}, \Cref{fig:thermal-battery-16x}, and \Cref{fig:thermal-battery-16x-umr-1} for the coarse, 16-fold refined, and uniform mesh refinement cases, respectively.
Results for the coarse mesh are shown in \Cref{fig:thermal-battery-1x}, we observe that smoothed aggregation {\soc} with pointwise or cut-drop dropping performs poorly until an aggressive drop tolerance ($\dropTol\geq 0.32$) is used.
On the other hand, the {\DistanceLaplacian} and material-based {\soc} converge much faster over a wide range of drop tolerances.
For sufficiently large drop tolerances $\dropTol\geq 0.02$, the material-based {\soc} achieves both the lowest iteration count and overall cost.
Despite the many order-of-magnitude differences across the material interfaces, the degree of mesh anisotropy is sufficiently large such that the {\DistanceLaplacian} {\soc} approach performs comparably well to the material-based {\soc}, provided $\dropTol\geq 0.32$.
However, even in the presence of a high degree of mesh anisotropy, incorporating \emph{both} material and mesh information into the {\soc} algorithm reduces the overall cost of the multigrid method across a wide range of drop tolerances.
For the refined cases with more isotropic meshes, mesh information alone is not sufficient and can lead to sub-optimal performance relative to smoothed aggregation dropping. However, material-based dropping helps to improve the solver performance.

\begin{figure}
\begin{subfigure}{0.48\textwidth}
\includegraphics[width=\textwidth]{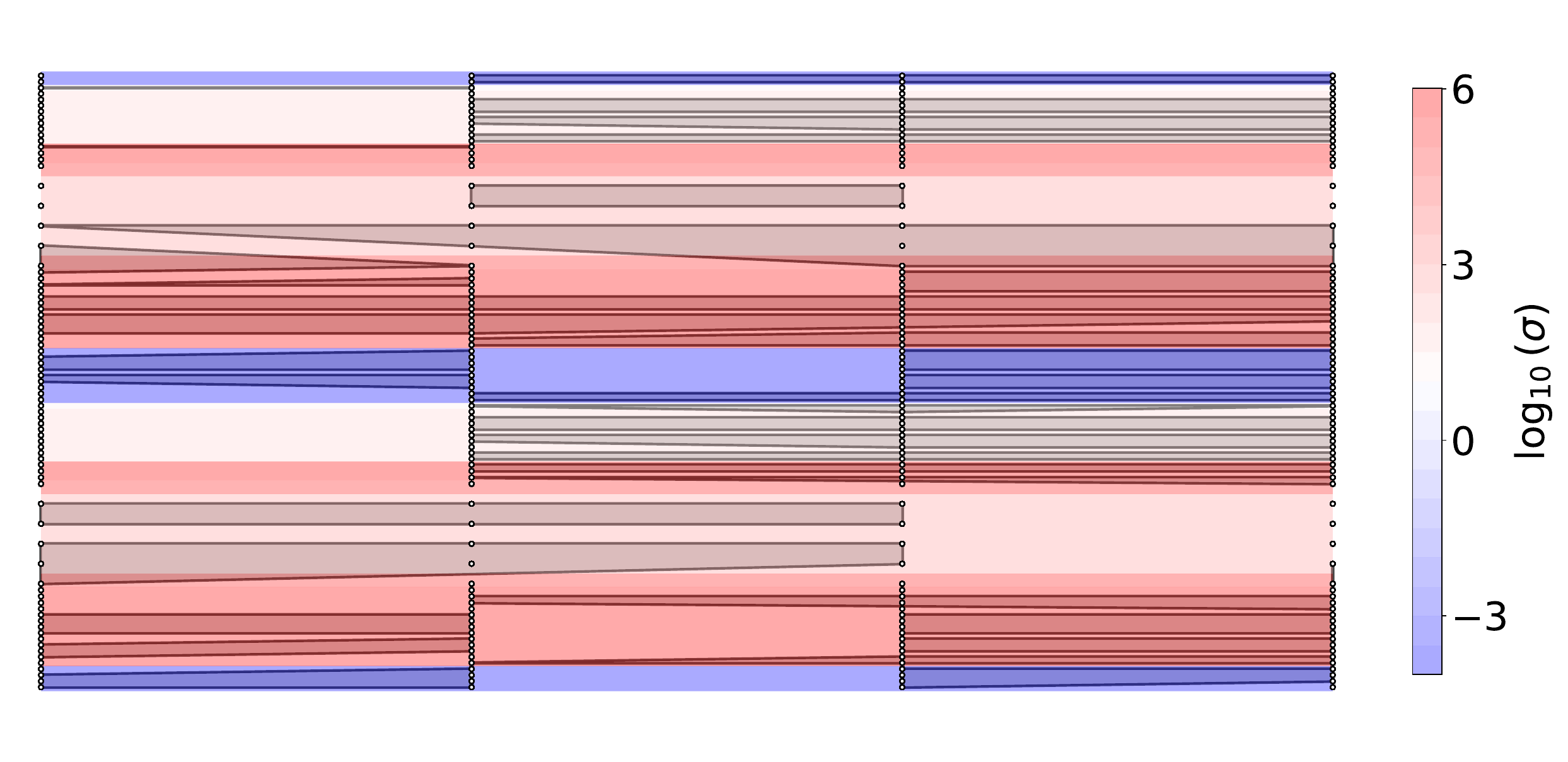}
\caption{SA-based pointwise dropping, $\dropTol=0.01$.}
\label{fig:battery-sa-agg}
\end{subfigure}
\hfill
\begin{subfigure}{0.48\textwidth}
\includegraphics[width=\textwidth]{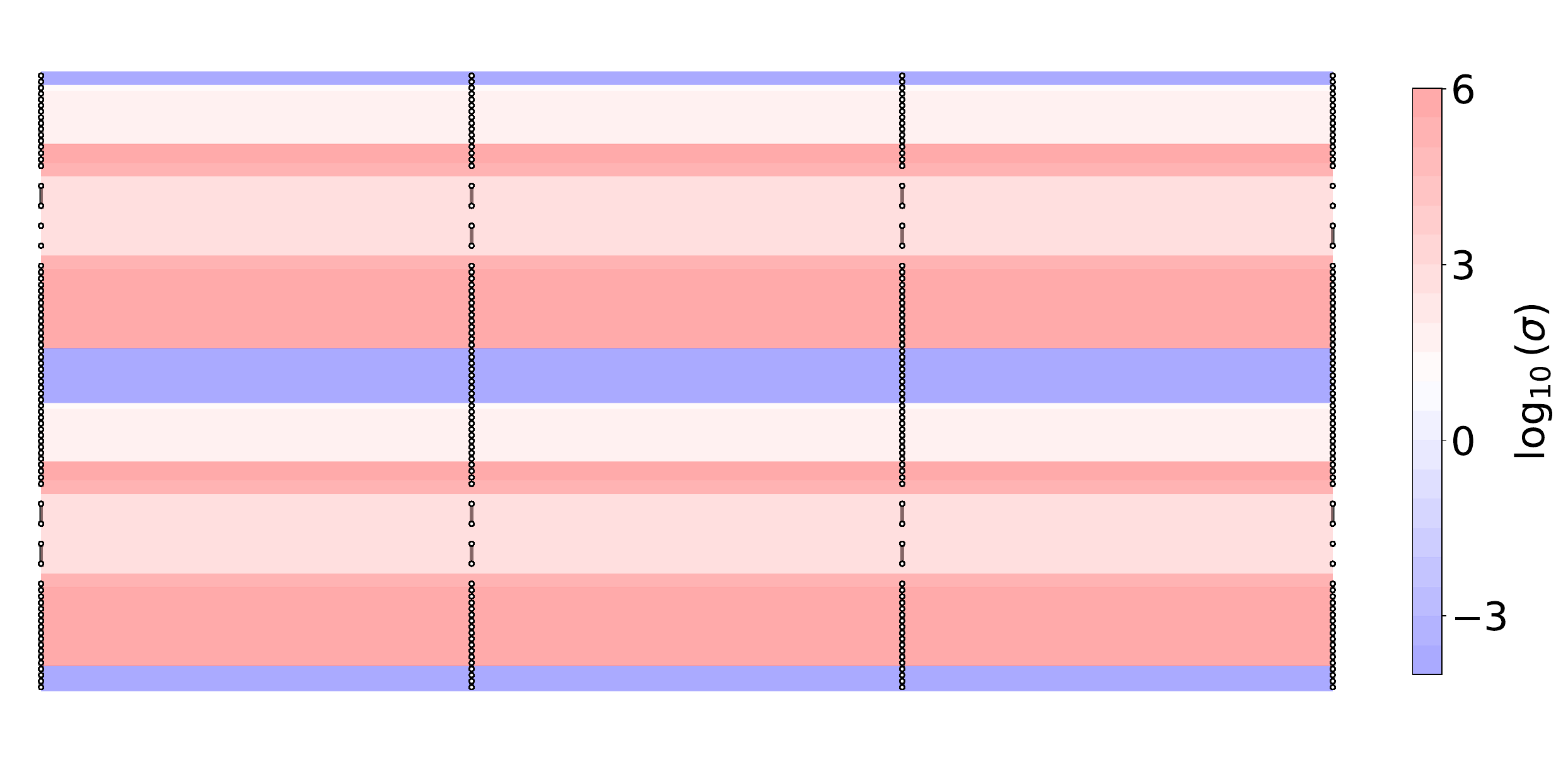}
\caption{Material-based pointwise dropping, $\dropTol=0.08$.}
\label{fig:battery-material-agg}
\end{subfigure}
\caption{Aggregates for the initial (coarse) thermal battery mesh. Singletons are represented by a single point.}
\label{fig:battery-agg}
\end{figure}

To further illustrate the inability of traditional smoothed aggregation based dropping schemes to account for mesh anisotropy and material interfaces, we show the aggregates for the initial coarse mesh resolution in \Cref{fig:battery-agg}.
Several aggregates for \(\dropCriterion^{\text{pw}}(\strength^{sa})\) span across material interfaces.
Further, pointwise smoothed aggregation does not effectively `semi-coarsen' the radially-stretched elements when constructing the coarser level.
The pointwise material dropping, however, cleanly separates the material interfaces in the coarse grid representation.
As a consequence, pointwise material dropping constructs higher quality, richer coarse spaces that function to effectively `semi-coarsen' the thin elements in the initial coarse mesh resolution.

We consider the 16-fold radially refined case in \Cref{fig:thermal-battery-16x}.
With the mesh refinement in the radial direction, the mesh for this case features elements with aspect ratios significantly closer to unity than the case in the preceding paragraph.
While the {\DistanceLaplacian} dropping scheme achieved low iteration count and cost at fairly low drop tolerances ($\dropTol\leq 0.01$) in the anisotropic case in \Cref{fig:thermal-battery-1x}, we observe in \Cref{fig:thermal-battery-16x} that the iteration count and cost do not improve over the no dropping option for all drop tolerances, irrespective of using either pointwise dropping or cut-drop methods.
Incorporating mesh information into the dropping scheme seems to negatively affect the convergence of the multigrid method in this instance.
Classical smoothed aggregation with pointwise dropping, on the other hand, performs reasonably well for all drop tolerances greater than zero, $\dropTol\geq0$.
This is the opposite result for the anisotropic mesh in the previous case; there, the {\DistanceLaplacian} based schemes provided significant improvement over the smoothed aggregation based dropping schemes.
One constant between the two cases, however, is the ability of the material-based approach to provide the best iteration count and cost.
Using mesh and material property information, pointwise material dropping achieves comparable iteration count and costs relative to pointwise smoothed aggregation dropping, provided $\dropTol\geq0.01$.
We observe that pointwise material dropping provides superior iteration count and cost relative to pointwise smoothed aggregation, provided a sufficiently high drop tolerance $\dropTol\geq0.08$ is used.
Moreover, pointwise material dropping with $\dropTol=0.08$ reduces the iteration count relative to pointwise SA dropping with $\dropTol=0.04$ from 34 to 17 while incurring only a modest increase in the smoother complexity from $1.32$ to $1.70$.
As a result, the best configuration for the material-based pointwise dropping reduces the overall solve cost relative to the best pointwise material SA dropping by a factor of $1.5$.
Finally, we note that while the iteration count is improved with larger drop tolerances, the \emph{cost} for the material-based pointwise dropping scheme is minimized at $\dropTol=0.16$ with $\dropTol=0.08$ performing comparably.

While material-based dropping provides better iteration counts and costs with sufficiently large drop tolerances, the same cannot be said regarding the cut-drop method.
\Cref{fig:thermal-battery-16x} shows that SA with cut-drop is less
costly than the material-based scheme with cut-drop for all drop tolerances considered.
The authors remark that, as the drop tolerance grows, both the
iteration count and overall computational cost consistently
decline. This trend implies that choosing even larger $\dropTol$
values
could yield additional improvements.

We now consider the 16-fold radially refined case with a single uniform mesh refinement in \Cref{fig:thermal-battery-16x-umr-1}.
As expected in the case of uniform mesh refinement, the results in \Cref{fig:thermal-battery-16x-umr-1} closely mirror those presented in \Cref{fig:thermal-battery-16x}.
The authors note that the material-based scheme with pointwise dropping provides the lowest iteration and cost that are minimized at $\dropTol=0.08$ with $\dropTol=0.16$ providing comparable performance.
Similar to the 16-fold radially refined case, we observe that the material-based scheme with pointwise dropping outperforms smoothed aggregation with pointwise dropping in iteration count and cost when $0.02\leq\dropTol\leq 0.16$.
The overall findings mostly coincide with the results discussed in \secref{subsec:Example_1}.
Pointwise SA-based dropping and material-based dropping outperform the {\DistanceLaplacian} {\soc}.
With exception to $\dropTol=0.32$ and $\dropTol=0.64$, the pointwise approaches easily outcompete the cut-drop based schemes.
Finally, material-based pointwise dropping yields the lowest iteration and cost observed with $\dropTol=0.08$.

\begin{figure}
\centering
\includegraphics[width=\textwidth]{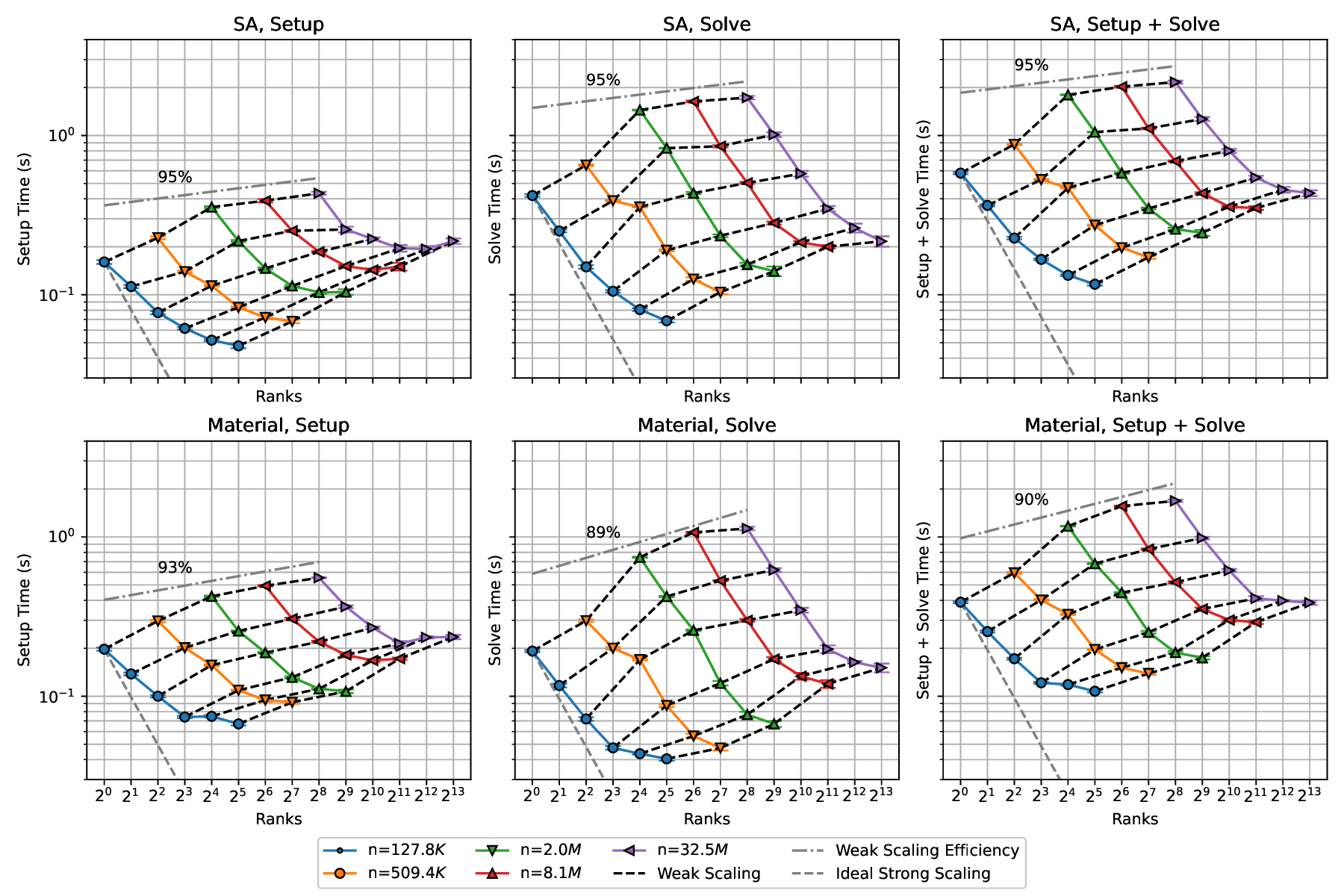}
\caption{Scaling results for the thermal battery problem, starting at 16x radial refinement and progressing through four levels of uniform mesh refinement.
smoothed aggregation-based pointwise dropping with $\dropTol=0.04$ is compared against material-based dropping with $\dropTol=0.08$.
Multigrid setup, solve, and combined times are reported. Ideal strong scaling is shown for reference, along with a line corresponding to the estimated weak scaling efficiency over the three largest problem scales. Ideal weak scaling is a horizontal line.}
\label{fig:thermal-battery-scaling}
\end{figure}

To assess the scalability of the pointwise smoothed aggregation and material-based algorithms, we perform a strong and weak scaling study using between one and \num{8192} processors on the Eclipse system at Sandia National Laboratories, which features a dual socket, 18 core Intel Broadwell E5-2695 processor per node.
The lowest cost configuration for pointwise SA-based dropping from \Cref{fig:thermal-battery-16x,fig:thermal-battery-16x-umr-1} is used with $\dropTol=0.04$.
Similarly, \Cref{fig:thermal-battery-16x,fig:thermal-battery-16x-umr-1} demonstrate that $\dropTol=0.08$ yields the lowest cost for pointwise material-based dropping.
We start with the 16-fold radially refined case with $1.27\times 10^5$ degrees of freedom as a baseline, and perform four levels of uniform mesh refinement up to $3.25\times 10^7$.
Each problem scale starts at an initial processor count that is then doubled a total of 5 times for strong scaling.
Dashed lines indicating constant degrees of freedom per processor are drawn to demonstrate weak scaling.
The setup and solve phases of the AMG solver are repeated 100 times for each configuration, with the first run discarded as an initial warmup.
The mean setup, solve, and combined setup plus solve times are reported in \Cref{fig:thermal-battery-scaling}, with the error bars representing one standard deviation from the mean.

The scaling results in \Cref{fig:thermal-battery-scaling} confirm that the material-based pointwise approach yields faster solution times than pointwise smoothed aggregation.
These results are in line with the cost predictions from \Cref{fig:thermal-battery-16x,fig:thermal-battery-16x-umr-1}.
The setup costs for pointwise smoothed aggregation, however, are smaller than the material-based scheme.
The higher setup costs for the material-based scheme may be mitigated in scenarios with successive right-hand side solves where the matrix remains fixed, such as in the solution to the incompressible Navier--Stokes equations~\cite{Fischer1998a,Austin2021a}.
In that case, the multigrid setup cost is only required once.
Despite the higher setup costs, the combined setup plus solve for the material-based approach outperforms pointwise smoothed aggregation across all problem sizes and resource configurations considered.
Past the combined setup and solve strong scaling limit ($n/P<\num{15000}$), setup related costs for material-based dropping start exceed the solve cost.
Despite this, material-based dropping scheme exhibits better end-to-end setup plus solve costs than the SA-based dropping scheme.
In terms of the solve time, moreover, the material-based dropping provides significant speed ups relative to SA-based dropping.
For example, smoothed aggregation based dropping only reaches sub-second solve times for the $n=3.25\times 10^7$ case
only after utilizing $2^{10}$ MPI ranks with a solve time of $0.58$s.
The sub-second solve time feat is nearly accomplished by the material-based scheme with as few as $2^8$ MPI ranks.
Under the same resource utilization of $2^{10}$ MPI ranks, the material-based approach reaches solve times of $0.35$s.

Strong scaling efficiencies at or above 50\% are achieved for the multigrid setup phase provided $n/P\sim\num{30000}$ for the material-based and SA pointwise dropping.
Despite the somewhat better strong scalability of the setup phase of the material-based approach, the overall time-to-setup at the strong scaling limit favors the SA-based dropping scheme by a small amount.
The solve phase, however, exhibits significantly better strong scalability with $n/P\sim\num{8000}$ being the 50\% efficiency point for material-based and smoothed aggregation dropping.
At the strong scaling limit, material-based dropping achieves faster time to solutions compared to the smoothed aggregation approach.
For example, material-based dropping achieves solves in $0.17$s for $n=3.25\times 10^7$ using $2^{12}$ ranks, whereas the SA-based approach takes $0.26$s for the same resource use and problem size.
For combined setup plus solve, the strong scaling limit for smoothed aggregation and material dropping is $n/P\sim\num{16000}$.
Both setup time and solve time exhibit problem size dependence despite fixing $n/P$ in the weak scaling lines.
To a point, this is expected.
For a two dimensional Poisson problem, a minimum of 9 processors is needed to fully saturate the communication stencil for the matrix-vector products required in the multigrid \Chebychev{} smoother and residual evaluation.
This explains the relatively poor weak scaling of the multigrid solve phase from $n=1.27\times 10^5$ to $n=2.03\times 10^6$.
The second expected performance impact in weak scaling comes from going from purely on-node communication to off-node communication, which occurs at 32 ranks for Eclipse.
Despite this, we observe good weak scaling near 95\% efficiency for setup and solve between $n=2.03\times 10^6$ and $n=3.25\times 10^7$ for the SA dropping scheme.
The material-based scheme maintains decent weak scaling at 93\% efficiency for setup and 89\% efficiency for the solve phase.

We conclude our discussion of the thermally activated battery cases by commenting on the relative performance of each dropping scheme across the three cases.
In the highly anisotropic mesh, we observe that {\DistanceLaplacian}-based dropping schemes greatly outperforms the smoothed aggregation based approaches.
After performing a 16-fold radial mesh refinement, however, the smoothed aggregation based methods now outperform the {\DistanceLaplacian}-based schemes.
The implication of this result is that, upon performing a mesh refinement, a user would need to change the solver settings to achieve optimal performance.
This requirement is avoided with the material-based pointwise method, however, which exhibits near best performance with a fixed $\dropTol=0.08$, for example.
While smoothed aggregation is robust with respect to the jumps in the material coefficient, it is not robust with respect to both jumps in the material coefficient and mesh anisotropy.
In contrary, {\DistanceLaplacian}-based methods exhibit the opposite behavior; they are robust to mesh anisotropy,
but not to jumps in the material properties.
The material-based pointwise dropping method, however, combines the best properties of the smoothed aggregation based and {\DistanceLaplacian}-based methods; material-based dropping is robust both to jumps in the material coefficient and mesh anisotropy.

\subsection{Solar cell example}

\begin{figure}
\centering
\begin{subfigure}[b]{0.49\textwidth}
\centering
\includegraphics[width=\textwidth]{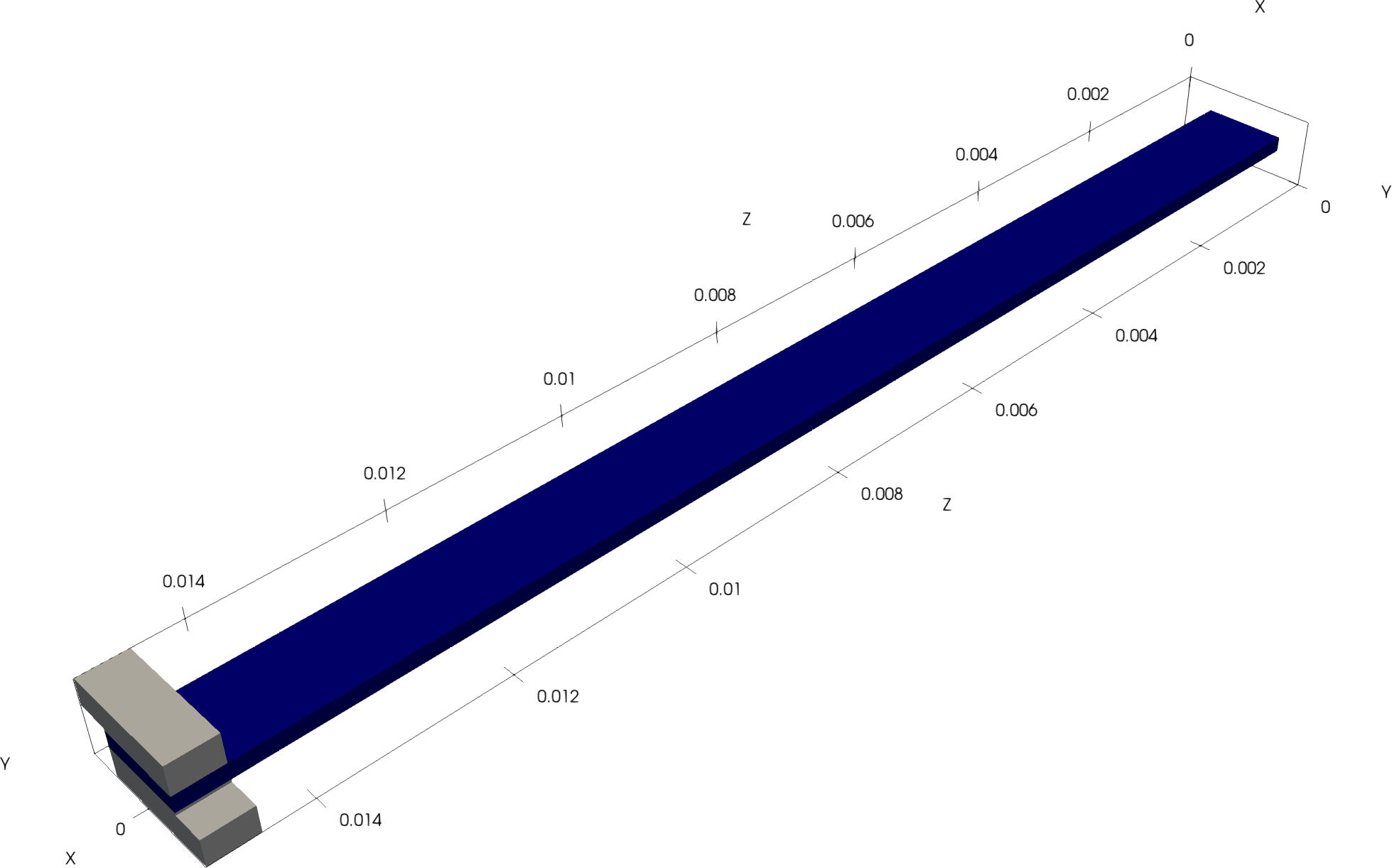}
\caption{Geometry of single solar cell model with interconnects.
Silver colored regions are the silver interconnects with isotropic electrical conductivity, $\material=6.21\times10^7$ \unit{\siemens\per\meter}.
The dark blue region is the cell with highly anisotropic electrical conductivity, $\material=\diag{0.05, 0.05, 3.6\times 10^6}$ \unit{\siemens\per\meter}.}
\label{fig:solar_cell_geometry}
\end{subfigure}
\hfill
\begin{subfigure}[b]{0.49\textwidth}
\centering
\input{figures/solar_cell_circuit.tex}
\caption{Equivalent circuit of a single diode solar cell.}
\label{fig:equivalent_circuit}
\end{subfigure}
\caption{Model of a unit-volume of a full solar cell.}
\label{fig:solar_cell}
\end{figure}

The performance of photovoltaic cells can be significantly impacted by the effects of cracking~\cite{goudelis2022review}, shading~\cite{ramezani2025shading}, and Joule heating~\cite{shang2017photovoltaic}.
To better understand the impact of these effects, detailed modeling of the governing physics is required.
We consider the model problem depicted in \Cref{fig:solar_cell}, which represents a simplified `unit volume' of a full solar cell.
The resulting mesh for the geometry in \Cref{fig:solar_cell_geometry} is of high quality; it features hexahedral elements that have a scaled Jacobian no smaller than 0.58 throughout the domain, and the majority of the elements have a unity scaled Jacobian.
The domain is comprised of two different materials: the first are the silver interconnects with large, isotropic electrical conductivities, $\material=6.21\times10^7$ \unit{\siemens\per\meter};
the second represents the solar cell bulk material with highly anisotropic electrical conductivites, $\material=\diag{0.05, 0.05, 3.6\times 10^6}$ \unit{\siemens\per\meter}.
We solve for the voltage, $V$, which is governed by Ohm's law
\begin{equation}\label{eq:voltage}
\nabla \cdot \left(-\material(\coord)\nabla V\right) = S,
\end{equation}
where $S$ is a source term.
The source term $S$ represents the output current from the equivalent circuit of a single diode solar cell as shown in \Cref{fig:equivalent_circuit}.
This current source term is provided by the characteristic equation
\begin{equation}\label{eq:solar_cell}
  S = I = I_L - I_0\left[\exp{\left(\dfrac{V+IR_s}{nV_T}\right)}\right] - \dfrac{V+IR_s}{R_{sh}},
\end{equation}
where $V$ is the voltage (\unit{\volt}), $I_L$ is the light-generated current (\unit{\A}), $I_0$ the diode reverse saturation current (\unit{\ampere}), $R_s$ is the series resistance ($\Omega$), $R_{sh}$ is the shunt resistance ($\Omega$), $n$ is the diode ideality factor (dimensionless), and finally $V_T$ is the thermal voltage given by
\begin{equation*}
  V_T = \dfrac{kT_c}{q}
\end{equation*}
with Boltzmann's constant $k=1.381\times 10^{-23}$ \unit{\joule\per\kelvin} and the elementary charge $q=1.602\times10^{-19}$ \unit{\coulomb}.
For a more complete introduction to the physics of solar cells, the reader is referred to the introduction provided by Gray~\cite{gray2011physics}.

\begin{figure}
\centering
\includegraphics[width=\textwidth]{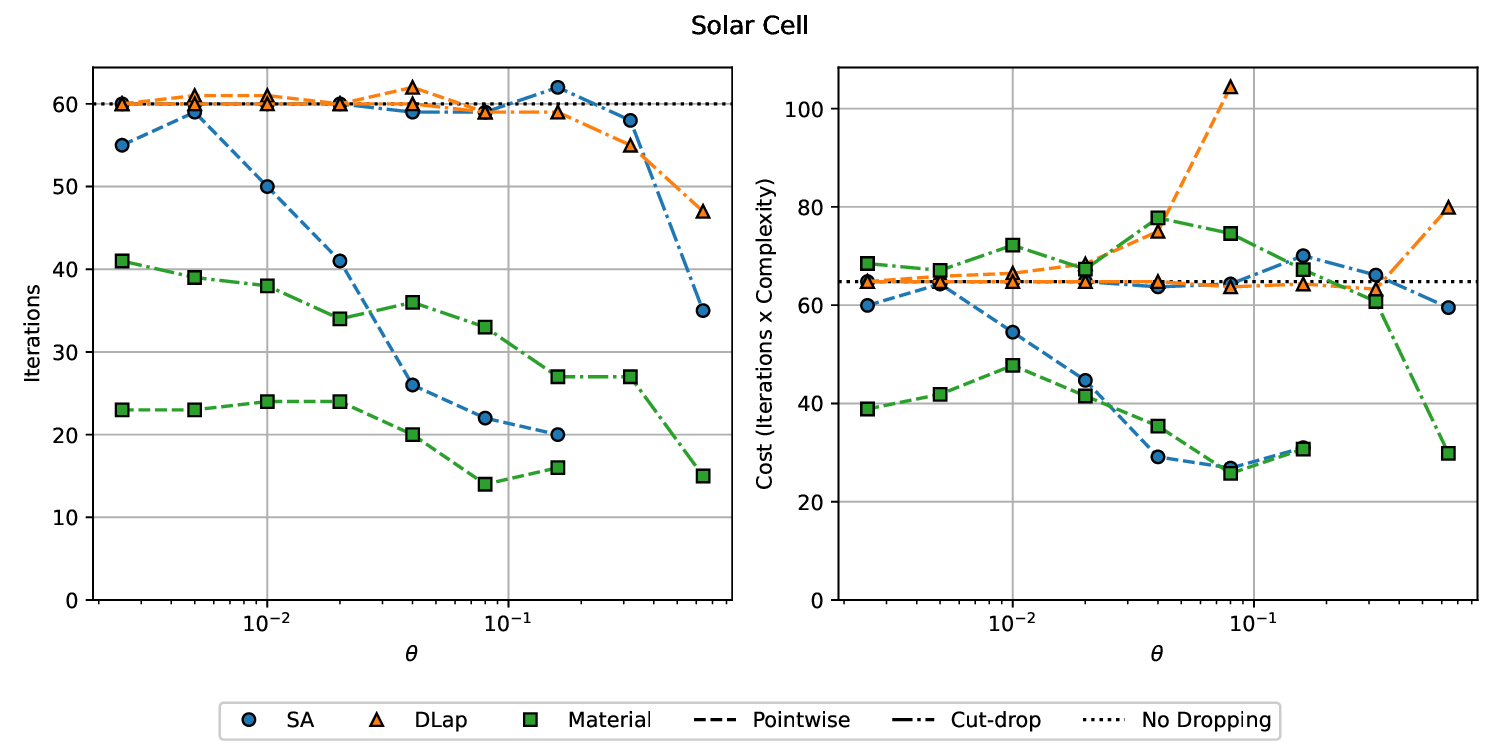}
\caption{Number of iterations and cost of application shown over different drop tolerances $\dropTol$ for
combinations of dropping criteria $\dropCriterion^{\text{pw}}$, $\dropCriterion^{\text{cut-drop}}$ and {\soc}
measures $\strength^{sa}$ , $\strength^{dlap}$ and $\strength_{\sigma}^{dlap}$ for the solar cell case.
Configurations that fail to reach the designated coarse grid size or fail to converge are not shown.
The {\DistanceLaplacian} {\soc} offers little improvement both in iteration count and cost over no dropping.
Material based dropping yields the lowest iteration count and cost across nearly all drop tolerances $\dropTol\leq 0.16$.}
\label{fig:solar_cell_results}
\end{figure}

For our experiments we terminate the linear solver with an absolute $l^2$-norm residual of $10^{-6}$.
We terminate multigrid coarsening when the number of unknowns drops
below $\num{5000}$ rows on a given level.
The drop tolerance is varied from $\dropTol=0$ (no dropping) to $\dropTol=0.64$.
Results for the solar cell case are shown in \Cref{fig:solar_cell_results}.
Provided $\dropTol\leq 0.16$, pointwise material dropping provides the lowest iteration count and cost solver in nearly every scenario,
reaching its optimal configuration with $\dropTol=0.08$.
The one exception observed is that SA-based pointwise dropping outperforms the material-based approach at $\dropTol=0.04$.
For $\dropTol\geq 0.32$,the multigrid coarsening
stagnates and fails to reach the desired maximum coarse grid size of 5000 within the allotted maximum of 10 levels.
Pointwise smoothed aggregation has slightly higher cost than pointwise material, but require more iterations.
Similar to pointwise material, pointwise smoothed aggregation reaches its minimum
cost at $\dropTol=0.08$ and stagnates when $\dropTol\geq 0.32$. The {\DistanceLaplacian} and cut-drop approaches
provide only marginal improvement to no dropping in a few limited cases. The former result is not surprising, given
the available high-quality mesh. Although material cut-drop provides solvers that improve the iteration count relative
to no dropping, the cost to apply the setup is generally prohibitively expensive compared to pointwise SA or material dropping.
One notable exception is material-based cut-drop with $\dropTol=0.64$, which provides comparable performance to the lowest cost pointwise material-based dropping with $\dropTol=0.08$.
We conclude that pointwise material dropping
provides good solvers that are robust with respect to the particular choice of $\dropTol$ with pointwise smoothed
aggregation as a second option, similar to the refined cases in \secref{sec:battery}.
Concluding, the newly introduced material-based {\DistanceLaplacian} {\soc} measure proves to be not only competitive with
existing methods, but shows to be superior for three-dimensional application cases featuring a complex material behavior.

\section{Concluding remarks}
\label{sec:conclusion}

In this paper, we have presented a new material-aware {\soc} measure for smoothed aggregation AMG
methods, which improves robustness for problems with highly heterogeneous material distributions
and strong anisotropies. These kind of systems commonly appear in engineering applications and severely degrade
algebraic multigrid performance. Explicitly incorporating material tensor information into the
coarsening process addresses shortcomings of classical {\soc} and distance-based measures, enabling
more accurate detection of weak connections across material interfaces and alignment with anisotropy
directions. This results in coarse grid hierarchies that better represent the underlying physics of
heterogeneous and anisotropic problems and therefore result in better convergence properties.

Through a series of academic tests and practical applications, including anisotropic thermal diffusion,
thermally activated batteries and solar cell units, we demonstrated that the method is robust with respect
to coefficient jumps, anisotropies, and mesh variations. Across these test cases, the material-based
{\soc} consistently outperformed classical smoothed aggregation and {\DistanceLaplacian} approaches,
both in terms of iteration counts and overall solver cost. Parameter studies showed that the method
remains effective over a wide range of drop tolerances, with different dropping criterion variants
offering flexibility depending on performance or robustness needs. Importantly, scalability and parallel
performance tests confirmed the suitability for large-scale high-performance computing simulations.

Overall, the proposed material-aware {\soc} combines the advantages of existing {\soc} measures while
improving the handling of material properties, providing a robust, scalable, and efficient algebraic multigrid
method for scalar partial differential equations with challenging material distributions and anisotropies.
Future work will focus on extending this approach to different types of partial differential equations such
as Maxwell equations or equation systems like they appear in computational solid mechanics in form of the
equations of elasticity. Solving multiphysics applications with block AMG methods in combination with
the presented {\soc} measure is another point of interest, as well as the construction of surrogate Green's
function like {\soc} measures. An open-source software implementation of these methods is available in
Trilinos' multigrid package {\muelu} \cite{Trilinos}.

\bmsection*{Data Availability Statement}

Building blocks of the AMG preconditioner developed and applied in this study are openly available in
Trilinos at https://github.com/trilinos/Trilinos\cite{Heroux2005a}\cite{Trilinos}. All other data that support
the findings of this study are available from the corresponding author upon reasonable request.

\bmsection*{Acknowledgments}

The authors thank James Yuan Hartley for providing the problem setup for the solar cell problem.

M. F., M. M. and A. P. acknowledge funding by \emph{dtec.bw - Digitalization and Technology Research Center of the Bundeswehr}
under the project ``hpc.bw - Competence Platform for High Performance Computing''.
dtec.bw is funded by the European Union – NextGenerationEU.

This work was supported by the Laboratory Directed Research and Development program (Project 236939) at Sandia National Laboratories, a multimission laboratory managed and operated by National Technology and Engineering Solutions of Sandia LLC, a wholly owned subsidiary of Honeywell International Inc. for the U.S. Department of Energy’s National Nuclear Security Administration under contract DE-NA0003525.
SAND2026-16734O

\bmsection*{Conflict of interest}

The authors declare that they have no known competing financial interests or personal relationships
that could have appeared to influence the work reported in this paper.

\bibliography{bibliography}

\end{document}

%% file: defines.tex
\newcommand{\todo}[1]{\textcolor{red}{{\normalfont\bfseries ToDo:} #1}}
\newcommand{\todomath}[1]{\text{\textcolor{red}{{\normalfont\bfseries ToDo:} #1}}}
\newcommand{\verify}[1]{\textcolor{magenta}{{\normalfont\bfseries Verify:} #1}}
\newcommand{\verifymath}[1]{\text{\textcolor{magenta}{{\normalfont\bfseries Verify:} #1}}}
\newcommand{\remove}[1]{\textcolor{blue}{{\normalfont\bfseries Remove?:} #1}}
\newcommand{\citecheck}[1]{\text{\textcolor{red}{(Check citation \cite{#1})}}}
\newcommand{\addcite}{\textcolor{red}{[add cit.]}}
\newcommand{\addCite}[1]{\textcolor{red}{[add cit. '#1']}}
\newcommand{\addref}{\textcolor{red}{[section]}}
\newcommand{\addRef}[1]{\textcolor{red}{[reference to '#1']}}
\newcommand{\headword}[1]{\textcolor{blue}{Headword: #1}\\}

\newcommand{\chapref}[1]{Chapter~\ref{#1}} %
\newcommand{\Chapref}[1]{Chapter~\ref{#1}} %
\newcommand{\secref}[1]{Section~\ref{#1}} %
\newcommand{\Secref}[1]{Section~\ref{#1}} %
\newcommand{\secsref}[2]{Sections~\ref{#1} and~\ref{#2}} %
\newcommand{\secssref}[3]{Sections~\ref{#1}, \ref{#2}, and~\ref{#3}} %
\newcommand{\secsrangeref}[2]{Sections~\ref{#1} -- \ref{#2}} %
\newcommand{\appref}[1]{\ref{#1}} %
\newcommand{\Appref}[1]{\ref{#1}} %
\newcommand{\defref}[1]{Definition~\ref{#1}} %
\newcommand{\Defref}[1]{Definition~\ref{#1}} %
\newcommand{\figref}[1]{Figure~\ref{#1}} %
\newcommand{\figsref}[2]{Figures~\ref{#1} and~\ref{#2}} %
\newcommand{\figssref}[3]{Figures~\ref{#1}, \ref{#2}, and~\ref{#3}} %
\newcommand{\Figref}[1]{Figure~\ref{#1}} %
\newcommand{\figsrangeref}[2]{Figures~\ref{#1} -- \ref{#2}} %
\newcommand{\tabref}[1]{Table~\ref{#1}} %
\newcommand{\Tabref}[1]{Table~\ref{#1}} %
\newcommand{\algsref}[2]{Algorithms~\ref{#1} and~\ref{#2}} %
\newcommand{\Algref}[1]{Algorithm~\ref{#1}} %
\newcommand{\remref}[1]{Remark~\ref{#1}} %
\newcommand{\Remref}[1]{Remark~\ref{#1}} %

\hyphenation{bio-medicine}

\newcommand{\teo}[1]{\ensuremath{\boldsymbol{#1}}} %
\newcommand{\mao}[1]{\ensuremath{\mathbf{#1}}} %
\newcommand{\tet}[1]{\ensuremath{\boldsymbol{#1}}} %
\newcommand{\mat}[1]{\ensuremath{\mathbf{#1}}} %
\newcommand{\tef}[1]{\ensuremath{\mathds{#1}}} %
\newcommand{\maf}[1]{\ensuremath{\underline{\mathds{#1}}}} %

\newcommand{\lagMultDiscVec}{\bm{\uplambda}} %

\newcommand{\id}{\ensuremath{I}} %
\newcommand{\idtet}{\tet{\id}} %
\newcommand{\idmat}{\mat{\id}} %
\newcommand{\idtef}{\tef{\id}} %

\newcommand{\mr}[1]{\ensuremath{\mathrm{#1}}} %
\newcommand{\script}[1]{\ensuremath{\eucal{#1}}}

\newcommand{\mbs}[1]{\ensuremath{\boldsymbol{#1}}} %
\newcommand{\define}[1]{\emph{#1}} %

\newcommand{\indexedTime}[2]{#1_{#2}} %
\newcommand{\indexedIter}[2]{#1^{#2}} %
\newcommand{\indexedTimeIter}[3]{#1_{#2}^{#3}} %
\newcommand{\indexedRow}[2]{#1_{#2}} %
\newcommand{\indexedRowCol}[3]{#1_{#2#3}} %
\newcommand{\indexedRowIter}[3]{#1_{#2}^{#3}} %
\newcommand{\indexedDom}[2]{#1^{\left(#2\right)}} %
\newcommand{\indexedNode}[2]{#1_{#2}} %

\newcommand{\indexArbitraryOne}{a}
\newcommand{\indexArbitraryTwo}{b}

\newcommand{\indexedDomain}[2]{#1^{#2}}

\newcommand{\inv}[1]{{#1}^{-1}} %
\newcommand{\trans}[1]{{#1}^{\mr{T}}} %
\newcommand{\invTrans}[1]{{#1}^{\mr{-T}}} %

\newcommand{\diagElement}{d}
\newcommand{\diagEntry}[1]{\diagElement_{#1}}
\newcommand{\diag}[1]{\mathrm{diag}\left(#1\right)}
\newcommand{\graph}{\mathcal{G}}
\newcommand{\graphOf}[1]{\graph\left(#1\right)}

\newcommand{\Disc}{\mr{h}} %
\newcommand{\Exact}{\mr{ex}} %
\newcommand{\disc}[1]{#1_{\Disc}} %

\newcommand{\textfrac}[2]{#1/#2} %

\newcommand{\Der}[2]{#1,_{#2}} %
\newcommand{\pDer}[2]{\frac{\partial #1}{\partial #2}} %
\newcommand{\pDerText}[2]{\textfrac{\partial #1}{\partial #2}} %
\newcommand{\spDer}[2]{\frac{\partial^2 #1}{\partial #2^2}} %
\newcommand{\multipDer}[3]{\frac{\partial^{#3} {#1}}{\partial {#2}^{#3}}} %
\newcommand{\multipDerText}[3]{\textfrac{\partial^{#3} {#1}}{\partial {#2}^{#3}}} %
\newcommand{\tDer}[2]{\frac{\mr d #1}{\mr d #2}} %
\newcommand{\stDer}[2]{\frac{\mr d^2 #1}{\mr d #2^2}} %

\newcommand{\TimeDer}[2]{\frac{\partial^{#2} {#1}}{\partial {\ttime}^{#2}}} %
\newcommand{\TimeDerCompact}[2]{{#1}^{(#2)}\left(\ttime\right)} %
\newcommand{\TimeDerCompactAt}[3]{{#1}^{(#2)}\left(#3\right)} %

\newcommand{\RealSp}{\mathbb{R}} %
\newcommand{\REalSp}[1]{\RealSp^{#1}} %

\newcommand{\abs}[1]{\left |#1\right |}
\newcommand{\norm}[1]{\left\|#1\right\|} %
\newcommand{\normOne}[1]{\norm{#1}_{1}} %
\newcommand{\normTwo}[1]{\norm{#1}_{2}} %
\newcommand{\normInf}[1]{\norm{#1}_{\infty}} %
\newcommand{\normFrobenius}[1]{\norm{#1}_{\mathrm{F}}} %
\newcommand{\normMaterial}[1]{\norm{#1}_{\sigma}} %

\newcommand{\dx}{\,\mr{d}} %

\newcommand{\Tol}{\varepsilon} %
\newcommand{\TolNonLin}{\Tol^\mr{nln}} %
\newcommand{\TolNonLinRes}{\sub{\TolNonLin}{\mao\res}} %
\newcommand{\TolNonLinInc}{\sub{\TolNonLin}{\Delta\mao\sol}} %
\newcommand{\TolLin}{\Tol^\mr{lin}} %

\newcommand{\bsigma}{\mbs{\sigma}}
\newcommand{\btau}{\mbs{\tau}}
\newcommand{\bxi}{\mbs{\xi}}
\newcommand{\bchi}{\mbs{\chi}}
\newcommand{\bthe}{\mbs{\theta}}
\newcommand{\beps}{\mbs{\varepsilon}}
\newcommand{\boeta}{\mbs{\eta}}
\newcommand{\blamb}{\mbs\lambda}
\newcommand{\bmu}{\mbs\mu}
\newcommand{\bvth}{\mbs\vartheta}
\newcommand{\bvphi}{\mbs\varphi}
\newcommand{\bvpsi}{\mbs\psi}

\newcommand{\timeN}[1]{\indexedTime{#1}{\indTimeStep}} %
\newcommand{\timeNP}[1]{\indexedTime{#1}{\indTimeStep+1}} %
\newcommand{\ttime}{\ensuremath{t}} %
\newcommand{\ttimeZero}{\indexedTime{\ttime}{0}} %
\newcommand{\ttimeend}{\ensuremath{T}} %
\newcommand{\timeinterval}{\left(0,\ttimeend\right)} %
\newcommand{\Dt}{\Delta t} %

\newcommand{\anyQuantity}{\left(\bullet\right)} %
\newcommand{\anyMatrix}{H} %
\newcommand{\anyRhs}{f}
\newcommand{\pow}[2]{#1^{#2}} %
\newcommand{\HOT}[2]{\mathcal{O}\left(#1^{#2}\right)} %

\newcommand{\grad}{\mathrm{\nabla}}
\newcommand{\divergence}[1]{\mathrm{\nabla} \cdot #1} %

\newcommand{\ShapeFunc}{N} %
\newcommand{\ShapeFuncLM}{\varPhi} %
\newcommand{\stiffele}{k} %
\newcommand{\stiffglobal}{K} %
\newcommand{\hEle}{h} %
\newcommand{\ParamCoord}{\xi} %
\newcommand{\ParamCoordVec}{\mbs\bxi} %
\newcommand{\ParamCoordAlt}{\eta} %

\newcommand{\dom}{\ensuremath{\Omega}} %
\newcommand{\domi}{\ensuremath{\dom_{\matCoord}}} %
\newcommand{\domc}{\ensuremath{\dom_{\spatCoord}}} %
\newcommand{\domr}{\ensuremath{\dom_{{\bchi}}}} %
\newcommand{\Dom}[1]{\indexedDom{\dom}{#1}} %
\newcommand{\domDisc}{\dom_{\Disc}}
\newcommand{\boundary}{\ensuremath{\Gamma}}

\newcommand{\matConfig}[1]{#1_{0}} %
\newcommand{\spatConfig}[1]{#1} %

\newcommand{\trac}{h} %
\newcommand{\normal}{n} %
\newcommand{\lmb}{\blamb} %
\newcommand{\lm}{\lambda} %

\newcommand{\spatCoord}{\mathrm{x}} %
\newcommand{\matCoord}{X} %
\newcommand{\coord}{x}
\newcommand{\distance}{d}

\newcommand{\weakForm}{\delta\mathscr{W}} %
\newcommand{\WeakForm}[1]{\weakForm_{#1}} %

\newcommand{\variation}{\delta} %

\newcommand{\unknown}{u} %
\newcommand{\test}{v}    %
\newcommand{\material}{\sigma} %

\newcommand{\kronecker}[2]{\delta_{{#1}{#2}}} %

\newcommand{\average}[1]{\overline{#1}} %

\newcommand{\jacobian}{J} %

\newcommand{\indLinIter}{n} %
\newcommand{\indSweep}{k} %
\newcommand{\indSpaiSweep}{m}
\newcommand{\indRow}{i} %
\newcommand{\indCol}{j} %
\newcommand{\indBlockRow}{\iota} %
\newcommand{\indBlockCol}{\zeta} %
\newcommand{\indBaseVecOne}{\xi}
\newcommand{\indBaseVecTwo}{\eta}

\newcommand{\indexedNormal}[1]{#1_{\normal}}

\newcommand{\indDirichlet}{\mathrm{D}}
\newcommand{\indNeumann}{\mathrm{N}}
\newcommand{\indexedDirichlet}[1]{#1_\indDirichlet}
\newcommand{\indexedNeumann}[1]{#1_\indNeumann}

\newcommand{\numBlockRows}{N_{\mathrm{R}}}
\newcommand{\numBlockCols}{N_{\mathrm{C}}}
\newcommand{\indDim}{d} %
\newcommand{\nproc}{n^\mr{proc}} %
\newcommand{\ncore}{n^\mr{proc}} %
\newcommand{\nsubdomain}{M} %
\newcommand{\ndim}{\mathrm{d}} %
\newcommand{\ndof}{n^\mr{dof}} %
\newcommand{\ndofPerProc}{n^\mr{dof/proc}} %
\newcommand{\nnode}{n^\mr{nd}} %
\newcommand{\nNodesPerEle}{J} %
\newcommand{\nEle}{n^\mr{el}} %
\newcommand{\nEleMin}{n^\mr{el}_{\mr{min}}} %
\newcommand{\nRows}{M}
\newcommand{\nCols}{N}
\newcommand{\nNodesSlave}{\indexedDom{n}{1}} %
\newcommand{\nNodesSlaveWithLM}{\indexedDom{m}{1}} %
\newcommand{\nNodesMaster}{\indexedDom{n}{2}} %
\newcommand{\nEleSlave}{n^\mr{el,\MoSlave}} %
\newcommand{\nEleMaster}{n^\mr{el,\MoMaster}} %
\newcommand{\nGaussPoints}{n^\mr{gp}} %
\newcommand{\nGaussPointsSlave}{n^{\mr{gp},\MoSlave}} %

\newcommand{\greensFunc}{G}

\newcommand{\wctime}{T_{\mr{wct}}} %

\newcommand{\tSetup}{T_{\mathrm{setup}}}
\newcommand{\tSolve}{T_{\mathrm{solve}}}
\newcommand{\tTotal}{T_{\mathrm{total}}}

\newcommand{\linearOperator}{\mat{\mathcal{A}}} %
\newcommand{\sol}{x} %
\newcommand{\rhs}{b} %

\newcommand{\residualNonlinear}{f}
\newcommand{\residualLinear}{r}

\newcommand{\approxMat}[1]{\widehat{#1}}

\newcommand{\matIdentity}{\mat{I}}
\newcommand{\matZero}{\mat{0}}

\newcommand{\linOp}{\boldsymbol{\mathcal{A}}}

\newcommand{\matFactD}{\mathcal{D}}
\newcommand{\matFactL}{\mathcal{L}}
\newcommand{\matFactU}{\mathcal{U}}

\newcommand{\myPrec}{P}
\newcommand{\precMatrix}{\mat{\myPrec}}
\newcommand{\schur}{S}
\newcommand{\schurMatrix}{\mat{\schur}}
\newcommand{\schurMatrixApprox}{\approxMat{\schurMatrix}}

\newcommand{\opComplexity}{\mathcal{C}}
\newcommand{\level}{\ell}
\newcommand{\indexedLevel}[2]{{#1}^{(#2)}}
\newcommand{\noLevels}{L}
\newcommand{\aggregate}{\mathcal{A}}
\newcommand{\smoother}{\mathcal{S}}
\newcommand{\dropTol}{\theta}
\newcommand{\strength}{S}
\newcommand{\dropCriterion}{C}

\newcommand{\axisorigin}{O} %
\newcommand{\xaxis}{x} %
\newcommand{\yaxis}{y} %
\newcommand{\zaxis}{z} %
\newcommand{\radial}{r}
\newcommand{\tangential}{t}

\newcommand{\distributedLoad}{q} %

\newcommand{\apriori}{\emph{a priori}}
\newcommand{\Apriori}{\emph{A priori}}
\newcommand{\beamsolid}{beam\hyp{}solid}
\newcommand{\blockLU}{Block\hyp{}LU}
\newcommand{\bruteforce}{brute\hyp{}force}
\newcommand{\FESquare}{FE\textsuperscript{2}}
\newcommand{\fiberfluid}{fiber\hyp{}fluid}
\newcommand{\fibersolid}{fiber\hyp{}solid}
\newcommand{\Fibersolid}{Fiber\hyp{}solid}
\newcommand{\fixedpoint}{fixed\hyp{}point}
\newcommand{\inhouse}{in\hyp{}house}
\newcommand{\LTwo}{L\textsubscript{2}}
\newcommand{\meshtying}{meshtying}
\newcommand{\Meshtying}{Meshtying}
\newcommand{\mixeddimensional}{mixed\hyp{}dimensional}
\newcommand{\mortartype}{mortar\hyp{}type}
\newcommand{\multidimensional}{multi\hyp{}dimensional}
\newcommand{\multigrid}{multigrid}
\newcommand{\Multigrid}{Multigrid}
\newcommand{\multilevel}{multi\hyp{}level}
\newcommand{\Multilevel}{Multi\hyp{}level}
\newcommand{\multiphysics}{multi\hyp{}physics}
\newcommand{\Multiphysics}{Multi\hyp{}physics}
\newcommand{\MultiPhysics}{Multi\hyp{}Physics}
\newcommand{\multiscale}{multi\hyp{}scale}
\newcommand{\Multiscale}{Multi\hyp{}scale}
\newcommand{\nonconforming}{non\hyp{}conforming}
\newcommand{\Nonconforming}{Non\hyp{}conforming}
\newcommand{\nonlinear}{nonlinear}
\newcommand{\Nonlinear}{Nonlinear}
\newcommand{\nonmatching}{non\hyp{}matching}
\newcommand{\Nonmatching}{Non\hyp{}matching}
\newcommand{\nonsymmetric}{non\hyp{}symmetric}
\newcommand{\nonzero}{non\hyp{}zero}
\newcommand{\nonzeros}{{\nonzero}s}
\newcommand{\oneDthreeD}{1D/3D}
\newcommand{\oneDtwoD}{1D/2D}
\newcommand{\outofthebox}{out-of-the-box}
\newcommand{\pullout}{pull\hyp{}out}
\newcommand{\runtime}{run\hyp{}time}
\newcommand{\speedup}{speed\hyp{}up}
\newcommand{\Speedup}{Speed\hyp{}up}
\newcommand{\stateoftheart}{state-of-the-art}
\newcommand{\subblock}{sub\hyp{}block}
\newcommand{\subblocks}{{\subblock}s}
\newcommand{\submatrix}{sub\hyp{}matrix}
\newcommand{\threeDthreeD}{3D/3D}
\newcommand{\timetosolution}{time-to-solution}
\newcommand{\torsionfree}{torsion\hyp{}free}
\newcommand{\soc}{strength\hyp{}of\hyp{}connection}

\newcommand{\master}{master}
\newcommand{\Master}{Master}
\newcommand{\slave}{slave}
\newcommand{\Slave}{Slave}

\newcommand{\const}{\ensuremath{const.}} %

\newcommand{\Chebychev}{Chebychev}
\newcommand{\GaussSeidel}{Gau{\ss}--Seidel}
\newcommand{\GreenLagrange}{Green--Lagrange}
\newcommand{\Lagrangian}{Lagrangean}
\newcommand{\Lagrangean}{Lagrangean}
\newcommand{\Laplacian}{Laplacian}
\newcommand{\NewtonKrylov}{Newton--Krylov}
\newcommand{\PiolaKirchhoff}{Piola--Kirchhoff}
\newcommand{\KirchhoffLove}{Kirchhoff--Love}
\newcommand{\RugeStueben}{Ruge--St\"uben}
\newcommand{\SimoReissner}{Simo--Reissner}
\newcommand{\StVenantKirchhoff}{St.-Venant--Kirchhoff}
\newcommand{\DistanceLaplacian}{distance \Laplacian}

\newcommand{\SoftwarePackage}[1]{\textsc{#1}} %
\newcommand{\baci}{\SoftwarePackage{4C}}
\newcommand{\belos}{\SoftwarePackage{Belos}}
\newcommand{\meshpy}{\SoftwarePackage{MeshPy}}
\newcommand{\muelu}{\SoftwarePackage{MueLu}}
\newcommand{\superlu}{\SoftwarePackage{SuperLU}}
\newcommand{\teko}{\SoftwarePackage{Teko}}
\newcommand{\trilinos}{\SoftwarePackage{Trilinos}}
\newcommand{\zoltan}{\SoftwarePackage{Zoltan}}

\newcommand{\EleType}[1]{\texttt{#1}} %
\newcommand{\MeshID}[1]{\emph{#1}} %
\newcommand{\hardware}[1]{\emph{#1}} %

\newcommand{\ie}{i.e.,}
\newcommand{\eg}{e.g.,}
\newcommand{\Eg}{E.g.,}
\newcommand{\cf}{cf.}
\newcommand{\wrt}{w.r.t.}
\newcommand{\vs}{vs.}
\newcommand{\etc}{etc.}

%% file: figures/test_problem.tex
\begin{figure}
\centering
\begin{subfigure}[t]{0.49\textwidth}
\centering
\begin{tikzpicture}
\tikzstyle{every node}=[font=\footnotesize]
\tikzset{gridline/.style={very thick, draw=gray!90}}
\tikzset{meshnode/.style={very thick, draw=gray, fill=gray}}
\tikzset{relevantmeshnode/.style={very thick, draw=colUniBwOr, fill=colUniBwOr}}
\fill[fill=gray!20] (0,0) rectangle (1.5,4);
\fill[fill=gray!50] (1.5,0) rectangle (4,4);
\draw[gridline] (0,0) grid (4,4);
\fill[meshnode] (0,0) circle[radius=2pt] {};
\node[anchor=south west, text opacity=0.5] at (0,0) {1};
\fill[meshnode] (1,0) circle[radius=2pt] {};
\node[anchor=south west, text opacity=0.5] at (1,0) {2};
\fill[meshnode] (2,0) circle[radius=2pt] {};
\node[anchor=south west, text opacity=0.5] at (2,0) {3};
\fill[meshnode]  (3,0) circle[radius=2pt] {};
\node[anchor=south west, text opacity=0.5] at (3,0) {4};
\fill[meshnode]  (4,0) circle[radius=2pt] {};
\node[anchor=south west, text opacity=0.5] at (4,0) {5};
\fill[meshnode] (0,1) circle[radius=2pt] {};
\node[anchor=south west, text opacity=0.5] at (0,1) {6};
\fill[meshnode] (1,1) circle[radius=2pt] {};
\node[anchor=south west, text opacity=0.5] at (1,1) {7};
\fill[meshnode] (2,1) circle[radius=2pt] {};
\node[anchor=south west, text opacity=0.5] at (2,1) {8};
\fill[meshnode]  (3,1) circle[radius=2pt] {};
\node[anchor=south west, text opacity=0.5] at (3,1) {9};
\fill[meshnode]  (4,1) circle[radius=2pt] {};
\node[anchor=south west, text opacity=0.5] at (4,1) {10};
\fill[meshnode] (0,2) circle[radius=2pt] {};
\node[anchor=south west, text opacity=0.5] at (0,2) {11};
\fill[relevantmeshnode] (1,2) circle[radius=2pt] {};
\node[anchor=south west, text=colUniBwOr] at (1,2) {12};
\fill[relevantmeshnode] (2,2) circle[radius=2pt] {};
\node[anchor=south west, text=colUniBwOr] at (2,2) {13};
\fill[relevantmeshnode] (3,2) circle[radius=2pt] {};
\node[anchor=south west, text=colUniBwOr] at (3,2) {14};
\fill[meshnode]  (4,2) circle[radius=2pt] {};
\node[anchor=south west, text opacity=0.5] at (4,2) {15};
\fill[meshnode] (0,3) circle[radius=2pt] {};
\node[anchor=south west, text opacity=0.5] at (0,3) {16};
\fill[meshnode] (1,3) circle[radius=2pt] {};
\node[anchor=south west, text opacity=0.5] at (1,3) {17};
\fill[meshnode] (2,3) circle[radius=2pt] {};
\node[anchor=south west, text opacity=0.5] at (2,3) {18};
\fill[meshnode]  (3,3) circle[radius=2pt] {};
\node[anchor=south west, text opacity=0.5] at (3,3) {19};
\fill[meshnode] (4,3) circle[radius=2pt] {};
\node[anchor=south west, text opacity=0.5] at (4,3) {20};
\fill[meshnode] (0,4) circle[radius=2pt] {};
\node[anchor=south west, text opacity=0.5] at (0,4) {21};
\fill[meshnode] (1,4) circle[radius=2pt] {};
\node[anchor=south west, text opacity=0.5] at (1,4) {22};
\fill[meshnode] (2,4) circle[radius=2pt] {};
\node[anchor=south west, text opacity=0.5] at (2,4) {23};
\fill[meshnode]  (3,4) circle[radius=2pt] {};
\node[anchor=south west, text opacity=0.5] at (3,4) {24};
\fill[meshnode]  (4,4) circle[radius=2pt] {};
\node[anchor=south west, text opacity=0.5] at (4,4) {25};
\end{tikzpicture}
\caption{\footnotesize Discretization of the model problem with different material properties (light and dark gray) and with important nodes 12, 13, and 14 shown in orange.}
\label{fig:model_problem_discretization}
\end{subfigure}
\hfill
\begin{subfigure}[t]{0.49\textwidth}
\centering
\begin{tikzpicture}[scale=1.3]
\tikzstyle{every node}=[font=\footnotesize]
\tikzset{gridline/.style={dashed, draw=gray!70}}
\tikzset{meshnode/.style={thick, draw=gray, fill=gray}}
\tikzset{neighborhood/.style={very thick, draw=gray}}
\draw[gridline] (-1.5,-1.5) grid (1.5,1.5);
\draw[neighborhood] (-1,0) -- (0,0) -- (1,0);
\draw[neighborhood] (-1,-1) -- (0,0) -- (1,-1);
\draw[neighborhood] (0,0) -- (0,-1);
\fill[meshnode] (0,0) circle[radius=2pt] {};
\node[anchor=south] at (0,0) {self};
\fill[meshnode] (-1,0) circle[radius=2pt] {};
\node[anchor=south east] at (-1,0) {left node};
\fill[meshnode] (1,0) circle[radius=2pt] {};
\node[anchor=south west] at (1,0) {right node};
\fill[meshnode] (0,-1) circle[radius=2pt] {};
\node[anchor=north] at (0,-1) {bottom node};
\fill[meshnode] (-1,-1) circle[radius=2pt] {};
\node[anchor=north east] at (-1,-1) {left diag};
\fill[meshnode] (1,-1) circle[radius=2pt] {};
\node[anchor=north west] at (1,-1) {right diag};
\end{tikzpicture}
\caption{\footnotesize Important neighbors in terms of strength-of-connection of an interior node. The strength measure is symmetric and the upper neighborhood is therefore omitted.}
\label{fig:neighbor_node_naming}
\end{subfigure}
\caption{Display of the uniform, quadrilateral discretization of the test problem and the relevant neighborhood for the {\soc} measure.}
\end{figure}

%% file: figures/battery.tex
\begin{figure}
\centering
\begin{tikzpicture}

\def\width{4}
\def\height{0.5}
\def\collectorHeight{0.2}
\def\insulation{0.5}
\def\can{0.5}
\def\ambient{0.5}
\def\numLayers{20}

\draw[fill=red!50] (0,0) rectangle (\width,\height) node[midway] {Heat Pellet};

\draw[] (0,\height) rectangle (\width,\height+\collectorHeight);
\draw[fill=blue!50] (0,\height+\collectorHeight) rectangle (\width,2*\height+\collectorHeight) node[midway] {Anode};
\draw[fill=gray!50] (0,2*\height+\collectorHeight) rectangle (\width,3*\height+\collectorHeight) node[midway] {Separator};
\draw[fill=yellow!50] (0,3*\height+\collectorHeight) rectangle (\width,4*\height+\collectorHeight) node[midway] {Cathode};
\draw[] (0,4*\height+\collectorHeight) rectangle (\width,4*\height+2*\collectorHeight);

\draw[decorate,decoration={brace,amplitude=10pt,mirror,raise=4pt},yshift=0pt]
(0.7*\width+0.1, \height+0.1) -- (0.7*\width+0.1, 4*\height+2*\collectorHeight-0.1) node[midway,right, xshift=0.35cm] {\( N \)};

\draw[fill=red!50] (0,4*\height+2*\collectorHeight) rectangle (\width,4*\height+2*\collectorHeight+\height) node[midway] {Heat Pellet};

\draw[fill=gray!30] (\width,0) rectangle (\width+\insulation,4*\height+2*\collectorHeight+\height) node[midway,rotate=90] {Insulation};
\draw[fill=black!20] (\width+\insulation,0) rectangle (\width+\insulation+\can,4*\height+2*\collectorHeight+\height) node[midway,rotate=90] {Can};

\draw[] (\width+\insulation+\can,0) rectangle (\width+\insulation+\can+\ambient,4*\height+2*\collectorHeight+\height) node[midway,rotate=90] {Ambient};

\draw[dashed] (0,0) -- (0,4*\height+2*\collectorHeight+\height);
\node[rotate=90] at (-0.2,2*\height+\collectorHeight+\height/2) {Axis};

\draw[<-] (\width/2, \height/2+\height-\collectorHeight) -- (\width+1.5, \height/2+\height-\collectorHeight) node[right] {Collector};
\draw[<-] (\width/2, 4*\height+\collectorHeight+\height/2-0.5*\collectorHeight) -- (\width+1.5, 4*\height+\collectorHeight+\height/2-0.5*\collectorHeight) node[right] {Collector};

\end{tikzpicture}
\caption{2D axisymmetric simulation domain for multi-physics simulations (not to scale).
Note that the collector, anode, separator, and cathode layers are repeated $N=20$ times.}
\label{fig:thermal_battery_stack}
\end{figure}

%% file: figures/solar_cell_circuit.tex
\begin{circuitikz}
    \draw (0,0) to[isource, l=$I_L$] (0,2);
    \draw[->] (0,0) -- (0,2);

    \draw (0,2) -- (2,2)
        to[diode, l=$I_D$] (2,0) -- (0,0);
    \draw[->] (2,2) -- (2,0);

    \draw (2,2) -- (4,2)
        to[R, l=$R_s$] (6,2);

    \draw (2,2) -- (4,2)
        to[R, l=$R_{sh}$] (4,0) -- (2,0);
    \draw[->] (4,2) -- (4,0) node [pos=0.1, left] {$I_{sh}$};

    \draw (6,2) -- (7,2);
    \draw (2,0) -- (7,0);
    \draw[->] (6,2) -- (7,2) node[midway, above] {$I$};

    \node at (7,2) [anchor=west] {$+$};
    \node at (7,0) [anchor=west] {$-$};
\end{circuitikz}

%% file: bibliography.bib
@misc{4C,
  author = {{4C}},
  title = {{4C}: A {C}omprehensive {M}ultiphysics {S}imulation {F}ramework},
  howpublished = {\url{https://www.4c-multiphysics.org}},
  year = {2025},
  note = {Accessed: 29.08.2025}
}

@article{Bavier2012a,
  author = {Bavier, Eric and Hoemmen, Mark and Rajamanickam, Sivasankaran and Thornquist, Heidi},
  title = {Amesos2 and Belos: Direct and Iterative Solvers for Large Sparse Linear Systems},
  journal = {Scientific Programming},
  volume = {20},
  number = {3},
  pages = {243875},
  doi = {https://doi.org/10.3233/SPR-2012-0352},
  year = {2012}
}

@techreport{BergerVergiat2023a,
  address = {Albuquerque, NM (USA) 87185},
  author = {Berger-Vergiat, Luc and Glusa, Christian A. and Harper, Graham and Hu, Jonathan J. and Mayr, Matthias and Prokopenko, Andrey and Siefert, Christopher M. and Tuminaro, Raymond S. and Wiesner, Tobias A.},
  institution = {Sandia National Laboratories},
  number = {SAND2023-12265},
  title = {{MueLu User's Guide}},
  year = {2023}
}

@article{Bochev2003,
  author  = {Pavel B. Bochev and Jonathan J. Hu and Allen C. Robinson and Raymond S. Tuminaro},
  title   = {Towards robust 3D Z-pinch simulations: discretization and fast solvers for magnetic diffusion in heterogeneous conductors},
  journal = {Electron. Trans. Numer. Anal.},
  volume  = {15},
  year    = {2003},
  pages   = {186--210},
}

@article{Brandt2015a,
  author = {Achi Brandt and James Brannick and Karsten Kahl and Irene Livshits},
  title = {Algebraic distance for anisotropic diffusion problems: multilevel results},
  journal = {Electronic Transactions on Numerical Analysis},
  year = {2015},
  volume = {44},
  pages = {472--496}
}

@article{Brannick2006,
  author = {Brannick, J. and Brezina, M. and MacLachlan, S. and Manteuffel, T. and McCormick, S. and Ruge, J.},
  title = {An energy-based AMG coarsening strategy},
  journal = {Numerical Linear Algebra with Applications},
  volume = {13},
  number = {2-3},
  pages = {133-148},
  doi = {https://doi.org/10.1002/nla.480},
  year = {2006}
}

@article{Brezina2006a,
  author = {Brezina, M. and Falgout, R. D. and MacLachlan, S. and Manteuffel, T. and McCormick, S. F. and Ruge, J. W.},
  title = {Adaptive Algebraic Multigrid},
  journal = {SIAM Journal on Scientific Computing},
  volume = {27},
  number = {4},
  pages = {1261-1286},
  year = {2006},
  publisher = {Society for Industrial and Applied Mathematics},
  doi = {10.1137/040614402}
}

@article{Clement1975a,
  author = {Cl\'ement, Ph.},
  title = {Approximation by finite element functions using local regularization},
  journal = {Revue fran\c{c}aise d'automatique, informatique, recherche op\'erationnelle. Analyse num\'erique},
  pages = {77--84},
  publisher = {Dunod},
  address = {Paris},
  volume = {9},
  number = {R2},
  year = {1975}
}

@book{Cole2010,
  author = {Cole, Kevin and Beck, J.V. and Haji-Sheikh, A. and Litkouhi, Bahman},
  year = {2010},
  month = {01},
  pages = {1--645},
  title = {Heat conduction using green’s functions},
  publisher = {CRC Press, Taylor \& Francis Group},
  edition = {2}
}

@Inbook{Crompton1982,
  author={Crompton, T. R.},
  title={High-temperature thermally activated reserve batteries},
  bookTitle={Small Batteries: Volume 2 Primary Cells},
  year={1982},
  publisher={Macmillan Education UK},
  address={London},
  pages={202--212},
  isbn={978-1-349-06319-2},
  doi={10.1007/978-1-349-06319-2_10},
}

@article{Deveci2015,
author = {Deveci, Mehmet and Rajamanickam, Siva and Devine, Karen and Catalyurek, Umit},
year = {2015},
month = {01},
pages = {1-1},
title = {Multi-Jagged: A Scalable Parallel Spatial Partitioning Algorithm},
volume = {27},
journal = {IEEE Transactions on Parallel and Distributed Systems},
doi = {10.1109/TPDS.2015.2412545}
}

@article{Falgout2006a,
  author = {Falgout, R. D.},
  journal = {Computing in Science \& Engineering},
  title = {An introduction to algebraic multigrid},
  year = {2006},
  volume = {8},
  number = {6},
  pages = {24--33},
  doi = {10.1109/MCSE.2006.105}
}

@article{Fang2019,
  author = {Rui Fang and Martin Kronbichler and Maximilian Wurzer and Wolfgang A. Wall},
  title = {Parallel, physics-oriented, monolithic solvers for three-dimensional, coupled finite element models of lithium-ion cells},
  journal = {Computer Methods in Applied Mechanics and Engineering},
  volume = {350},
  pages = {803-835},
  year = {2019},
  doi = {https://doi.org/10.1016/j.cma.2019.03.017},
}

@article{Gee2009a,
  author = {Gee, Michael W. and Hu, Jonathan J. and Tuminaro, Raymond S.},
  title = {A new smoothed aggregation multigrid method for anisotropic problems},
  journal = {Numerical Linear Algebra with Applications},
  volume = {16},
  number = {1},
  pages = {19-37},
  doi = {https://doi.org/10.1002/nla.593},
  year = {2009}
}

@article{Green2024a,
  author = {Green, David and Hu, Xiaozhe and Lore, Jeremy and Mu, Lin and Stowell, Mark L.},
  title = {{An Efficient High-Order Solver for Diffusion Equations with Strong Anisotropy on Non-Anisotropy-Aligned Meshes}},
  journal = {SIAM Journal on Scientific Computing},
  volume = {46},
  number = {2},
  pages = {S199-S222},
  year = {2024},
  doi = {10.1137/22M1500162}
}

@article{Guidotti2006,
  author = {Ronald A. Guidotti and Patrick Masset},
  title = {Thermally activated (“thermal”) battery technology: Part I: An overview},
  journal = {Journal of Power Sources},
  volume = {161},
  number = {2},
  pages = {1443-1449},
  year = {2006},
  issn = {0378-7753},
  doi = {https://doi.org/10.1016/j.jpowsour.2006.06.013},
}

@article{Sheikh2003,
  author = {A Haji-Sheikh and J.V Beck and D Agonafer},
  title = {Steady-state heat conduction in multi-layer bodies},
  journal = {International Journal of Heat and Mass Transfer},
  volume = {46},
  number = {13},
  pages = {2363-2379},
  year = {2003},
  issn = {0017-9310},
  doi = {https://doi.org/10.1016/S0017-9310(02)00542-2},
}

@article{Heinlein,
  editor = {Ronny Ramlau, Lothar Reichel (Hg.)},
  issn = {1068-9613},
  publisher = {Verlag der Österreichischen Akademie der Wissenschaften},
  copyright = {Österreichische Akademie der Wissenschaften},
  pages = {562-591},
  title = {A frugal FETI-DP and BDDC coarse space for heterogeneous problems},
  author = {Heinlein, Alexander and Klawonn, Axel and Lanser, Martin and Weber, Janine},
  URL = {https://epub.oeaw.ac.at/?arp=0x003c1ad0},
  note = {Online available: https://epub.oeaw.ac.at/?arp=0x003c1ad0 - Last access:8.12.2025},
  address = {Wien},
  journal = {Electronic Transactions on Numerical Analysis},
  year = {2020}
}

@article{Heroux2005a,
  author = {Heroux, Michael A. and Bartlett, Roscoe A. and Howle, Vicki E. and Hoekstra, Robert J. and Hu, Jonathan J. and Kolda, Tamara G. and Lehoucq and Long, Kevin R. and Pawlowski, Roger P. and Phipps, Eric T. and Salinger, Andrew G. and Thornquist, Heidi K. and Tuminaro, Ray S. and Willenbring, James M. and Williams, Alan and Stanley, Kendall S.},
  doi = {10.1145/1089014.1089021},
  journal = {ACM Transactions on Mathematical Software},
  number = {3},
  pages = {397--423},
  title = {{An Overview of the Trilinos Project}},
  volume = {31},
  year = {2005}
}

@article{Hu2022a,
  author = {Hu, Jonathan J. and Siefert, Christopher M. and Tuminaro, Raymond S.},
  title = {{Smoothed aggregation for difficult stretched mesh and coefficient variation problems}},
  journal = {Numerical Linear Algebra with Applications},
  volume = {29},
  number = {6},
  pages = {e2442},
  doi = {https://doi.org/10.1002/nla.2442},
  year = {2022}
}

@article{Kim2017,
  author = {Hyea Hyun Kim and Eric Chung and Junxian Wang},
  title = {BDDC and FETI-DP preconditioners with adaptive coarse spaces for three-dimensional elliptic problems with oscillatory and high contrast coefficients},
  journal = {Journal of Computational Physics},
  volume = {349},
  pages = {191-214},
  year = {2017},
  issn = {0021-9991},
  doi = {https://doi.org/10.1016/j.jcp.2017.08.003}
}

@misc{Mayr2025a,
  title={Trilinos: Enabling Scientific Computing Across Diverse Hardware Architectures at Scale},
  author={Matthias Mayr and Alexander Heinlein and Christian Glusa and Siva Rajamanickam and Maarten Arnst and Roscoe Bartlett and Luc Berger-Vergiat and Erik Boman and Karen Devine and Graham Harper and Michael Heroux and Mark Hoemmen and Jonathan Hu and Brian Kelley and Kyungjoo Kim and Drew P. Kouri and Paul Kuberry and Kim Liegeois and Curtis C. Ober and Roger Pawlowski and Carl Pearson and Mauro Perego and Eric Phipps and Denis Ridzal and Nathan V. Roberts and Christopher Siefert and Heidi Thornquist and Romin Tomasetti and Christian R. Trott and Raymond S. Tuminaro and James M. Willenbring and Michael M. Wolf and Ichitaro Yamazaki},
  year={2025},
  eprint={2503.08126},
  archivePrefix={arXiv},
  primaryClass={cs.MS},
  url={https://arxiv.org/abs/2503.08126},
}

@article{Olson2010a,
  author = {Olson, Luke N. and Schroder, Jacob and Tuminaro, Raymond S.},
  title = {A new perspective on strength measures in algebraic multigrid},
  journal = {Numerical Linear Algebra with Applications},
  volume = {17},
  number = {4},
  pages = {713-733},
  doi = {https://doi.org/10.1002/nla.669},
  year = {2010}
}

@inbook{Robinson,
  author = {Allen Robinson and Thomas Brunner and Susan Carroll and Richard Drake and Christopher Garasi and Thomas Gardiner and Thomas Haill and Heath Hanshaw and David Hensinger and Duane Labreche and Raymond Lemke and Edward Love and Christopher Luchini and Stewart Mosso and John Niederhaus and Curtis Ober and Sharon Petney and William Rider and Guglielmo Scovazzi and O Strack and Randall Summers and Timothy Trucano and V Weirs and Michael Wong and Thomas Voth},
  title = {ALEGRA: An Arbitrary Lagrangian-Eulerian Multimaterial, Multiphysics Code},
  booktitle = {46th AIAA Aerospace Sciences Meeting and Exhibit},
  chapter = {},
  pages = {},
  doi = {10.2514/6.2008-1235},
  URL = {https://arc.aiaa.org/doi/abs/10.2514/6.2008-1235},
  eprint = {https://arc.aiaa.org/doi/pdf/10.2514/6.2008-1235}
}

@article{Robinson2024,
  author = {Allen C. Robinson and Richard R. Drake and Christopher B. Luchini and Ramón J. Moral and John H.J. Niederhaus and Sharon V. Petney},
  title = {An MPMD approach coupling electromagnetic continuum mechanics approximations in ALEGRA},
  journal = {Computer Methods in Applied Mechanics and Engineering},
  volume = {429},
  pages = {117164},
  year = {2024},
  issn = {0045-7825},
  doi = {https://doi.org/10.1016/j.cma.2024.117164},
}

@misc{Trilinos,
  author = {{Trilinos}},
  howpublished = {https://trilinos.github.io},
  year = {2025},
  note = {Accessed: 29.08.2025},
  title = {{The Trilinos Project}}
}

@article{Vanek1996a,
  author = {Van\v{e}k, Petr and Mandel, Jan and Brezina, Marian},
  journal = {Computing},
  pages = {179--196},
  title = {{Algebraic Multigrid By Smoothed Aggregation For Second And Fourth Order Elliptic Problems}},
  volume = {56},
  year = {1996}
}

@article{Vanek2001a,
  author = {Van\v{e}k, Petr and Brezina, Marian and Mandel, Jan},
  journal = {Numerische Mathematik},
  pages = {559--579},
  title = {{Convergence of algebraic multigrid based on smoothed aggregation}},
  volume = {88},
  year = {2001}
}

@article{Voskuilen2021a,
  title = {Multi-fidelity electrochemical modeling of thermally activated battery cells},
  author = {Voskuilen, Tyler G and Moffat, Harry K and Schroeder, Benjamin B and Roberts, Scott A},
  journal = {Journal of Power Sources},
  volume = {488},
  pages = {229469},
  year = {2021},
  publisher = {Elsevier}
}

@unpublished{Phillips2026a,
  title = {{Multi-physics Preconditioning for Thermally Activated Batteries}},
  author = {Phillips, Malachi Timothy},
  institution = {Sandia National Laboratories},
  note = {In preparation}
}

@article{Cyr2016a,
  title = {{Teko: A block preconditioning capability with concrete example applications in Navier--Stokes and MHD}},
  author = {Cyr, Eric C and Shadid, John N and Tuminaro, Raymond S},
  journal = {SIAM Journal on Scientific Computing},
  volume = {38},
  number = {5},
  pages = {S307--S331},
  year = {2016},
  publisher = {SIAM}
}

@article{gray2011physics,
  title = {The physics of the solar cell},
  author = {Gray, Jeffery L},
  journal = {Handbook of photovoltaic science and engineering},
  volume = {2},
  pages = {82--128},
  year = {2011},
  publisher = {Wiley Online Library}
}

@article{goudelis2022review,
  title = {A review of models for photovoltaic crack and hotspot prediction},
  author = {Goudelis, Georgios and Lazaridis, Pavlos I and Dhimish, Mahmoud},
  journal = {Energies},
  volume = {15},
  number = {12},
  pages = {4303},
  year = {2022},
  publisher = {MDPI}
}

@article{shang2017photovoltaic,
  title = {Photovoltaic devices: opto-electro-thermal physics and modeling},
  author = {Shang, Aixue and Li, Xiaofeng},
  journal = {Advanced Materials},
  volume = {29},
  number = {8},
  pages = {1603492},
  year = {2017}
}

@article{ramezani2025shading,
  title = {Shading impact modeling on photovoltaic panel performance},
  author = {Ramezani, Faeze and Mirhosseini, Mojtaba},
  journal = {Renewable and Sustainable Energy Reviews},
  volume = {212},
  pages = {115432},
  year = {2025},
  publisher = {Elsevier}
}

@incollection{BrMcRu84,
  author = {A. Brandt and S. McCormick and J. Ruge},
  title = {Algebraic multigrid ({AMG}) for sparse matrix equations},
  editor = {D. Evans},
  booktitle = {Sparsity and its applications},
  publisher = {Cambridge University Press},
  address = {Cambridge},
  year = {1984},
  pages = {257--284}
}

@incollection{RuSt85,
  author = {J. Ruge and K. St\"{u}ben},
  editor = {S. McCormick},
  title = {Algebraic multigrid ({AMG})},
  booktitle= {Multigrid Methods},
  series = {Frontiers in Applied Mathematics},
  volume = {3},
  publisher= {SIAM},
  address = {Philadelphia},
  year = {1985},
  pages = {73--130}
}

@inproceedings{ddproc06,
  author = {M. Sala and P. Lin and R. Tuminaro and J. Shadid},
  title = {Algebraic Multilevel Preconditioners for non-symmetric {PDE}s on Stretched Grids},
  booktitle = {Domain Decomposition Methods in Science and Engineering XVI},
  series = {Lecture Notes in Computational Science and Engineering},
  editor = {Olof B. Widlund and David E. Keyes},
  volume = {55},
  publisher = {Springer-Verlag},
  pages = {739--746},
  year = {2006}
}

@TechReport{ML,
  author = {M.W. Gee and C.M. Siefert and J.J. Hu and R.S. Tuminaro and M.G. Sala},
  title = {{ML} 5.0 Smoothed Aggregation User's Guide},
  institution = {Sandia National Laboratories},
  year = {2006},
  number = {SAND2006-2649}
}

@article{Patankar1972a,
  author = {Patankar, S. V. and Spalding, D. B.},
  journal = {International Journal of Heat and Mass Transfer},
  number = {10},
  pages = {1787--1806},
  title = {A calculation procedure for heat, mass and momentum transfer in three-dimensional parabolic flows},
  volume = {15},
  year = {1972}
}

@article{Chorin1968a,
  author = {Chorin, Alexander Joel},
  journal = {Mathematics of Computation},
  pages = {745--762},
  title = {{Numerical solution of the Navier-Stokes equations}},
  volume = {22},
  year = {1968}
}

@article{Temam1969a,
  author = {T\'emam, R.},
  doi = {10.1007/BF00247678},
  journal = {Archive for Rational Mechanics and Analysis},
  number = {2},
  pages = {135--153},
  title = {Sur l'approximation de la solution des {\'e}quations de Navier-Stokes par la m{\'e}thode des pas fractionnaires (I)},
  volume = {32},
  year = {1969}
}

@article{Austin2021a,
  title = {Initial guesses for sequences of linear systems in a GPU-accelerated incompressible flow solver},
  author = {Austin, Anthony P and Chalmers, Noel and Warburton, Tim},
  journal = {SIAM Journal on Scientific Computing},
  volume = {43},
  number = {4},
  pages = {C259--C289},
  year = {2021},
  publisher = {SIAM}
}

@article{Fischer1998a,
  title = {Projection techniques for iterative solution of Ax=b with successive right-hand sides},
  author = {Fischer, Paul F},
  journal = {Computer Methods in Applied Mechanics and Engineering},
  volume = {163},
  number = {1-4},
  pages = {193--204},
  year = {1998},
  publisher = {Elsevier}
}

@misc{cut-drop,
      title={Improving Smoothed Aggregation AMG Robustness on Stretched Mesh Applications}, 
      author={Chris Siefert and Raymond Tuminaro and Daniel Sunderland},
      year={2026},
      eprint={2601.20119},
      archivePrefix={arXiv},
      primaryClass={math.NA},
      url={https://arxiv.org/abs/2601.20119}, 
}
